\documentclass[aps,pra,epsf,superscriptaddress,amsmath,amssymb,amsfonts,showpacs,nofootinbib,notitlepage]{revtex4-1}

\usepackage{graphicx}        

\usepackage{amsmath,amssymb,bm}
\usepackage{hyperref}
\usepackage{xcolor}

\usepackage{url} 

\begin{document}

\title{Extreme (Rogue) Waves: From Theory to Experiments\\ in Ultracold Gases and Beyond}

\author{Amin Chabchoub}
\affiliation{Marine Physics and Engineering Unit, Okinawa Institute of Science and Technology, Onna-son, Okinawa 904-0495, Japan}

\author{Peter Engels}
\affiliation{Department of Physics and Astronomy, Washington State University, Pullman, Washington 99164-2814}

\author{P.~G. Kevrekidis}
\affiliation{Department of Mathematics and Statistics, University of Massachusetts Amherst, Amherst, MA 01003-4515, USA}

\author{Simeon I. Mistakidis}
\affiliation{Department of Physics and LAMOR, Missouri University of Science and Technology, Rolla, MO 65409, USA}

\author{Garyfallia C.~Katsimiga}
\affiliation{Department of Physics and LAMOR, Missouri University of Science and Technology, Rolla, MO 65409, USA}

\author{Maren E.~Mossman}
\affiliation{Department of Physics and Biophysics, University of San Diego, San Diego, CA 92110}

\author{Sean Mossman}
\affiliation{Department of Physics and Biophysics, University of San Diego, San Diego, CA 92110}

\begin{abstract}
{In this Chapter, we review key theoretical and experimental advances in the study of extreme nonlinear wave events, called rogue waves (RWs), in both single-component attractively interacting and two-component repulsive mixtures of ultracold quantum gases. 
Starting from the exact rational solutions of the integrable focusing nonlinear Schr\"odinger model, the hierarchy of RW solutions is exemplified. These range from the Peregrine soliton (PS) and, related to it, the destabilization into a multi-peak cascade of PSs dubbed “Christmas-tree’’, to the Akhmediev breather, and Kuznetsov–Ma soliton as well as higher-order RWs. 
Emphasis is placed on their controllable dynamical emergence and characteristics in non-integrable quantum many-body systems described by Gross-Pitaevskii models and extensions thereof through different protocols such as modulational instability, gradient catastrophe, and dam-break flows. 
We further discuss how immiscible particle-imbalanced repulsive mixtures can be cast into effective attractive single-component environments capable of hosting RWs. 
Next, state-of-the-art experimental techniques are summarized within the ultracold realm that can be utilized to realize solitary waves, modulational instability, dispersive shock waves and RWs including the very recent first experimental observation of the PS, enabled through engineered effective focusing interactions and precise dynamical triggering.
Observations of these extreme events in water waves,
nonlinear optics and beyond are also outlined, highlighting their broader relevance and potential of emergence in disparate physical settings. 
Our exposition aims at showcasing ultracold atomic gases as versatile platforms for controllably generating and probing extreme nonlinear events, among others, in the quantum realm across integrable and non-integrable settings.
}
\end{abstract}


\maketitle



\section{Introduction}

Rogue waves (RWs) are extreme, large-amplitude nonlinear excitations that 
``appear from nowhere and disappear without a trace''~\cite{akhmediev2009waves}. 
Their defining traits include i) an amplitude several times larger than the ambient medium, and ii) a sharp spatial/temporal localization. 
These characteristics made RWs a central theme of modern nonlinear science. 
Although they have been argued to exist for much longer,
and even appear in artistic renderings such as Hokusai's famous
Great Wave off Kanagawa, created in the 1830s,
RWs were first thrust into prominence by the 1995 Draupner measurement in an offshore platform in the North Sea~\cite{kharif2008rogue}. This so-called New Year's Wave (measured by a laser wave sensor on 1/1/1995) provided incontrovertible evidence of extremely large ocean waves. 
Since then, controlled laboratory observations have emerged
in hydrodynamic wave tanks~\cite{Chabchoub_water,Chabchoub_water1}, optical fibers~\cite{kibler2010peregrine,solli2007optical,dudley2014instabilities}, plasma systems~\cite{Bailung_plasmas}, superfluid helium~\cite{Ganshin}, capillaries~\cite{Shats}, and very recently in ultracold atoms~\cite{Romero_Ros_2024,bougas2025observation} all confirming that these RW extreme events can 
arise naturally in nonlinear dispersive media. 
Their universal appearance in diverse fields has led to intense efforts toward unfolding the corresponding principles that govern RW formation, morphology and mathematical description. 

While a more philosophical discussion is ongoing regarding
the linear vs. nonlinear character of these RW events, 
herein we will focus predominantly on the nonlinear perspective
that has been a driving force behind their ultracold atomic
system realization. 
Accordingly, the major theoretical model describing weakly dispersive nonlinear waves in a plethora of  physical platforms is the so-called focusing (attractive) nonlinear Schr\"odinger equation (NLS)~\cite{sulem2007nonlinear,fibich2015nonlinear}. NLS admits  a complete analytical hierarchy of the RW family as exact rational solutions and also possesses families of 
temporally/spatially periodic solutions of which these 
rational solutions are special members. 
These waveforms include i) the fundamental Peregrine soliton (PS)~\cite{Peregrine1983}, namely a doubly localized in space and time waveform, ii) the spatially periodic Akhmediev (AB) breather~\cite{akhmediev2009waves} and iii) the time-periodic Kuznetsov–Ma (KM) soliton~\cite{kuznetsov1977solitons,ma1979perturbed}. 
Interestingly, the AB and KM structures represent finite period precursors whose infinite period limit yields the PS. 
Higher-order RW (HORW), namely  rational solutions featuring a multi-lobe structure along with larger peak amplitudes have also been predicted in different contexts~\cite{dudley2014instabilities,dubard2010multi,dubard2011multi,kedziora2012second,kedziora2013classifying}. 
A systematic characterization of these structures has been 
developed in the works  of~\cite{bilman2020extreme,bilman2022broader}
building up to the infinite order RW  that first
arose in the work of~\cite{suleimanov}.

RW formation is inherently related to the phenomenon of modulational instability (MI)~\cite{zakharov2009modulation,zakharov2013nonlinear,biondini2016universal}, also known as the Benjamin-Feir instability~\cite{benjamin1967disintegration}, that naturally appears in the focusing NLS model~\footnote{It should be noted here that although MI is generically expected to be a 
necessary condition for the formation of RWs in this class
of dispersive nonlinear models, it is
not clear at the moment whether it is also a sufficient one.}. 
MI refers to the exponential growth of weak perturbations on a finite background of a focusing medium and is responsible for the generation of numerous nonlinear excitations, such as bright solitons,  breathers, shock waves, and RWs. 
In addition to MI, which generates RWs on a finite background,
these waveforms have been argued to spontaneously {\it locally}
emerge via a variety of other mechanisms in NLS-like 
settings. These include the gradient catastrophe, originally discussed in dispersive hydrodynamics of integrable systems~\cite{dubrovin2008universality,bertola2013universality}, as well as the interfering dam-break flows generated by a pair of celebrated Riemann problems~\cite{el2016dam} leveraging step-like initial conditions. 
In the former mechanism, practically speaking, sufficiently
wide initial conditions undergo focusing, eventually resulting
in the local formation of a PS and more broadly in the formation
of a cascade of PS structures (the so-called Christmas tree)
at the poles of the famous 
 tritronqu\'ee solution of the Painlev\'e~\(\mathrm{I}\) equation~\cite{bertola2013universality}.
On the other hand, a pair of 
 dam-breaks arising from Riemann problems produces two dispersive shock waves (DSWs) whose interference
also leads to PS formation~\cite{el2016dam}.
These mechanisms have been experimentally realized in optics~\cite{kibler2010peregrine,solli2007optical}, water waves~\cite{Chabchoub_water,Chabchoub_water1}, and ultracold atoms~\cite{Romero_Ros_2024,Mossman_nonlinearMI,Shashwat}, and together with MI constitute
the backbone of modern RW theory in both integrable and non-integrable models.

Ultracold atoms, particularly Bose–Einstein condensates (BECs), constitute intriguing settings
for exploring RW phenomena.  
Indeed, their exquisite controllability in terms of system parameters such as the interaction strength tuned via Feshbach resonances~\cite{chin2010feshbach}, dimensionality~\cite{wolswijk2025trapping}, external confinement~\cite{henderson2009experimental}, and realization of different internal states~\cite{Kawaguchi2012} facilitates engineering of  RWs. 
Early experiments demonstrated MI~\cite{strecker2002formation}, bright solitons and trains thereof~\cite{khaykovich2002formation,marchant2013controlled,khaykovich2008bright,strecker2003bright}, 
and numerous refinements of both aspects have emerged
more recently. In particular, the interaction of
bright solitons was quantified in~\cite{Nguyen2014Collisions},
the MI~\cite{Robbins} and the resulting soliton wavetrains~\cite{nguyen2017formation} from it were
further studied, while the excitation modes of bright
solitons~\cite{di_carli_excitation_2019} and the
emergence of matter-wave breathers~\cite{luo_creation_2020}
were also explored. 
In the last few years, a sequence of landmark experiments has reported the observation of the Townes soliton~\cite{Chen_MI,ChenHung2021ScaleInvPRL,Bakkali_realization_2021,bakkali_townes_2022}, but also 
the PS RW and vector variants thereof in two- and multi-component immiscible mixtures have been generated~\cite{Romero_Ros_2024,bougas2025observation}.
While some of these works have examined directly 
the attractive, self-focusing regime~\cite{Chen_MI,ChenHung2021ScaleInvPRL} and have
even leveraged MI towards the creation of such 
higher-dimensional solitons~\cite{Chen_MI}, others
have produced ``effective attraction'',
by
exploiting the reduction of the underlying repulsive model to a suitable set of effective attractive ones~\cite{Bakkali_realization_2021,bakkali_townes_2022,Romero_Ros_2024,bougas2025observation}; see Chapter 4 for detailed discussions on this scheme.  
Hence, ultracold dilute gases are capable of bridging classical NLS dynamics with quantum many-body effects, materializing the generation of RWs in regimes far beyond those accessible in fluids and optics.

Our presentation of some of the key developments
in this blossoming field within
this chapter is organized as follows. 
In Section~\ref{sec:theory}, we survey the theoretical frontiers of RW physics in general NLS type systems and  ultracold quantum gases. 
We begin by reviewing the hierarchy of RW solutions of the 1D focusing NLS model and the mechanisms that dynamically seed them, including MI, gradient catastrophes, and dam-break flows. 
Moreover, a brief discussion on first attempts on the stability properties of these structures through Floquet analysis are discussed. 
Next, the RW framework is extended to repulsive two-component mixtures, where effective focusing behavior enables RW formation. 
Additionally, we elaborate on recent developments on HORWs within both the standard and the extended Gross-Pitaevskii models. 
Motivated by the theory developments, Section~\ref{sec:experiment} elucidates the current experimental landscape that led to the observation of RW dynamics in a two-component BEC, extending earlier observations in water waves and nonlinear optics. These experiments continue a long line of research into solitonic excitations in a BEC, a topic that has been of high interest to the BEC community since the early days of 
BEC studies~\cite{PitaevskiiStringari2016}. 
Finally, Section~\ref{sec:Generalizations} provides a broader perspective on RWs and extreme events across diverse physical platforms, ranging from optics, hydrodynamics and plasmas to other nonlinear dispersive systems. 
A summary along with a discussion on future perspectives and long-term goals of the field is provided in Section~\ref{sec:conclusions}.

\section{Theoretical Considerations of RWs in ultracold quantum gases}
\label{sec:theory} 

\subsection{RW solutions of the 1D NLS model and their dynamical seeding}

As discussed above, different members of the RW family are the temporal KM breather~\cite{kuznetsov1977solitons,ma1979perturbed}, the spatial AB  breather~\cite{akhmediev2009waves} and the PS~\cite{Peregrine1983}, namely a spatiotemporal (thus doubly) localized structure.
All of the above waveforms exist as analytical solutions of the 1D focusing (genuinely attractive), integrable NLS equation, see also Chapter 2 for further discussion of the relevant model. In dimensionless form ($\hbar=m=1$) it reads 
\begin{equation}
i \frac{ \partial \Psi}{\partial t} = -\frac{1}{2}\frac{\partial^2 \Psi}{\partial x^2} - g|\Psi|^2 \Psi, 
\label{eq:NLS}
\end{equation}
with $\Psi(x,t)$ denoting the 1D mean-field wave function. 
In the context of BECs, Eq.~\eqref{eq:NLS} is referred to as the mean-field Gross-Pitaevskii model featuring  attractive interactions in the absence
of an external potential. 
Below, we set $g=1$ without loss of generality unless stated otherwise. 
The aforementioned RW solutions of the NLS model have the explicit analytical expressions~\cite{karjanto2021peregrine}: 
\begin{subequations}
\begin{gather}
\Psi_{\mathrm{KM}}(x,t)=
\left[
1 +\,\frac{A^3 \cos(\Omega_1 t)+2iA \Omega_1 \sin(\Omega_1 t)}
 { 2 A \cos(\Omega_1 t)-2\Omega_1 \cosh(A x)}
\right]e^{it},  
\label{eq:KM}
\\
\Psi_{\mathrm{AB}}(x,t)=
\left[
1 -\,\frac{B^3 \cosh(\Omega_2 t)+2iB \Omega_2 \sinh(\Omega_2 t)}
 { 2 B \cosh(\Omega_2 t)-2\Omega_2 \cos(B x)}
\right]e^{it}, 
\label{eq:AB}
\\
\Psi_{\mathrm{PS}}(x,t)
= 
\left[
1 - \frac{4(1+2i t)}{1+4x^2+4t^2}
\right]e^{it},
\label{eq:Peregrine}
\end{gather}
\end{subequations}
 where $\Omega_1=2A \sqrt{4+A^2 }$ and $\Omega_2 =2B \sqrt{4- B ^2 }$ represent the modulation growth rates~\cite{karjanto2021peregrine} of the KM and AB breathers respectively. 
Intriguingly, these solutions can be written
in the form of a single family through the
analytic continuation $A=i B$ with 
the PS representing the special limit of
$\Omega_i \rightarrow 0$, in which the 
computation of the application of l'Hospital's rule
gives rise to the algebraic localization of the PS. 

\begin{figure*}[t]
\centering
\includegraphics[trim={0cm 12cm 0cm 10cm}, clip,width=1.0\textwidth]{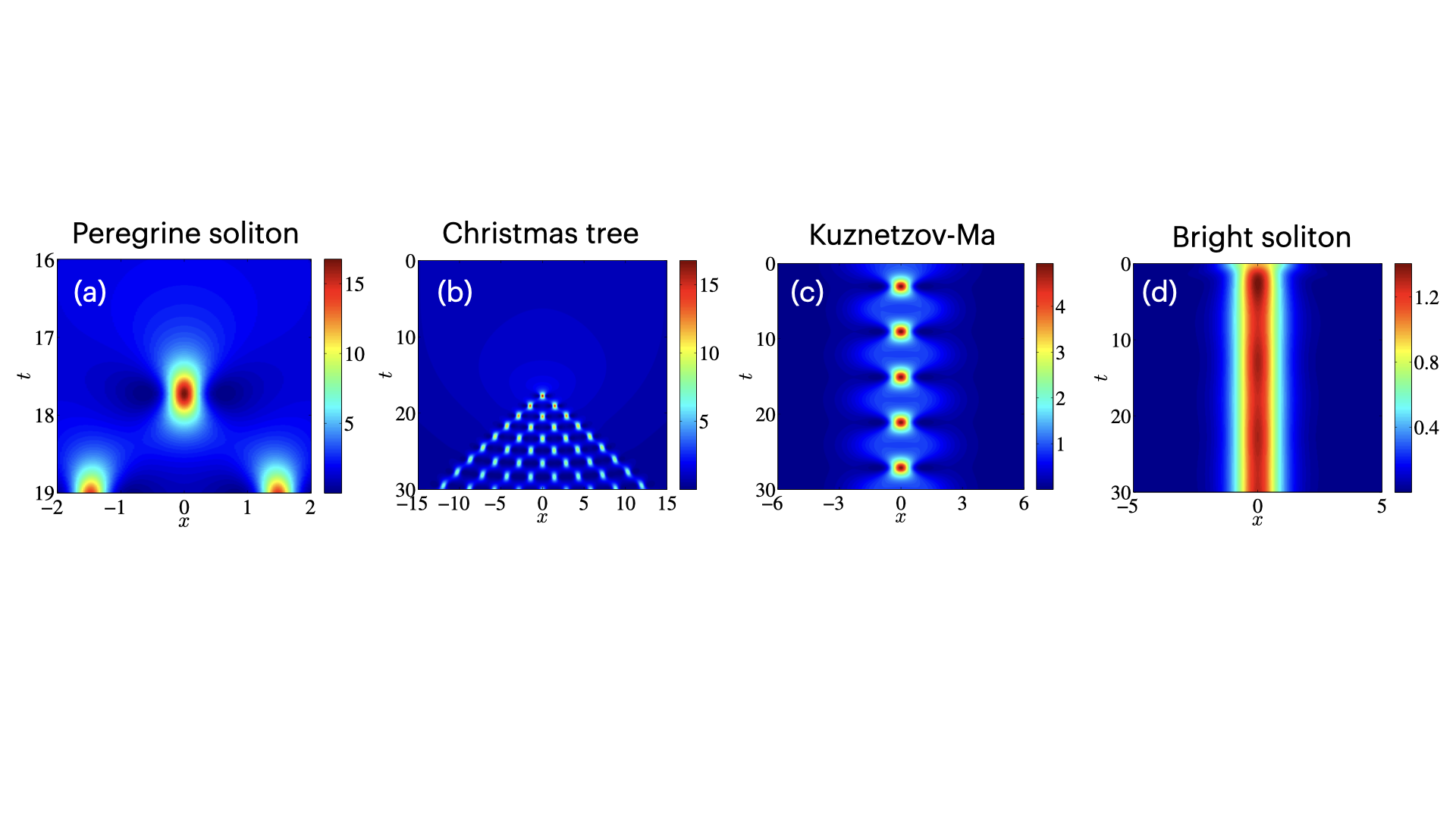}
\caption{Spatiotemporal mean-field density evolution using the Gaussian initial condition of Eq.~(\ref{eq:Gaussian_cond}) for fixed $\mathcal
{A}=1$ and different widths (a) $\sigma=30$, (b) $\sigma=30$, (c) $\sigma=2.5$ and (d) $\sigma=1.3$. Panel (a) is a magnification of (b) depicting the first PS formation before its destabilization due to gradient catastrophe into a Christmas tree configuration. Reducing the width of the Gaussian ansatz yields either (c) a time-periodic KM breather or (d) to a bright soliton. Adapted from~\cite{charalampidis2016rogue}. }
\label{Fig:RW_configurations}
\end{figure*}

Even though bright solitons~\cite{strecker2002formation,khaykovich2002formation,Cornish_2006} and multi-soliton extensions thereof~\cite{nguyen2017formation} as well as collapse phenomena have been experimentally demonstrated in 
numerous attractive condensate experiments, the 
laboratory realization of RWs was an open problem until very recently~\cite{Romero_Ros_2024,bougas2025observation} due to the challenge of preparing the MI background. 
While tailored optical potentials provide the versatility to prepare initial conditions, one has to still ``combat'' the (controllability of) unstable MI
dynamics towards observing PSs. 
For this reason, it was
realized that mechanisms such as the above mentioned
gradient catastrophe or dam-break-inducing interfering
DSWs, producing local PS events were more likely to be
successful. It is worth noting that
prior theoretical studies  often used idealized conditions to explore RW phenomena in single-component~\cite{wen2011matter,he2014rogue,LoombaRW,ManikandanRW}, binary~\cite{bludov2010vector,BaronioRWs,Mareeswaran_bin_RWs,Manikandan_bin_RWs,mareeswaran2016superposed}, and spinor BECs~\cite{Qin_spinRWs,Zhao_spinRWs,Shen_spinRWS}. 
With a view towards more experimentally tractable 
settings, it was
demonstrated in the works  of~\cite{charalampidis2016rogue,Romero_theor} that an efficient way to dynamically generate different RW events is to prepare the gas utilizing generic Gaussian initial conditions of the form: 
\begin{equation}
\Psi(x,0)= \mathcal{A} \,\exp\!\left(-\frac{x^2}{2\sigma^2}\right),
\label{eq:Gaussian_cond}
\end{equation}
with $\mathcal{A}$ and $\sigma$ denoting the amplitude and width of the Gaussian wave function. 
This Gaussian approach connects with the so-called
semiclassical NLS theory~\cite{bertola2013universality}, wherein gradient-catastrophe dynamics give rise to universal patterns and a distinctive hierarchical ``Christmas-tree'' cascade of Peregrine recurrences.

Based on the integrable NLS, a plethora of distinct regimes has been showcased~\cite{charalampidis2016rogue} under monoparametric variations of the Gaussian's width. 
In particular, these involve: i) A PS like waveform [Eq.~(\ref{eq:Peregrine})] for large $\sigma$ where the Gaussian resembles a nearly flat background (at least in the central region) followed by the nucleation of a cascade of an expanding multi-peak bright $N$-soliton pattern dubbed a “Christmas-tree’’, see Fig.~\ref{Fig:RW_configurations}(a), (b).  
ii) For intermediate $\sigma$'s the ensuing dynamical response entails the  transition toward a time periodic breather-like
pattern, see Fig.~\ref{Fig:RW_configurations}(c). 
iii) For small $\sigma$'s the Gaussian initial condition yields a single bright soliton configuration [Fig.~\ref{Fig:RW_configurations}(d)]. 
This path marks the ``Peregrine-to-soliton'' transition.
Notably, persistence of the PS structure was inferred through inclusion of random noise, and integrability breaking either by higher-order nonlinearity or consideration of higher-dimensions. In the latter, it remains intriguing to
explore the interplay of the gradient catastrophe mechanism
with the formation of scale-invariant Townes-like
waveforms~\cite{ChenHung2021ScaleInvPRL,Bakkali_realization_2021}.

\subsection{Stability analysis of the PS through the KM}

A key open question when dealing with RWs concerns their underlying spectral stability. 
This is a notorious problem even for the fundamental PS structure since it is aperiodic and  localized in time. 
Indeed, early on, various attempts more akin to MI/soliton linearization were utilized~\cite{VanGorder2014Peregrine,AlKhawaja2014PeregrineMI}, which were, arguably, less tailored to the special 
features of the PS.
However, in the work of Ref.~\cite{Cuevas_stability_PS} exploiting the connection of the PS with the KM breather (being temporally periodic) in the infinite-period limit allowed for the usage of Floquet theory for the KM as a way to tackle the stability of the PS. 
The latter involves the stability of periodic orbits,
via the solution of the so-called variational equation~\cite{GuckenheimerHolmes1983}.
The  fundamental solution at the
period is the so-called monodromy matrix, 
and the eigenvalues of this matrix at the
period of the orbit, the so-called Floquet multipliers,
determine the stability of the periodic orbit.

This analysis was conducted within the focusing NLS model of Eq.~(\ref{eq:NLS}) which admits the exact KM solution of Eq.~(\ref{eq:KM}) whose temporal frequency $\Omega_1$ determines the period. Importantly, this lends itself
to a limiting process connection to the PS:
recall that in the limit of $\Omega_1 \to 0$, the KM solution  converges to the PS described by Eq.~(\ref{eq:Peregrine}). 
Linearizing around the time-periodic KM and computing the Floquet multipliers associated with the relevant monodromy matrix~\cite{coddington1956theory,arnold1992ordinary} reveals that for all $\Omega_1$, there exist in the spectrum real pairs of unstable Floquet modes rooted in the MI of the plane-wave background. 
In the associated computational implementation, 
the monodromy matrix was computed via numerical integration of the variational equations using an exponential time-differencing fourth-order Runge-Kutta method (ETDRK4), well suited to the stiffness arising near the breather core. 
The MI of the plane-wave background is dictated by the dispersion relation $\Omega_0 = \pm \frac{1}{2} k\sqrt{k^2 - 4}$ which predicts destabilization for wavenumbers $|k|<2$. 
Here, $\Omega_0$ is related to the Floquet multipliers ($\lambda$) according to $\lambda=\exp(2i \pi \Omega_0/\Omega_1)$. 
Moreover, in the Floquet spectrum there are three multipliers reflecting the symmetries of the system, i.e., global phase invariance, spatial translation, and temporal periodicity. 
All unstable modes identified originate from the MI of the background. 
This outcome persists even for $\Omega_1 \to 0$ where the PS is retrieved. 
As a result, the destabilization of the PS which becomes stronger around its core stems solely from its interaction with the inherently unstable plane-wave background
(rather than from localized eigenmodes reflecting the
instability of the Peregrine). This is a sort of 
``mixed'' result implying that on the
one hand, the PS is not responsible for additional instabilities. 
Yet, on the other hand, its presence
enhances the growth rates of the existing instabilities
associated with the modulationally unstable background,
i.e., it catalyzes the associated destabilization which is seeded
at its location.

The above-discussed Floquet analysis predictions were confirmed  via direct numerical simulation of the 
focusing NLS
model.
Specifically, it was found that upon perturbing the KM breather with the unstable spatially even  eigenmodes led to modulations in the breather's amplitude and temporal frequency. 
On the other hand, odd-symmetry perturbations induced small oscillations of the breather around the origin. 
In all cases, the breather remained structurally recognizable, while the background deformation was observed to
grow, in accordance to what was expected on the basis
of MI.

\subsection{RWs in 1D repulsive mixtures}\label{1D_repul_RWs}

Similar semi-classically-motivated  Gaussian initial conditions have been employed to generate the PS solution but in repulsively interacting  immiscible two-component BECs~\cite{Romero_theor} avoiding in this way complications stemming from wave collapse within attractive condensates. 
This work demonstrated that multicomponent repulsive BECs offer viable and experimentally accessible/controllable platforms for exploring extreme nonlinear wave phenomena traditionally associated with focusing NLS models. 
In particular, wide Gaussian wave packets but also direct imprinting of the analytical waveform of Eq.~(\ref{eq:Peregrine}) were proven to be sufficient for PS production. 

Two-component 1D BECs, again in the absence of
an external potential, are described by the dimensionless coupled set of GPEs 
\begin{subequations}
\begin{gather}
i \frac{ \partial \Psi_1}{\partial t} = -\frac{1}{2m_1} \frac{\partial^2 \Psi_1}{\partial x^2} + \left( g_{11}|\Psi_1|^2 + g_{12}|\Psi_2|^2 \right) \Psi_1,
\label{eq:2comp_gpe1}
\\
i \frac{ \partial \Psi_2}{\partial t} = -\frac{1}{2m_2} \frac{\partial^2 \Psi_2}{\partial x^2} + \left( g_{22}|\Psi_2|^2 + g_{12}|\Psi_1|^2 \right) \Psi_2,
\label{eq:2comp_gpe2}
\end{gather}
\end{subequations}
where $\Psi_i(x,t)$, $m_i$ are the 1D mean-field wave function and mass of each component ($i=1,2$) of the mixture. 
Additionally, $g_{ij}$ refer to the intracomponent ($i=j$) and intercomponent ($i \neq j$) 1D effective interaction strengths. 
This model supports a variety of nonlinear structures, including dark-bright solitons, domain walls, and 
immiscibility-induced patterns~\cite{kevrekidis2015defocusing,KevrekidisFrantzeskakis2016Review}.  
Crucially, within the immiscible regime of interactions ($g_{12}^2>g_{11}g_{22}$) and under strong intercomponent particle imbalance, the above system can be reduced to an effective attractive single-component mean-field model  describing the minority species $\Psi_2(x,t)$, see also Chapter 4 for an elaborated discussion on the general reduction scheme for $N$ components. 
This effective single-component model, describing the evolution of the minority species, acquires the form
\begin{equation}
i \frac{\partial \Psi_2}{\partial t} = -\frac{1}{2} \frac{\partial^2 \Psi_2}{\partial x^2} + g_{\mathrm{eff}} |\Psi_2|^2 \Psi_2,
\label{NLSEq}
\end{equation}
where the effective coupling (in this simplest
case where one goes from two- to one-component) reads 
\begin{equation}
g_{\mathrm{eff}} = g_{22} - \frac{g_{12}^2}{g_{11}}.
\label{eq:geff}
\end{equation}
Interestingly, for $g_{\mathrm{eff}} < 0$, despite the fact that all interactions are repulsive, the minority component experiences  \emph{effective focusing} interactions. 
It is important to appreciate that this condition
is the same as that of {\it immiscibility} between
the two components~\cite{AoChui1998Miscibility},
a feature that has been realized initially
by~\cite{DuttonClark2005} and was leveraged by~\cite{Bakkali_realization_2021,bakkali_townes_2022} towards the 
creation of the Townes soliton 
(see also Chapters 3 and 4 of this volume)
and by~\cite{Romero_Ros_2024}
towards the creation of the PS.

\begin{figure*}[t]
\centering
\includegraphics[trim={0cm 1cm 0cm 0cm}, clip,width=1.0\textwidth]{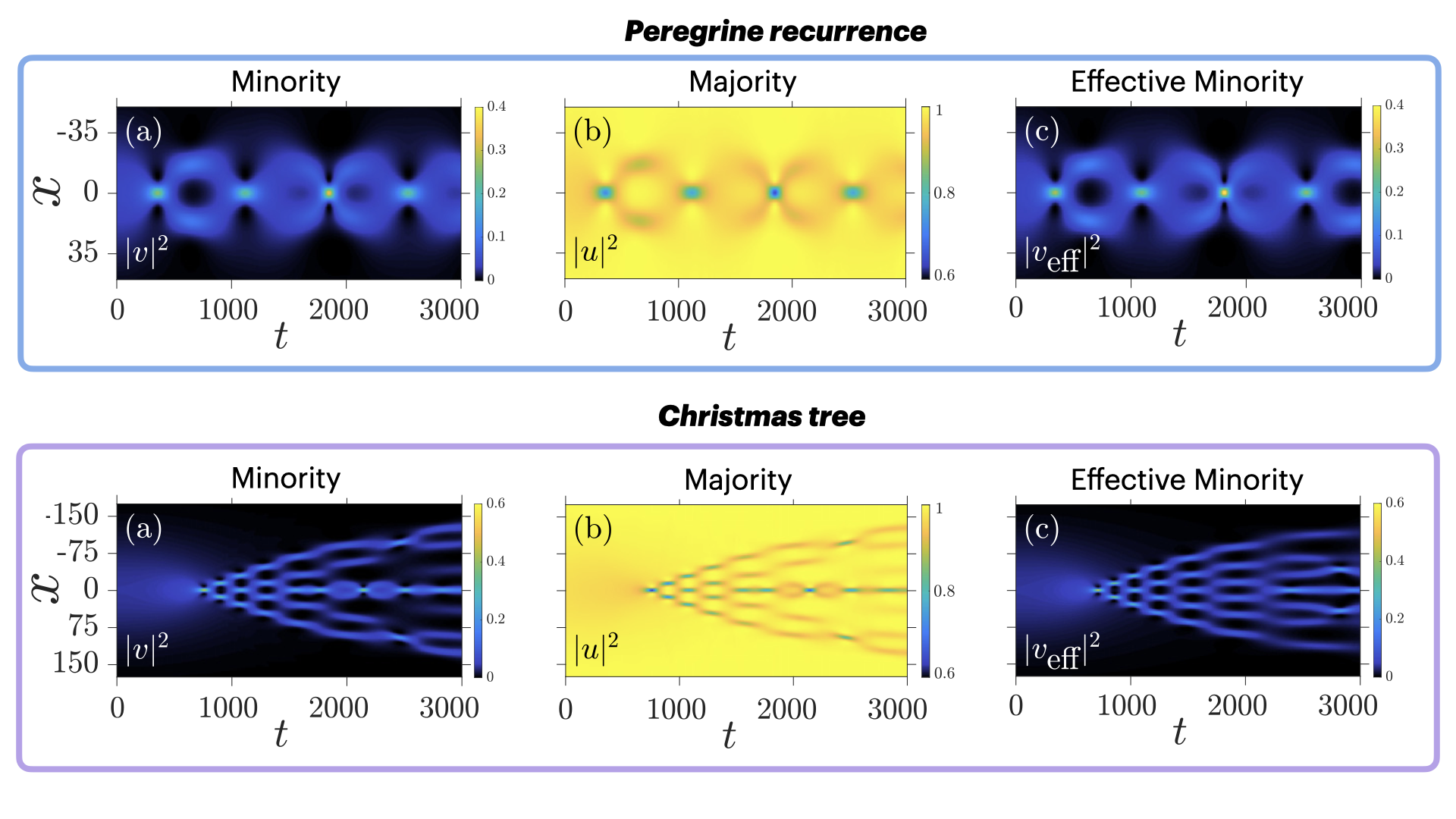}
\caption{(Upper block of panels) Time-evolution of the mean-field density of a two-component highly imbalanced mixture using an initial Gaussian wave packet having amplitude $\mathcal{A}= 0.2$ and width $\sigma = 50$ presenting the dynamics of the (a) minority and (b) majority species with $\mu = 1$. 
(c) Evolution of the effective single-component minority species. 
PS formation occurs at (a) $t_{{\rm two}} = 1846$ and (c) $t_{{\rm single}} = 1808$. 
(Lower block of panels) Density dynamics of (a), (b) the mixture and (c) its effective reduction for significantly broader Gaussian ansatz with $\mathcal{A}= 0.2$ and width $\sigma = 150$. 
Here, PS nucleation takes place at $t_{{\rm two}} = 758$ and $t_{{\rm single}} = 716$, respectively before its destabilization into the Christmas-tree configuration. 
In both cases an overall agreement is evident between the reduced single-component and the actual minority species of the two-component mixture, with the majority exhibiting complementary structures to the minority species. Adapted from~\cite{Romero_theor}.}
\label{Fig:Per_2comp}
\end{figure*}

Similarly to the work of Ref.~\cite{charalampidis2016rogue}, wide Gaussian wavepackets give rise to the PS structure [see top panels of Fig.~\ref{Fig:Per_2comp}], whilst even wider Gaussians result in the aforementioned "Christmas tree" destabilization [bottom panels of Fig.~\ref{Fig:Per_2comp}]. 
Mass imbalance modifies the timescales and onset conditions for PS and ``Christmas tree" structures. 
For instance, if the majority component is heavier, PS formation takes place for narrower Gaussians (compared to the mass balance case) and the "Christmas tree" structure is absent. 
However, for heavier minority species both phenomena can occur with appropriate Gaussian widths. 
On the other hand, finite temperature effects being modeled by a set of dissipative GPEs reduce the amplitudes of the waveforms and above a critical threshold suppress the generation of higher-order RWs.

\subsection{PS in 3D two-component mixtures}\label{PS_3D_mixture}

The first experimental observation of the PS in a 
BEC was only very recently reported~\cite{Romero_Ros_2024}. 
To achieve this, a two-component repulsively interacting BEC featuring high-particle imbalance among different $^{87}$Rb hyperfine states was placed in the immiscible regime and a highly elongated trap geometry was considered, see also below for experimental details. 
These conditions, as explained in the previous section~\ref{1D_repul_RWs}, are sufficient for the creation of an effective single-component self-focusing medium for the minority component. 
This setup allows for the dynamic creation of a PS, facilitated by a Gaussian optically induced attractive potential  well of the form 
\begin{equation}
V_G(\textbf{r})=-V_0 e^{-2\left[\left(\frac{x}{s_x}\right)^2+\left(\frac{y}{s_y}\right)^2\right]}. \label{eq:PS_potential}  
\end{equation}
In this expression, ($s_x$, $s_y$) denote the Gaussian widths and $V_0$ represents the potential depth. 
The transverse spatial profile of the Gaussian potential does not significantly affect PS nucleation as long as its width is larger than the vertical extension of the BEC. 
This external potential is responsible for seeding the MI of the minority component, eventually leading to the formation of the PS. 

The dynamics of the system are adequately modeled using the set of coupled 3D GPEs describing the time-evolution of the 3D mean-field wave functions ($\Psi_F(\mathbf{r}, t)$) of each hyperfine component  
\begin{equation}
i\hbar \frac{\partial \Psi_F(\mathbf{r}, t)}{\partial t} = \left( -\frac{\hbar^2}{2m} \nabla^2 + V(\mathbf{r}) + V_G(\mathbf{r}) + \sum_{F'} g_{FF'} |\Psi_{F'}(\mathbf{r}, t)|^2 \right) \Psi_F(\mathbf{r}, t). \label{eq:3D_GPEs}
\end{equation}
Here, $F = 1, 2$ labels each hyperfine state, $\mathbf{r} = (x, y, z)$ is the position vector, and  $m$ is the atomic mass.  Additionally, $V(\mathbf{r})=m (\Omega_{x}^2 x^2+\Omega_{y}^2 y^2+\Omega_{z}^2 z^2)/2$ is the external trapping potential, $V_G(\mathbf{r})$ is the Gaussian potential, and $g_{FF'}$ are the intra- ($F=F'$) and inter-component ($F \neq F'$) interaction strengths. 
These are given by $g_{FF'}=4 \pi N_{F'} \hbar^2 a_{FF'}/m$, with $a_{FF'}$ designating the 3D s-wave scattering lengths, and $N_F$ is the atom number in the $F$ hyperfine state. 
Here, the scattering lengths correspond to the $|1 \rangle =\lvert 1,0\rangle $ and $|2 \rangle = \lvert 2,0\rangle$ states characterized by $a_{11}=100.86a_0$, $a_{22}=94.57a_0$, and $a_{12}=a_{21}=98.9a_0$, with $a_0$ being the Bohr radius.

\begin{figure*}[t]
\centering
\includegraphics[width=1.0\textwidth]{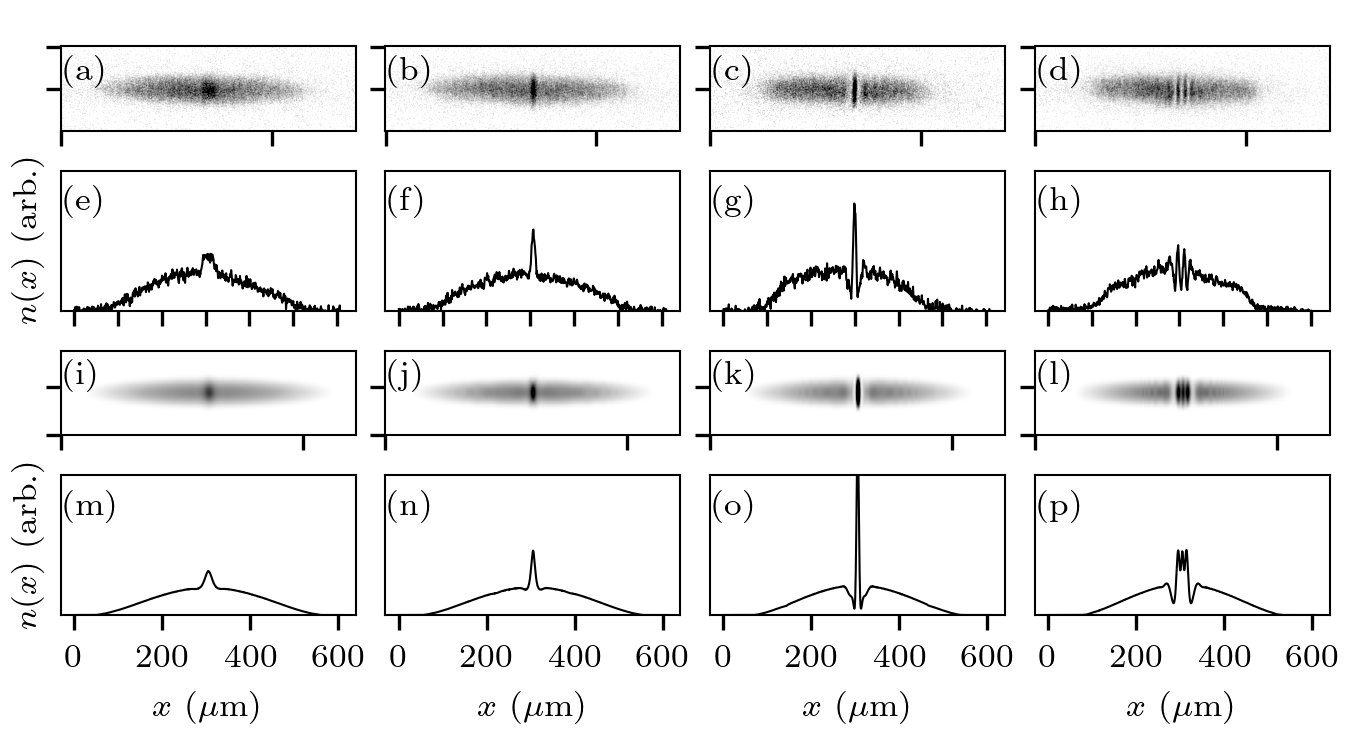}
\caption{Dynamical emergence of the PS within (a)–(h) the experiment and (i)–(p) 3D numerical simulations. 
(a)–(d) Integrated density profiles of the absorption images shown in (e)–(h) at $t=10, 30, 65, 85$ ms with an additional 9 ms of free expansion. 
(i)–(l) Integrated density profiles of the 2D density contours depicted in (m)–(p) stemming from the 3D mean-field simulations emulating the experimental process. Adapted from~\cite{Romero_Ros_2024}.}
\label{Fig:PS_experiment}
\end{figure*}

The procedure to dynamically produce the PS is as follows: 
\begin{itemize}
\item All $N$ atoms are initially placed at the ground state of the $|1,-1 \rangle$ hyperfine level under the optical well.  
\item Subsequently, an instantaneous particle transfer of typically $15 \%$ ($85 \%$) atom fraction to the $|2,0 \rangle$ ($|1,0\rangle$) hyperfine state is performed which 
reflects the radiofrequency experimental process. 
\item The two-component highly particle imbalanced mixture is allowed to evolve.     
\end{itemize}
This leads to the PS nucleation, see Fig.~\ref{Fig:PS_experiment}(i)-(p), which manifests as a density spike localized in both space and time flanked by two side density dips and having two $\pi$ phase jumps between the core and the tail of the configuration. 
This is further corroborated by fitting the numerical solution to the analytical one given by Eq.~(\ref{eq:Peregrine}) using the effective interaction strength $g_{\text{eff}}$ of Eq.~(\ref{eq:geff}). 
The PS dissolves after a few ms into three equidistant solitonic entities.  
Excellent agreement with the experimental results is observed, see Fig.~\ref{Fig:PS_experiment}. 
The formation of this RW is highly reproducible in the two-component repulsive mixture, with this feature being vital for creating an effectively self-focusing medium. 
In sharp contrast, a realization in a miscible two-component setting
or with a single-component (repulsive) cloud does not result in PS formation, 
as was illustrated in~\cite{Romero_Ros_2024}.

\subsection{Nonlinear stage of MI in repulsive BECs}
\label{subsec:nonlinearMI}

As explained in the Introduction, MI is a general mechanism driving the growth of long-wavelength perturbations in nonlinear dispersive media and is responsible, among others, for producing solitary waves in different dimensions~\cite{Chen_MI} and importantly RWs. 
It consists of two ``dynamical stages", namely the linear and the nonlinear one. 
The former is characterized by an exponential growth of perturbations featuring small wave numbers
(since it is a long-wavelength instability)~\cite{KivsharAgrawal2003}, as can be determined by linearizing the governing equations around a
uniform background. The latter refers to the amplification of these perturbations during the complex dynamical evolution that leads to their reorganization into coherent structures becoming comparable to the background medium~\cite{zakharov2009modulation,Everitt_MI}.
Describing this nonlinear stage of MI requires slow-modulation frameworks such as the so-called Whitham theory~\cite{whitham1965non} or the inverse scattering method~\cite{biondini2014inverse}, being capable of predicting the expanding oscillatory waveforms with well-defined edge velocities. 
Recall that such a phenomenon is inherently related to focusing NLS models and has been experimentally confirmed also in optical fibers~\cite{Kraych_MI,copie2020physics} and deep-water waves~\cite{Bonnefoy}.

However, in the recent work of~\cite{Mossman_nonlinearMI}, leveraging the effective reduction scheme of a two-component immiscible $^{87}$Rb bosonic gas, it was showcased that 
in a scenario strongly reminiscent of the pair
of dam breaks of~\cite{el2016dam}, DSWs  
and the nonlinear stage of MI can arise also in repulsive BECs utilizing a repulsive potential well. 
The latter in practice emulates Riemann type (i.e. step-like) initial conditions that cleanly trigger the onset of MI in a deterministic manner through the emergence of counterpropagating coherent nonlinear oscillatory structures, i.e., the DSWs. 
Note that the latter have been independently experimentally observed in repulsive single-component BECs~\cite{Hoefer_DSWs,Chang_DSW}.

To theoretically describe the dynamical manifestation of MI, the protocol delineated in the previous Section~\ref{PS_3D_mixture} is invoked within the set of coupled 3D GPEs of Eq.~(\ref{eq:3D_GPEs}) but with the crucial difference of a potential barrier (i.e., $V_0<0$ in Eq.~(\ref{eq:PS_potential})) instead of a well. 
This Gaussian barrier splits the condensate initially into two disconnected spatial regions enforcing ---upon its release--- a steep spatial discontinuity that is the analogue of a dispersive hydrodynamic dam-break problem~\cite{el2016dam,gurevich1993modulational,whithamlinear}. 

To analytically tackle the MI dynamics, the system is reduced to an effective 1D model governed by a set of coupled defocusing NLS equations given by Eq.~(\ref{eq:2comp_gpe2}) setting $m_1=m_2=1$. 
In this context, immiscibility drives the MI despite the repulsive nature of the overall system. Based on this coupled NLS model an analytical expression is derived for the MI envelope expansion speed. 
The latter refers to the characteristic velocity marking the boundary of the nonlinear oscillatory zone produced by MI and reads 
\begin{equation}
V_{2c} = 4\sqrt{
-g_{+} + \sqrt{\,g_-^2 + 4 Q_1^2 Q_2^2 g_{12}^2\,}},
~~~\textrm{with}~~~
g_{\pm} = g_{11} Q_1^2 \pm g_{22} Q_2^2. 
\label{eq:MI_Vel}
\end{equation}
In this expression, $Q_j$ denotes the background amplitude of each component. 
Moreover, this formula generalizes the classical focusing NLS result $V = 4 q_0\sqrt{-2g}$~\cite{biondini2016universal,gurevich1993modulational} to arbitrary two-component repulsive bosonic mixtures.

\begin{figure*}[t]
\centering
\includegraphics[width=1.0\textwidth]{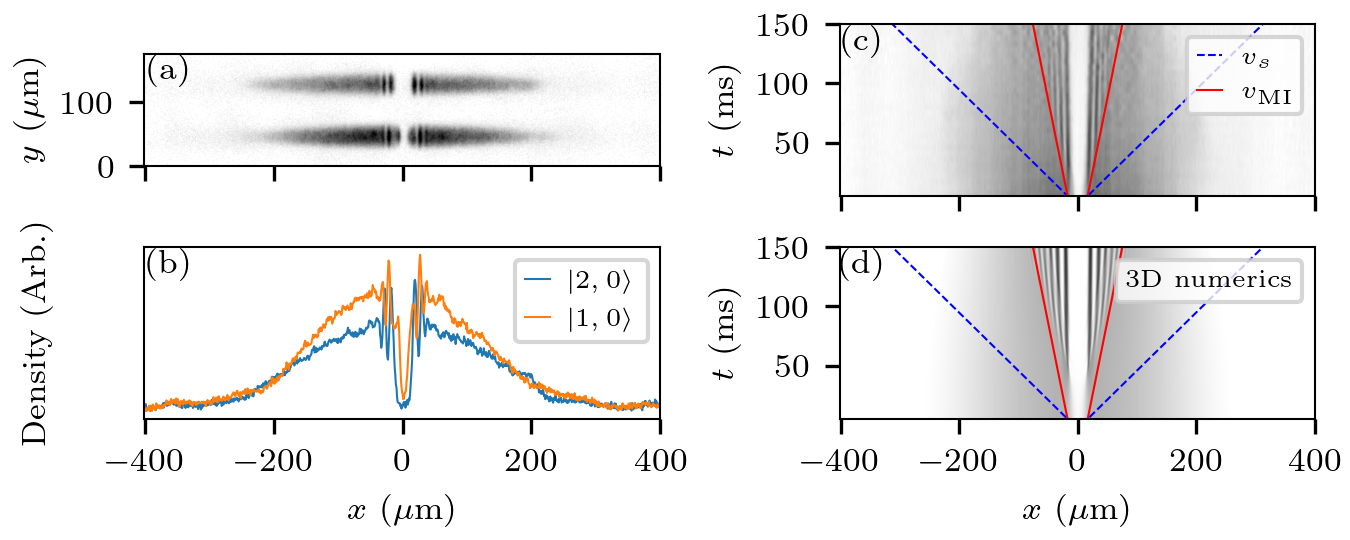}
\caption{ Nonlinear stage of MI triggered by a repulsive barrier for a particle balanced two-component mixture of $^{87}$Rb atoms in the $|2,0 \rangle$ and $|1,0 \rangle$ hyperfine states.
(a) Experimental single-shot absorption image taken after $60$~ms of evolution time, showing the $|2,0 \rangle$ (top) and the $|1,0 \rangle$ (bottom) state. (b) Corresponding integrated cross sections, averaged over 20 independent realizations. 
(c,d) Time-evolution of the integrated $|2,0 \rangle$ cross sections obtained from the experiment, shown in panel (c), and 3D numerical simulations [Eq.~(\ref{eq:3D_GPEs})], in panel (d). 
Dashed lines indicate the experimentally determined speed of sound and solid ones the theoretically predicted MI expansion rate [Eq.~(\ref{eq:MI_Vel})]. Adapted from~\cite{Mossman_nonlinearMI}.}
\label{Fig:MI_experiment}
\end{figure*}

Indeed, the corresponding envelope speed known for focusing single-component NLS models is retrieved in the highly particle imbalanced limit $Q_2 \ll Q_1$ by expanding Eq.~(\ref{eq:MI_Vel}) yielding 
\begin{equation}
V_c^{(2)} = 4 Q_2 \sqrt{-2g_{\mathrm{eff}}} + \mathcal{O}(Q_2^2),~~~\textrm{where}~~~
g_{\mathrm{eff}} = g_{22} - \frac{g_{12}^2}{g_{11}}. 
\label{eq:MI_Vel_singlecomp}
\end{equation}
This provides a unified picture spanning particle balanced to strongly imbalanced mixtures. 

The aforementioned analytical predictions are corroborated by direct numerical simulations of the full 3D GPE system [Eq.~(\ref{eq:3D_GPEs})] and importantly experimental observations, see the detailed discussion in the corresponding experimental Section~\ref{sec:experiment} and Fig.~\ref{Fig:MI_experiment}.  
In particular, for balanced mixtures (50:50), the system develops two symmetric DSWs whose solitonic inner edge 
(which corresponds to a gray soliton)
and harmonic (linear) outer edge match the expected MI-induced DSW structure, see Fig.~\ref{Fig:MI_experiment}(d). 
The measured envelope expansion agrees quantitatively with the prediction of  Eq.~(\ref{eq:MI_Vel}). 
As the population imbalance is increased (e.g., to 85:15 or 96:4), the MI onset is delayed and the envelope propagates in a slower manner, converging toward the single-component prediction of Eq.~(\ref{eq:MI_Vel_singlecomp}).  This phenomenology demonstrates the transition between fully coupled and effectively single-component MI dynamics.

\subsection{Higher-order RW solutions in different dimensions}

\subsubsection{HORWs in GPE models}\label{HORWs_GPE}

\begin{figure*}[t]
\centering
\includegraphics[trim={0cm 1cm 0cm 1cm}, clip,width=1.0\textwidth]{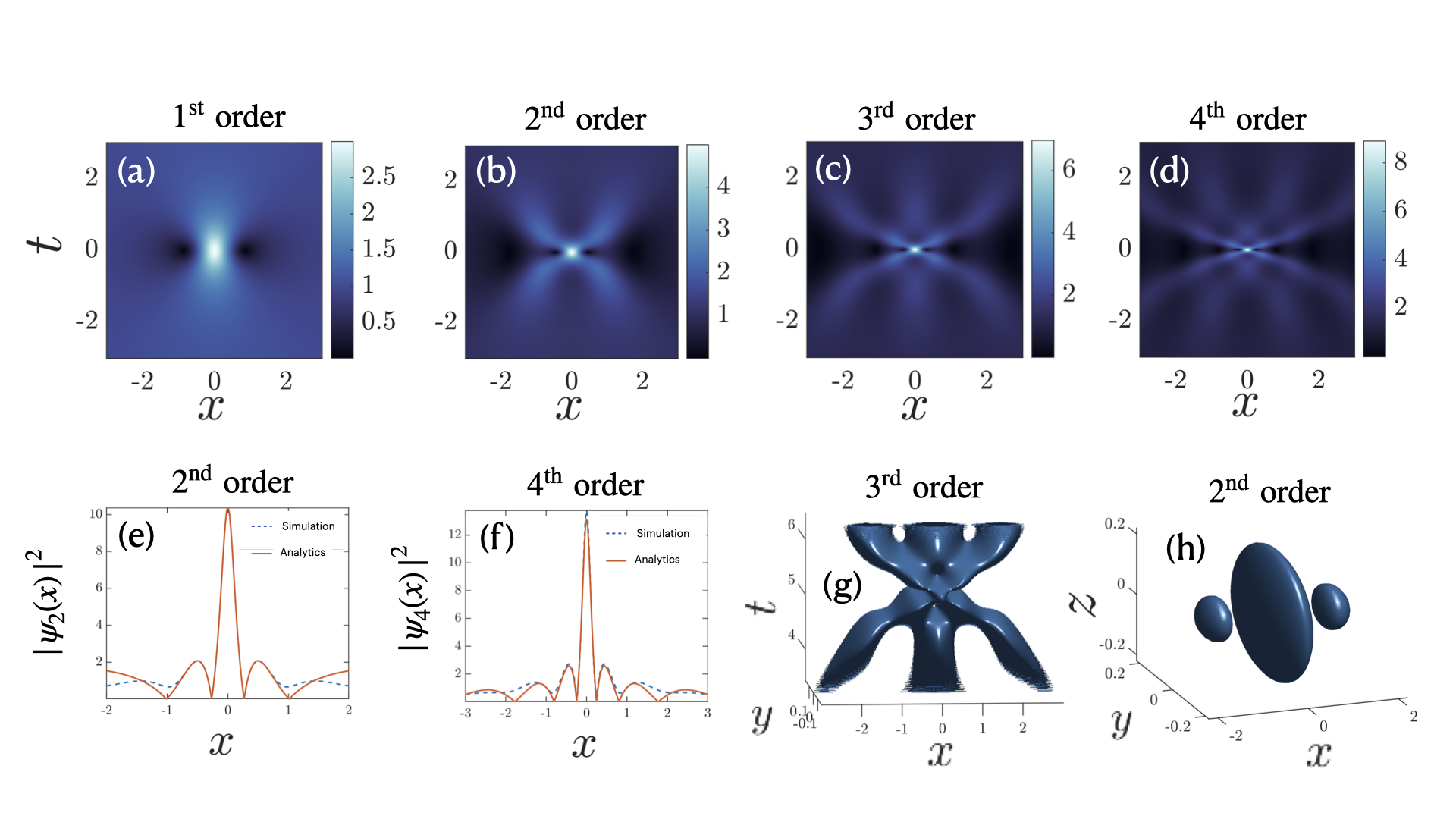}
\caption{(a)-(d) Wave function magnitude, $|\psi_k|$ with $k=1,2,3,4$ of the first four NLS RWs whose solutions are described by Eq.~(\ref{eq:HORW}). 
(e), (f) Profiles taken from (b), (d) respectively, depicting HORWs of $2^{{\rm nd}}$ and $4^{{\rm th}}$ order. Dashed lines correspond to 1D NLS simulations and solid lines provide a fit of the relevant analytical solution. 
Numerical parameters used for panel (e) correspond to trap frequency $\Omega=1.23$, prequench interaction strength $g_i=0.891$ and postquench one $g_f=-0.701$ and chemical potential $\mu=6.19$. 
Similarly for panel (f), $\Omega=1.93$, $g_i=1.84$, $g_f=-0.483$, $\mu=18.2$. 
(g) A 2D simulation illustrating the iso-surface of the wave function magnitude of a $3^{{\rm rd}}$ order RW with parameters $\Omega_x = 0.75$, $\Omega_y = 100$, $\mu =71.1$, $g_i = 1$, and $g_f =-0.05$. 
(h) A 3D simulation depicting the iso-surface of a $2^{{\rm nd}}$ order RW with $\Omega_x = 1$, $\Omega_y = \Omega_z = 100$, $\mu =143$, $g_i = 1$, and $g_f = -0.05$. Adapted from~\cite{adriazola2024experimentally}.
}
\label{Fig:HORWs}
\end{figure*}

While the fundamental first-order RW of the NLS model --the PS-- has been realized in several physical platforms, including water waves, optics and BECs, far less attention has been placed to HORWs, also at least
in part because the systematic theory of such
entities is much more recent~\cite{bilman2020extreme,bilman2022broader},
as is the development of the RW of infinite order~\cite{suleimanov}.
HORWs refer to rational solutions of Eq.~(\ref{eq:NLS}), that are constructed e.g. via Darboux transformations~\cite{He_Darboux,bilman2022broader}.   
Their fully explicit and compact representation is given in terms of determinants of $k\times k$ matrices derived from the generating functions of~\cite{bilman2020extreme}. 
The $k$th-order RW is expressed as
\begin{equation}
\psi_k(x,t) = (-1)^k \frac{\det\!\left(\mathcal{K}^{(k)}(x,t) + H^{(k)}(x,t)\right)}{\det \mathcal{K}^{(k)}(x,t)},
\label{eq:HORW}
\end{equation}
where the $k \times k$ matrices $\mathcal{K}^{(k)}$ and $H^{(k)}$ are constructed from the expansion coefficients $F_l(x,t)$ and $G_l(x,t)$. 
Specifically, the matrices $\mathcal{K}^{(k)}$ and  $H^{(k)}$ are defined through  
\begin{subequations}
\begin{gather}
\mathcal{K}^{(k)}_{pq}(x,t)
=
\sum_{\mu=0}^{p-1}
\sum_{\nu=0}^{q-1}
\binom{\mu+\nu}{\mu}
\big(
F_{q-\nu-1}^* F_{p-\mu-1}
+
G_{q-\nu-1}^* G_{p-\mu-1}
\big),~~ 1\le p,q\le k,\\ 
H^{(k)}_{pq}(x,t)
=
-2\big(F_{p-1}(x,t)+G_{p-1}(x,t)\big)
\big(F_{q-1}^*(x,t)-G_{q-1}^*(x,t)\big).\label{eq:KH_matrices}
\end{gather}
\end{subequations}
Here, the symbol $*$ stands for complex conjugation. Moreover, the two families of coefficient functions, ${F_l}$ and ${G_l}$, are obtained from generating functions expanded around the spectral point $\lambda = i$ using the following series expansions
\begin{subequations}
\begin{gather}
\frac{(1 - i\lambda)\,\sin\big((x + \lambda t)\sqrt{\lambda^2+1}\big)}
{\sqrt{\lambda^2+1}}
=
\sum_{l=0}^{\infty}
\left(\frac{i}{2}\right)^l
F_l(x,t)(\lambda-i)^l,\\
\cos\big((x+\lambda t)\sqrt{\lambda^2+1}\big)
=
\sum_{l=0}^{\infty}
\left(\frac{i}{2}\right)^l
G_l(x,t)(\lambda-i)^l.\label{eq:Gener_coef}
\end{gather}
\end{subequations}
For instance, the PS of Eq.~(\ref{eq:Peregrine}), i.e. the first-order member of this hierarchy, is retrieved by setting $k=1$ in Eq.~(\ref{eq:HORW}). 
In contrast, HORWs are increasingly more complex multi-lobed configurations with peak amplitudes scaling as $ \sim 2k+1$, see Fig.~\ref{Fig:HORWs}(a)-(f). This provides
a natural, systematic theory towards 
building RWs of order $k$.

In the work of Ref.~\cite{adriazola2024experimentally} the authors exploit a non-generic but physically realizable wave-breaking scenario, namely ---nearly--- semi-circular (Talanov-type) initial conditions~\cite{talanov1973certain}. These are known to lead to the so-called Painlev{\'e}-III type singularity in the dispersionless limit~\cite{Demontis}.   
Importantly, such a semicircular wavefunction can be ---approximately--- engineered within the Thomas-Fermi (TF) large density limit in defocusing BECs under the presence of a 1D harmonic trap $V(x) =(1/2)\Omega^2 x^2$ of frequency $\Omega$. 
The ground-state density profile has the well-known form $|\psi_{\rm TF}(x)|^2 = \mu - V(x)$, with $\mu > V(x)$ being an excellent approximation as long as $\mu \gg \Omega$ where $\mu$ is the chemical potential.
This is an ideal initial condition, having an
approximately semicircular profile, to seed HORWs upon considering an interaction quench from repulsive ($g>0$) to attractive couplings ($g<0$). 
Rigorous mathematical efforts have sought to quantify
the diminishing ---as the chemical potential
$\mu$ is increased--- effect of dispersion,
towards forming a ``boundary layer'' at the edge
of the condensate, where the ground state deviates
from the TF inverted parabola~\cite{Gallo2009OnTT,karali}.
In the work of~\cite{adriazola2024experimentally}, it was shown that adjusting $\Omega$, $\mu$, and $g$ can lead to dynamical generation of HORWs of order $k=2$, $3$, $4$. 
This is corroborated by comparing the numerical findings to the analytical predictions of Eq.~(\ref{eq:HORW}) indicating excellent agreement near the core of the emergent waveform although deviations are present in the relevant asymptotic regions. 
Notably, HORWs of $k=2$, $3$ were also demonstrated to
persist  in the presence of highly anisotropic 2D and 3D external confinement in the transverse directions, see Fig.~\ref{Fig:HORWs}(g), (h).  This, in turn, suggests
that this mechanism can be leveraged towards the observation
of HORWs in realistic atomic gases.

\subsubsection{HORWs in extended GPE models}

The study of RWs, and of course HORWs, beyond the standard cubic NLS/GPE models which can provide an analytically integrable description of RWs, is challenging and still remains an open issue. 
Indeed, real physical systems generally include non-integrable perturbations and effects that go beyond a mean-field treatment highlighting the necessity for systematic investigations of RW characteristics in more complex environments. At the moment, this is far less well
mathematically and even computationally understood,
as these non-integrable models face the absence
of the inverse scattering mathematical machinery
(and of associated methods, including, e.g., the 
Darboux transformation etc.) that allow one to 
analytically construct such waveforms. One then
has to resort to different types of methods
towards analytically constructing or numerically
computing such structures. For instance, one
such approach that has been proposed towards numerical
identification of RWs is that of considering time
as a ``spatial'' dimension and seeking to identify
localized waveforms in spatio-temporal domains
(possibly through continuation from integrable
limits)~\cite{WARD20192584}. Similarly, a
proposal towards identifying RWs mathematically has
been that of considering self-similar solutions
on top of a finite background in such dispersive
wave problems, upon appreciating that the
PS constitutes such a waveform~\cite{Ward2019}.

A particularly intriguing non-integrable setting arises in ultracold atomic mixtures featuring repulsive intracomponent interactions and attractive intercomponent ones supporting a new phase of matter called quantum droplets, see Refs.~\cite{cheiney2018bright,semeghini2018self,cabrera2018quantum,Errico_droplets,burchianti2020dual} for corresponding experimental realizations and Chapter 8 for elaborated descriptions on droplets. 
The latter are known to be stabilized due the competition between average mean-field interactions and quantum fluctuations which to first order are customarily described by the famous Lee–Huang–Yang (LHY) correction~\cite{lee1957eigenvalues}. 
This results in an extended Gross–Pitaevskii equation (eGPE) description of these mixtures containing  1D competing cubic (repulsive) and quadratic (effectively attractive LHY) nonlinear couplings~\cite{Petrov_2016}, rendering this theory framework qualitatively distinct from its classical focusing counterpart. 
We remark that in 3D the role of competing nonlinearities is reversed and it is the repulsive LHY effect that prevents attractive mean-field features from causing collapse by stabilizing the condensates in the form of the quantum droplet~\cite{Petrov_2015,LuoPangLiuLiMalomed2021_Qdroplets}. 

\begin{figure*}[t]
\centering
\includegraphics[trim={0cm 4cm 0cm 4cm}, clip,width=1.0\textwidth]{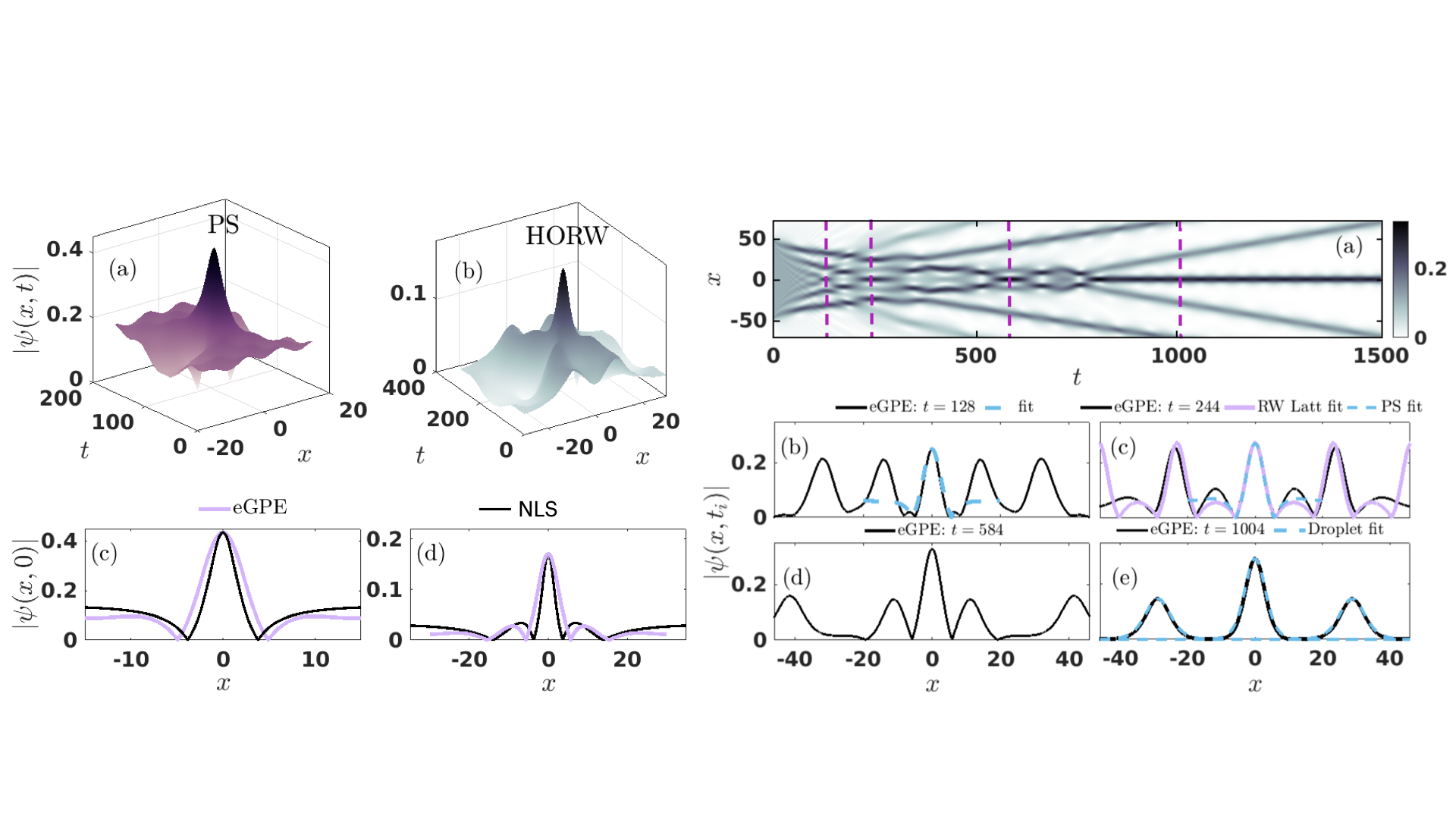}
\caption{(Left block of panels): (a) Space-time profile of a PS for $\mu \approx -0.12$, and (b) a $2^{{\rm nd}}$ RW with $\mu \approx -0.02$ obtained via a spatiotemporal fixed-point iteration scheme within the eGPE model using box sizes $L= 30$ and $L= 60$ respectively.  
Snapshot of (c) the PS and (d) the $2^{{\rm nd}}$ RW moduli profiles, $|\psi(x,t)|$, within the eGPE and the NLS models (see legend). 
(Right block of panels): (a) Interfering dam-break flows resulting in a modulated RW lattice and PS generation using $\rho_0= 0.0125$ in Eq.~(\ref{step_IC}). 
Snapshot of $|\psi|$ showing the dynamical formation of (b) an eGPE PS, (c) a RW lattice, (d) higher order solitonic waveform and (e) a breathing quantum droplet. 
Vertical dashed lines in panel (a) mark the times at which the relevant profiles of panels (b)-(e) are depicted. 
The PS and RW lattice are compared to the ones obtained via the fixed point method for $\mu \approx -0.0689$ and $\mu \approx -0.085$ respectively.  
The central breather of panel (c) is also contrasted to an eGPE PS for $\mu \approx -0.0767$. 
In panel (e) inner and outer peaks are compared to the analytical droplet solution using $\mu \approx -0.1518$ and $\mu \approx -0.087$ respectively. 
Adapted from~\cite{Chandramouli2025rogue}.
}
\label{Fig:HORWs_eGPE}
\end{figure*}

A first attempt toward addressing this apparent gap in the literature is made in the work of Ref.~\cite{Chandramouli2025rogue} that investigates both the existence and dynamical generation of RWs within the 1D eGPE model. 
Here, the physical system under consideration comprises of two hyperfine states (e.g., of $^{39}K$) that have the same atom number ($N_1=N_2\equiv N$), experience equal intra-component repulsion, $g_{1}=g_{2}\equiv g>0$, and inter-component attraction, $g_{12}<0$, thus entering the 1D droplet regime for $\delta g=g_{12}+g>0$. 
It is known that under these assumptions, the two components become equivalent, and the droplet is effectively described by a reduced single-component 1D eGPE which in dimensionless units reads 
\begin{equation}
\label{GP-eqn}
i\frac{\partial \psi}{\partial t}=-\frac{1}{2}\frac{\partial^2 \psi}{\partial x^2} + \left(|\psi|^2 -|\psi| \right )\psi. 
\end{equation}
Here, $\psi(x,t)$ represents the 1D droplet wave function, while the average mean-field and LHY strengths, respectively, within the nonlinear
terms have been rescaled to unity without loss of generality~\cite{tylutki}. 
Moreover, it has been shown in~\cite{Chandramouli_DSWs} that the competition between LHY attraction and mean-field repulsion yields a density-dependent transition between focusing and defocusing dynamical regimes. 
A Taylor expansion around a constant background density $\rho_0=|\psi_0|^2$ results in an approximate cubic GPE model 
\begin{equation}
\label{cubic-GP}
i\frac{\partial \psi}{\partial t}+\frac{1}{2}\frac{\partial^2 \psi}{\partial x^2}+\left(f(\rho_0)-\rho_0f'(\rho_0)\right)\psi+f'(\rho_0)|\psi|^2\psi=0,
\end{equation}
provided that $f'(\rho_0)>0$ with $f(\rho_0)=-\rho_0+\sqrt{\rho_0}$. 
Here, vanishing $f'(\rho_0)$ dictates the transition from attractive ($\rho_0< 1/4$) to repulsive  
($\rho_0>1/4$) dynamics, with $\rho_0=1/4$ defining the so-called hyperbolic to elliptic threshold as argued in Ref.~\cite{Chandramouli_DSWs}. 
In view of this NLS type reduction, the original eGPE model can now admit rational solutions such as the PS of Eq.~(\ref{eq:Peregrine}). 
[Notice that the 3rd term of Eq.~(\ref{cubic-GP}) can
always be absorbed through a gauge transformation,
as can the prefactor $f'(\rho_0)$ through
scaling, thus retrieving Eq.~(\ref{eq:NLS})].
This reduction can, in turn, provide useful initial 
{\"A}nsatze towards the numerical identification of eGPE RWs based on a Newton-conjugate-gradient fixed-point scheme that incorporates a doubly periodic basis, in the
same spirit as the above-mentioned non-integrable
systems approach of~\cite{WARD20192584}. 
Two distinct families of eGPE RWs are found upon parametric variations of the system's chemical potential, $\mu$. 
Specifically, family I exhibits a PS-like configuration [Fig.~\ref{Fig:HORWs_eGPE}(a), (c) of the left panels] that broadens and deforms as $|\mu|$ increases, displaying bifurcations and transitions into multi-hump structures. 
On the other hand, family II emanates from periodic waves, forming spatially modulated cnoidal-like backgrounds that evolve into RW-like waveforms for negative $\mu$. 
Interestingly, eGPE HORWs [Fig.~\ref{Fig:HORWs}(b), (d) of the left panels] resemble corresponding second-order NLS rational solutions (see also Section~\ref{HORWs_GPE}) with observed deviations stemming from the non-integrability of the model.

Experimentally relevant dynamical protocols for generating these entities have been elucidated utilizing two different techniques. 
Namely, the interference of dam-break flows~\cite{el2016dam} and the gradient catastrophe mechanism~\cite{bertola2013universality}, both
of which were discussed earlier in this Chapter.
In the former setting, one uses box-type initial conditions
\begin{equation}
\label{step_IC}
\psi(x,0)=\begin{cases}
\sqrt{\rho_0}, ~~|x|<L\\ 
0,\;\;\;\;\; ~~|x|>L
\end{cases}, 
\end{equation}
with $2L$ being the considered box size. 
The focal point of these investigations corresponds to densities that lie below the hyperbolic-to-elliptic threshold ($\rho_0=1/4$) where the LHY (attractive) contribution dominates, providing
an effectively focusing character to the system dynamics.
Specifically, with the initial condition of Eq.~(\ref{step_IC}) 
in the low-density regime $\rho_0 \ll 1/16$, two counterpropagating dam-break flows arise and via their interaction RW lattices are produced closely resembling AB trains, see right panels (a)-(e) of Fig.~\ref{Fig:HORWs_eGPE}.  
However, for higher densities broadened RW events are observed and unseen features caused by the droplet environment are found such as self-evaporation of matter-wave jets and persistent breathing droplet cores.

Turning to gradient catastrophe events, the system is initiated using localized initial conditions of the form 
\begin{equation}
\label{Sech_waveform}
\psi(x,0)=A {\rm sech}(x),
\end{equation}
where $A$ is the amplitude of the localized configuration. 
The competing attractive and repulsive interactions with the former dominating in the course of the evolution lead to the dynamical nucleation of a PS structure, followed by second-order RW events accompanied by long-lived breathing multi-soliton states.

\section{Experiment}
\label{sec:experiment}
Dilute-gas BECs provide unparalleled experimental access to the study of superfluid hydrodynamics. 
Rapid progress has been enabled by a versatile and sophisticated experimental toolbox that allows precise manipulation and observation of quantum gases \cite{kevrekidis2008emergent, Straten2016}.
Due to the prominent role that solitons play as hallmark phenomena of nonlinear dynamics, their  formation in BECs has been of particular interest since the early days of BEC research.
Dilute-gas BECs have since been shown to support a wide variety of RW-like features and solitons. 
While this topic is extensive, the following discussion will focus specifically on aspects relevant for the first realization of a Peregrine soliton (PS) in this system, and related ramifications of this work.

\subsection{Realization of a PS}

A first realization of the PS in the context of ultracold quantum gases was reported in Ref.~\cite{Romero_Ros_2024}. This work used a highly elongated BEC with repulsive interactions. Since PSs emerge in self-focusing systems (i.e., systems with attractive interactions), at first sight this seems to preclude the formation of these features. Furthermore, an additional experimental difficulty was posed by the fact that systems with attractive interactions are prone to modulational instability (MI), which can mask the formation of a PS. Furthermore, BEC experiments frequently use destructive imaging techniques, requiring the observation of dynamics over time to be achieved by stitching together the results of many independent experimental iterations. Hence, reliable realization of a PS required a way to trigger the dynamics in a highly reproducible way. In Ref.~\cite{Romero_Ros_2024}, these difficulties were successfully circumvented. First, to engineer an effectively self-focusing (attractive) system, a two-component mixture of two hyperfine states was chosen in which a minority component was embedded in a large bath of a majority component. The hyperfine states were chosen to result in an immiscible two-component mixture, which, as already shown in Chapters 3-4, can be theoretically reduced to an effective one-component system with attractive interactions. The resultant effective interaction strength was sufficiently weak so that the MI of the background density did not mask the dynamics over a time scale of approximately 100~ms, sufficiently long to observe Peregrine dynamics. Finally, the formation dynamics of the Peregrine was started in a repeatable way by the rapid creation of the two-component mixture from an initial single-component cloud, leading to a highly reproducible time evolution that could be followed over a sequence of experimental runs.

The experimental setting consisted of a highly elongated BEC of $^{87}$Rb atoms held in an optical dipole trap, providing essentially harmonic confinement in all spatial directions. The trap geometry was highly elongated with an aspect ratio of 100:1, ensuring effectively one-dimensional dynamics even though the BEC was not in a strictly 1D regime. A weak additional laser was sent transversely through the center of the elongated BEC to create a small attractive Gaussian-shaped optical potential in the center, leading to a slight initial density hump in the BEC and setting the BEC on its course towards the formation of a Peregrine. After the creation of the mixture, the minority component evolved into a PS characterized by its iconic shape: a central peak flanked by a density dip on either side, consistent with theoretical predictions. These dynamics are showcased in Fig.~\ref{Fig:PS_experiment}. The soliton appeared after around 65~ms of evolution, and then decayed into multiple peaks before broad-spread MI of the overall system took over the dynamics. The observed dynamics were highly reproducible and were also confirmed through 3D simulations of the Gross-Pitaevskii equation as well as 1D numerical and analytic modeling. The numerical studies also confirmed the phase structure of the PS--the formation of a distinct $\pi$ phase jump between the dips formed on either side of the central peak coincident with the fully formed Peregrine soliton.

Various experimental checks were performed to show that such dynamics are not observed in single component BECs, and that the initial Gaussian potential only served to trigger the dynamics but otherwise was inconsequential. For example, after initializing the dynamics, the Gaussian potential could be turned off, and still the system evolved further into a PS. This, together with the matching numerics, solidly established the observation of the PS.

\subsection{Nonlinear stage of MI and matter-wave cavity}

Underlying the dynamics that drive the system towards a PS is the notion that the minority component of an immiscible, particle-imbalanced mixture can form an effectively attractive medium. Such a system supports the emergence of MI. Accordingly, the experimental platform described in the previous section is also a powerful testbed for the study of MI--a nonlinear phenomenon that is prevalent in many attractive (focusing) nonlinear systems. As described in the theoretical Sec.~\ref{subsec:nonlinearMI}, the evolution of MI typically can be divided into several stages: a first  stage, where excitations quickly grow in amplitude according to the linearization prediction, followed by a nonlinear stage where excitations spread out through the system. 
A detailed experimental study of this phenomenon with cold atoms is complicated by the fact that, as an instability, MI has an onset time that can vary between experimental runs. Experiments with ultracold atoms often use destructive imaging techniques, so that time evolutions of a system are acquired through multiple, independent runs of the experiment, each one advancing the evolution by one timestep. A variable onset time of dynamics can obscure the detected evolution. 
Fortunately, the onset of the nonlinear stage of MI can be triggered by the creation of an effectively attractive mixture in the presence of a strong repulsive barrier, as demonstrated in Ref.~\cite{Mossman_nonlinearMI}. Additionally, the barrier also fixes a point in space from which the MI spreads outwards, allowing for the measurement of the propagation speed of the wave fronts. In the experiments described in Ref.~\cite{Mossman_nonlinearMI}, the procedure began by creating a BEC of approximately one million $^{87}$Rb atoms confined in an optical dipole trap. As in the case for the Peregrine experiments described above, this trap was generated by an infrared laser traveling horizontally and focused to a waist of $20~\mu\text{m}$. It leads to a pencil-shaped BEC with an aspect ratio of approximately 100:1. A strong static repulsive barrier, formed by a blue detuned laser, separated the cloud along the long axis into two halves, as shown in Fig.~\ref{Fig:MI_experiment}. For the initial single-component BEC, generated with all atoms in the $|1,-1\rangle$  hyperfine state and characterized by repulsive (defocusing) interactions, this is a stationary situation and no time evolution of the density was observed. However, MI could be triggered by employing brief microwave and radiofrequency pulses to generate a mixture of minority atoms in the $|2,0\rangle$ state embedded in a majority of atoms in the $|1,0\rangle$ state, similar to what was done for the Peregrine experiments. As before, this specific mixture can be effectively described as an attractively interacting system for the minority. The creation of the mixture triggered the onset of the nonlinear stage of MI in time, while the repulsive barrier served as a ``nucleation point" from which the excitation spread outwards. The role of the barrier is to provide a region where the density of the fluid must abruptly go to zero, with this increased curvature region enhancing the onset of MI. The time evolution was then accumulated in successive experimental runs, each one advancing the time between the creation of the mixture and the imaging by one time step. Integrated cross sections were stacked in time to form the space-time-plots shown in Fig.~\ref{Fig:MI_experiment}. Matching numerics supported the observation and, together with the experimental images, formed an important benchmark for the analytic studies described in Sec.~\ref{subsec:nonlinearMI}.  

\begin{figure*}[t]
    \centering
    \includegraphics[width=1.0\textwidth]{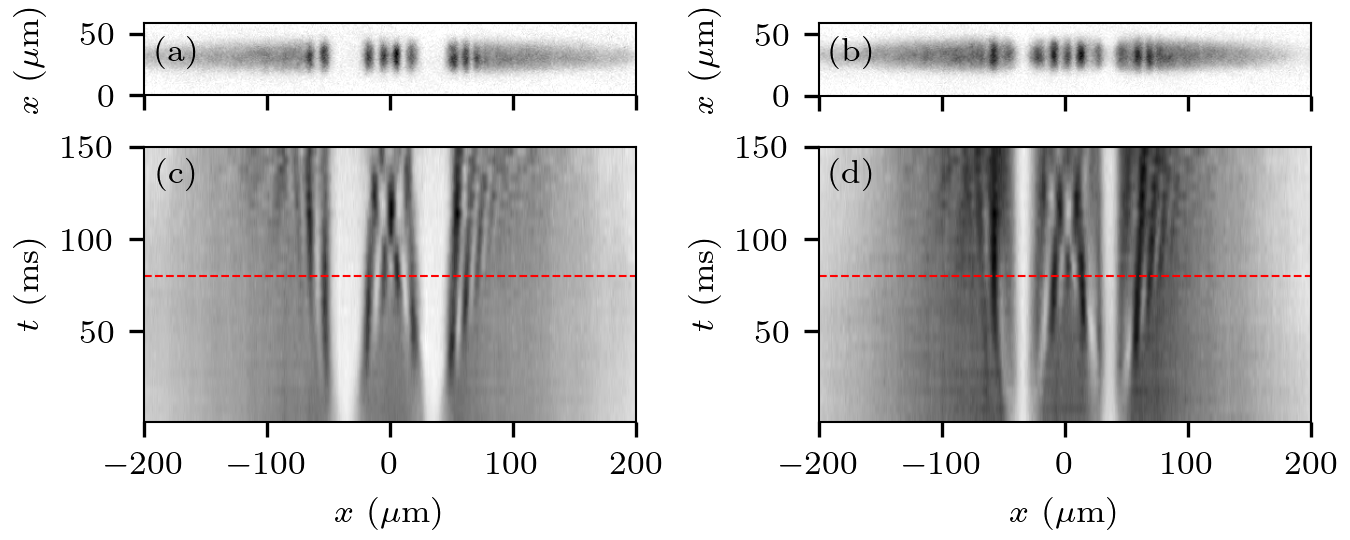}
    \caption{Experimental images of nonlinear waves colliding within a matter wave cavity produced with two repulsive barriers in an immiscible mixture of atoms in the $|1,0\rangle$ and $|2,0\rangle$ state. The two different states are shown in panels (a,c) and (b,d), respectively. (a,b) Absorption images of the $|1,0\rangle$ and $|2,0\rangle$ states, respectively, averaged over 15 independent experimental realizations with 80 ms of evolution time. (c,d) Spacetime diagrams for the density where each row of pixels represents an integrated cross section of the density averaged over 15 realizations at 5 ms intervals. The red dashed line indicates the time segment corresponding to the images in panels (a) and (b). A Peregrine-like feature peaks at the center around 120 ms.}
    \label{fig:dual_barrier}
\end{figure*}

In an interesting extension of this setup, shown in Fig.~\ref{fig:dual_barrier}, two repulsive barriers, separated along the long axis of the pencil-shaped BECs, were employed to form a matter-wave cavity for the nonlinear waves of the MI. Each edge of both barriers served as the origin of the nonlinear stage of MI when the immiscible mixture was created. In the area between the two barriers, the spreading waveforms approached each other and, 
upon interference at the center between the barriers, formed a Peregrine-like structure. This feature is seen in Fig.~\ref{fig:dual_barrier}(c,d) at time $t \approx 120\,\text{ms}$ and $x = 0\,\mu\text{m}$. This is a highly intriguing setup: the observed wave dynamics in the two-component system are marked by long wavelengths that can easily be optically resolved and followed in space and time as they spread out. The nonlinearity of wave interactions, evidenced by the formation of a Peregrine-like structure in the center, implies nontrivial wave dynamics that are at the forefront of contemporary studies.

\subsection{Extension to three components}

\begin{figure*}
    \centering
    \includegraphics[width=1.0\textwidth]{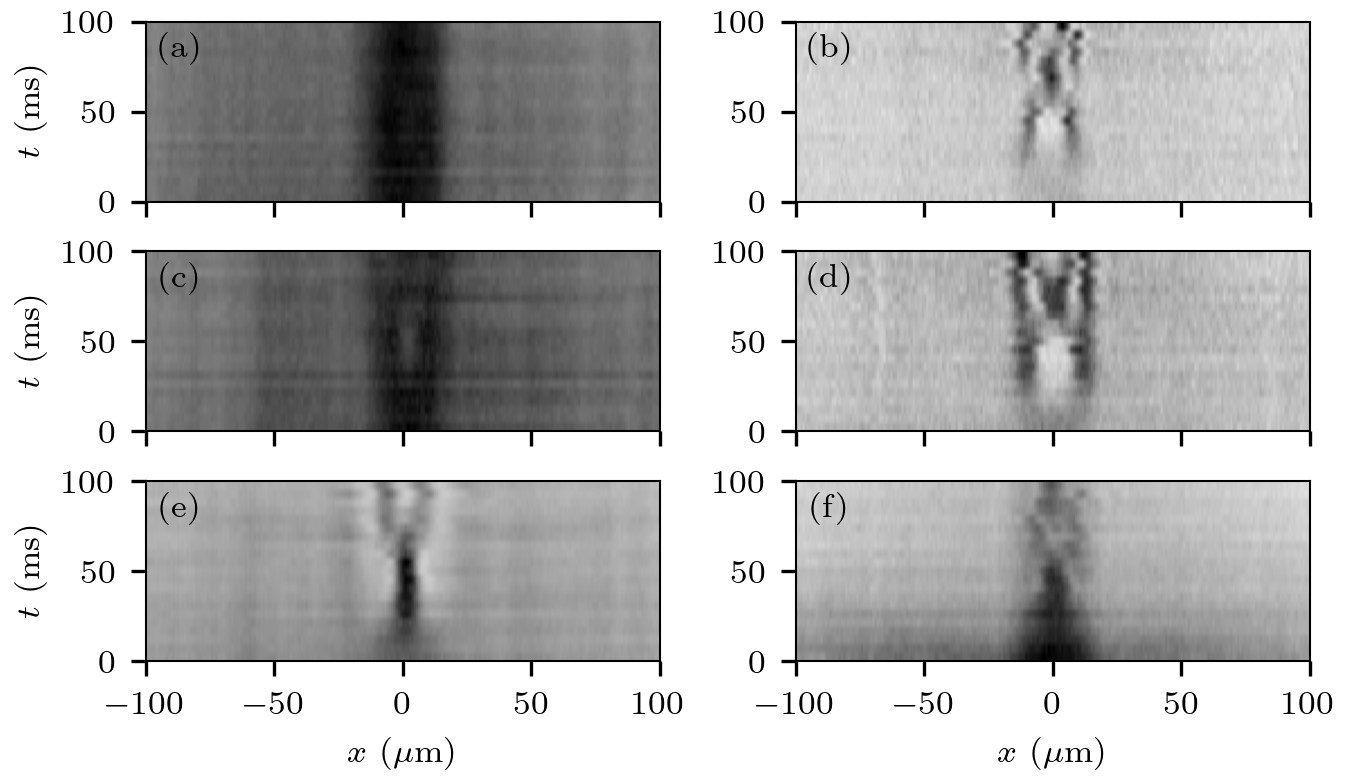}
    \caption{Experimental realizations of vector PSs in three-component mixtures. Three-component mixtures are constructed from the states $|1,-1\rangle$, $|1,0\rangle$, and $|2,0\rangle$, shown in panels (a,b), (c,d) and (e,f), respectively. On the left (a,c,e), the system is prepared in a 10\%/40\%/40\% mixture with the $|1,-1\rangle$ state as the minority while, on the right (b,d,f), the system is prepared in a 42.5\%/42.5\%/15\% mixture with the $|2,0\rangle$ state as a minority.}
    \label{fig:3component_exp}
\end{figure*}

As discussed above, the two-component mixture utilized in~\cite{Romero_Ros_2024}  can be described by a reduced, effective single-component model for the minority, featuring attractive interactions despite the fact that all inter- and intra-component interactions in the system are repulsive. The great versatility of ultracold quantum gases allows to expand this scheme even further. Particle-imbalanced mixtures of more than two components can be reduced to effective models with attractive interactions. This opens a path to the experimental generation of vector RWs and provides access to a much larger range of nonlinear excitations. For example, in a three component mixture with pairwise repulsive interactions, one or two minority species can be embedded in a majority background giving rise to single PSs, vector Peregrines with one Peregrine per minority component, or twin Peregrine structures composed of two adjacent PSs in each minority
component. Examples of vector Peregrines are shown in Fig.~\ref{fig:3component_exp}. The experimental procedure is an extension of the scheme described in the context of the Peregrine formation in two components. The experiments begin with a single-component BEC held in an optical dipole trap with an additional weak attractive Gaussian potential at the center. The trap provides spin-independent confinement, so that all hyperfine states of the atoms can be trapped simultaneously. A homogeneous 10~G magnetic bias field is applied to separate the energy levels of the hyperfine states by virtue of the Zeeman effect. This allows for controlled state transfer between different hyperfine states using brief radiofrequency and microwave pulses. After the preparation of the desired mixture, the system is allowed to evolve, and state-selective absorption imaging is used to record the spatial profiles for each of the components.

In the example shown in Fig.~\ref{fig:3component_exp}, three different hyperfine states were simultaneously involved.
In principle, the scheme described above can be extended to even more than three components. In practice, while the addition of more radiofrequency and microwave pulses does not pose any problem, care must be taken during the initial preparation of the mixture when nearly degenerate transitions are present that could be driven at the same time during the preparation of the mixture. For example, the first order Zeeman effect leads to an energy spacing between the
$|F=1, m_F =-1\rangle$ and $|1,0\rangle$ state that is equal to the spacing between the $|1,0\rangle$ and $|1,1\rangle$ state, making it more difficult to selectively drive only one of the two transition. Often, these complications can be overcome by using sufficiently large magnetic fields so that the quadratic Zeeman effect becomes noticeable, leading to further splittings of the transitions. A more severe problem is posed by collisional dynamics between different hyperfine states, in particular when states of different hyperfine manifolds (such as the $F=1$ and $F=2$ manifold of $^{87}\mathrm{Rb}$) are populated at the same time. Here, the collisional dynamics depend crucially on the specific chosen states and can lead e.g. to a rapid depopulation of a state in the $F=2$ state over a time scale of tens or hundreds of milliseconds, while states from the $F=1$ manifold might be nearly constant over the same time frame. Such dynamics may not necessarily preclude the observation of vector RWs but must be taken into account in the analysis.

The investigation of nonlinear dynamics in BECs is an exciting frontier of research, with many ramifications to the work described in this chapter (e.g., in the form of further solitons, breather solutions, shock wave structures and more). Most closely related 
to the setup presented herein (of effectively attractive condensates arising from 
majority-minority settings of repulsive multi-component BECs)
is the observation of a Townes soliton by the Dalibard group, described in
Chapter~3, which was also based on the engineering of effectively attractive interactions in a two-component mixture.

\section{Broader perspective: RWs and Extreme Events in other systems }\label{sec:Generalizations} 

RWs have drawn significant scientific attention in recent years, particularly because of the strong analogies that can be established across various dispersive media governed by the NLS equation \cite{onorato2013rogue,dudley2019rogue}. This section aims to place the topic in a broader context by discussing the origins of scientific interest in RW phenomena within oceanography, and their subsequent connection to analogous extreme wave events in Kerr media. Moreover, different aspects of the nonlinear stage of MI will also be introduced, thereby motivating future experimental efforts to confirm similar unstable and nonlinear wave patterns in BECs.

\subsection{Historical context: From an oceanography concept to nonlinear dispersive media}

\begin{figure}[t]
\centering
\includegraphics[width=0.7\textwidth]{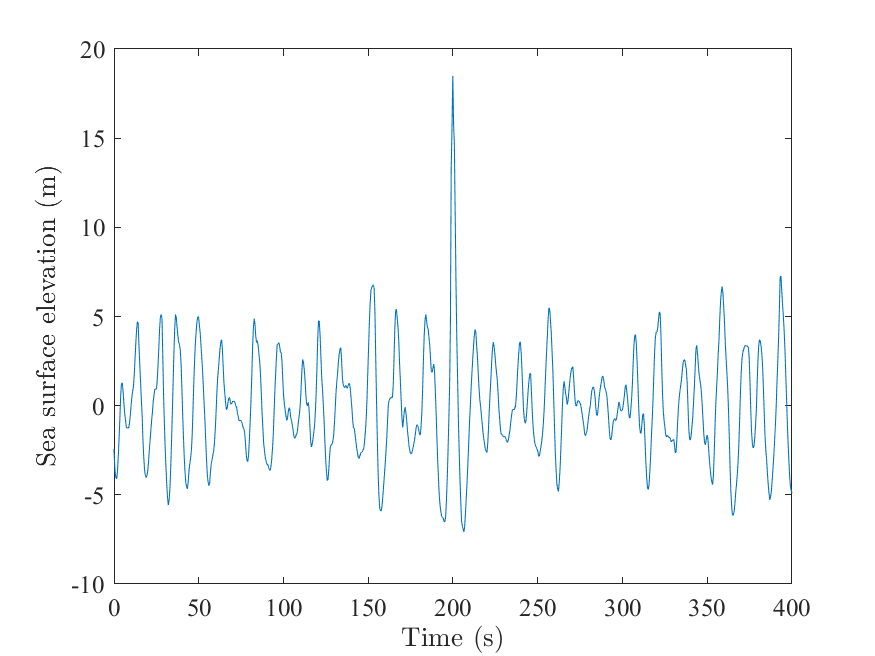}
\caption{First scientific record of an ocean RW, measured on 1 January 1995 at the Draupner platform in the North Sea.}
\label{Draupner}
\end{figure}

The first observations of RWs, and thus the initial interest and sense of mystery surrounding the phenomenon, originated from ocean waves, which are directly and mechanically observable at large scales. Such phenomena have been experienced and reported by mariners for centuries, however, they have rarely found a trusting ear in society willing to acknowledge their existence. As mentioned in the Introduction, the first scientific recording of the Draupner, or New Year, wave on January 1, 1995~\cite{haver2004possible} marked a paradigm shift in ocean engineering; see Figure~\ref{Draupner} for an excerpt of the associated time-series. 
Although there had previously been indications of severe wave loading inferred from shipwrecks and structural damage, this direct measurement provided the first unequivocal evidence of the existence of extreme or rogue waves.

Subsequently, research on extreme ocean waves intensified rapidly within the ocean engineering community, and discussions regarding the adaptation of design standards for ships and offshore structures are still ongoing \cite{mori2023science}. A major focus of this research has been the identification of the key physical mechanisms responsible for extreme wave formation and the correlation of these mechanisms with the observed increase in extreme events in the tail of exceedance probability distributions, which deviate from the Rayleigh distribution that serves as the distribution model of linear wave theory. 

\begin{figure}[t]
\centering
\includegraphics[width=0.7\textwidth]{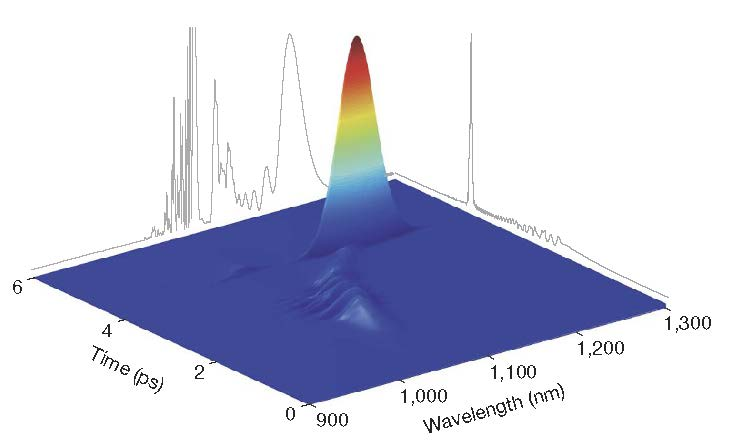}
\caption{Short-time Fourier transform–based time–wavelength profile of an optical RW, measured in a Crystal Fiber NL-5.0-1065 as from  \cite{solli2007optical}.}
\label{opticalRW}
\end{figure}

Indeed, different studies highlighted the role of MI in the increase of extreme wave event frequency as a result of modulation, or Benjamin-Feir instability or quasi four-wave resonant interaction, in the hydrodynamic context \cite{tulin1996breaking,fujimoto2019impact,malila2023statistical}; the higher the value of the wave steepness and/or the smaller the initial bandwidth of the wave field, the larger the deviation from Rayleigh in the tail of distribution \cite{onorato2006extreme}. Such characteristics have been defined in the so-called Benjamin-Feir index \cite{janssen2003nonlinear}. Experimental data confirmation of the MI dynamics was consistent in qualitative agreement with the NLS equation \cite{yuen1982nonlinear} and in quantitative agreement with the modified NLS (MNLS) equation \cite{dysthe1979note}. Another milestone in RW research is the demonstration that optical RWs from supercontinuum generation, extreme-intensity pulses, lead to heavy-tailed intensity statistics far from a Gaussian behavior. This is in analogy to oceanic RWs, since such dynamics can be described by the generalized nonlinear Schr\"odinger equation (GNLS) \cite{solli2007optical} 
\begin{eqnarray}
i\frac{\partial \psi}{\partial \xi}
+\sum_{m=2}^{6} \frac{i^{\,m}\beta_m}{m!}\,\frac{\partial^{m} \psi}{\partial \tau^{m}}+ \left[
|\psi|^{2}\psi
+ \frac{i}{\omega_{0}} \frac{\partial}{\partial t}\!\left(|\psi|^{2}\psi\right)
- T_{R} \psi \frac{\partial |\psi|^{2}}{\partial \tau}
\right]=0. 
\end{eqnarray}
Here, $\psi$ denotes the slowly varying dimensional envelope of the electric field, $\beta_m=\frac{\partial^m k}{\partial\omega^m}$ are the dispersion coefficients of the fiber, and $T_R$ represents 
the fiber's Raman response parameter~\cite{agrawal2013nonlinear}. Since electromagnetic 
waves propagate along the fiber, that is, through space, the corresponding time-like
NLS equation is written with respect to dimensional space $\xi$ and time $\tau$ evolution variables. Please note that the evolution variables are reversed in the BEC formulation, which will be elaborated upon in the next subsection. A measured RW observation in a crystal fiber is shown in Figure \ref{opticalRW}. 
Following this latter work, a strong focus has been on understanding the critical role of exact breather solutions of the NLS equation in the formation of extreme events in a deterministic or random setup, particularly in optics and hydrodynamics \cite{onorato2013rogue,dudley2014instabilities,dudley2019rogue,tlidi2022rogue}.

\subsection{The NLS equation and its foundational role in the study of unstable waves across various physical settings} 

Considering the derivation of the NLS equation as a universal modulation model in 
areas other than BEC, 
a heuristic approach to proceed is to approximate the nonlinear dispersion relation of the respective wave medium. Whether one expands the wave frequency $\omega$ in powers of the wavenumber $k$ and the wave intensity $|\psi|^{2}$, or conversely expands $k$ in terms of $\omega$ and $|\psi|^{2}$, depends on the aimed weakly nonlinear equation describing wave propagation either in time and or space. As for electromagnetic waves propagating in an optical fiber, water waves evolving in hydrodynamic facilities require the use of wave makers fixed at a particular and driven by times-series (boundary conditions). The wave evolution is then measured by the use of wave gauges, which are placed along the tank. 
Therefore, it is more appropriate and convenient to consider a space evolving (time-like) NLS equation form, which is derived by expanding $k$ in terms of $\omega$ and $|\psi|^{2}$, leading to 
\begin{eqnarray}\label{timenlse}
i(\frac{\partial \psi}{\partial \xi} + \beta_1 \frac{\partial \psi}{\partial \tau})-\dfrac{1}{2}\beta_2\frac{\partial^2\psi}{\partial \tau^2}+\gamma|\psi|^2\psi=0,
\end{eqnarray}
with $\beta_1=\dfrac{\partial k}{\partial\omega}=\dfrac{1}{c_g}$, $\beta_2=\dfrac{\partial^2 k}{\partial\omega^2}$, and $\gamma=\dfrac{\partial k}{\partial |\psi|^2}$, $c_g$ being the group velocity, see \cite{chabchoub2015nonlinear} for more details also regarding the specific parameters for water waves and/or Kerr media for corresponding nonlinear dispersion relations $k(\omega,|\psi|^2)$. This is different from BEC experiments, which are initialized starting from a well-defined state at an initial time. For the water wave problem for instance, interchanging the spatial and temporal co-ordinate in the NLS is straightforward \cite{osborne2010nonlinear}; since at leading-order of approximation $\dfrac{\partial\psi}{\partial\xi} + \beta_1 \dfrac{\partial\psi}{\partial\tau}=0$ holds, thus Eq. (\ref{timenlse}) can be transformed to a space-like equation similar to BEC 
\begin{eqnarray}
i(\frac{\partial \psi}{\partial \tau} + c_g \frac{\partial \psi}{\partial \xi})+\delta\frac{\partial^2 \psi}{\partial \xi^2}+\nu|\psi|^2\psi=0,
\label{spaceNLSEhydro}
\end{eqnarray}
with $\delta=-\dfrac{c_g^3\beta_2}{2}$, and $\nu=\gamma c_g$. In deep-water, these coefficients admit a simple parametrization and correspond to $c_g=\dfrac{\partial \omega}{\partial k}=\dfrac{1}{\beta_1}=\dfrac{\omega_0}{2k_0}$, $\delta=-\dfrac{\omega_0}{8k_0^2}$, and $\nu=-\dfrac{\omega_0 k_0^2}{2}$, where 
$\omega_0$ and $k_0$ represent the frequency and wavenumber of the carrier wave under
consideration. Note that the latter space-like representation can be rigorously derived from the Euler equations of motion for an incompressible, inviscid, and irrotational fluid  using the multiple scales approach \cite{mei2005theory}, or equivalently from the Zakharov equation when the spectrum is assumed to be narrowband \cite{zakharov1968stability}. This formulation naturally allows the initialization of numerical experiments through an initial conditions, as is also commonly done \cite{kharif2008rogue}. 

The choice between temporal and spatial evolution to study the water wave problem in an experiment, is dictated by the available measurement techniques. Visual camera-based measurements in the ocean capture the wave field over a broad spatial domain and therefore naturally favor a space-like evolution framework. In contrast, laboratory experiments in wave flumes or basins require wave makers driven by prescribed time-series, while the common use of point-wise wave gauges. This necessitates a time-like representation of the wave dynamics~\cite{osborne2010nonlinear}.

A heuristic derivation is not limited to hydrodynamic and electromagnetic waves, but can also be extended to nonlinear transmission lines \cite{remoissenet2013waves}. 
It is well-known that for $\delta\nu>0$ (or $-\beta_2\gamma>0$), the NLS equation is of the focusing type and a wave train may undergo MI when subjected to a long-wave wave perturbation. When considering more broadband wave processes with spectral bandwidth $\Delta k$, it is useful to define a so-called Benjamin-Feir Index (BFI), which is a ratio of wave steepness $\sqrt{\langle|\psi|^2\rangle k_0^2}$ to the normalized and relative spectral bandwidth $\dfrac{\Delta k}{k_0}$ \cite{janssen2003nonlinear,chabchoub2015nonlinear} to predict unstable wave growth in such conditions. 
Thus, for an irregular wave field in BECs, and as inspired from the water wave problem, this translates to 
\begin{eqnarray}
BFI_{BEC}=\dfrac{\sqrt{\langle|\psi|^2\rangle k_0^2}}{\dfrac{\Delta k}{k_0}}=\dfrac{\sqrt{\sigma^2k^4_0}}{\Delta k},
\end{eqnarray} 
$\sigma^2$ being the variance of
the carrier wave. 
It is also worth noting that this one-to-one analogy of wave dynamics between nonlinear dispersive media holds at third-order of approximation, i.e., at the NLS  level. For instance, when wave nonlinearity, expressed through the water wave steepness which also serves as the perturbation expansion parameter, becomes significant, it is more appropriate to consider the MNLS for wave modeling purposes \cite{dysthe1979note,trulsen2001spatial,kit2002spatial,goullet2011numerical} 

\begin{eqnarray} i\frac{\partial \psi}{\partial \xi} -\,\frac{k_0}{\omega_0^{2}}\frac{\partial^{2}\psi}{\partial \tau^{2}} -k_0^{3}\lvert \psi\rvert^{2}\psi - i\frac{k_0^{3}}{\omega_0}\left( 6\lvert \psi\rvert^{2}\frac{\partial \psi}{\partial \tau} + 2\psi\frac{\partial \lvert \psi\rvert^{2}}{\partial \tau} - 2i\psi\mathcal{H}\!\left[\frac{\partial \lvert \psi\rvert^{2}}{\partial \tau}\right] \right)=0,\nonumber \\ \end{eqnarray}

Here, $\mathcal{H}$ denotes the Hilbert transform, which accounts for the wave-induced mean flow. The MNLS arises at the fourth-order of the asymptotic expansion and provides a different evolution framework compared with its optical GNLS counterpart (for $m=2$). Nevertheless, similar soliton-related phenomena, such as soliton fission, can be observed and investigated \cite{dudley2006supercontinuum,chabchoub2013hydrodynamic}. Overall, employing this higher-order NLS framework is important in optics (GNLS) and hydrodynamics (MNLS), as it offers a more quantitative description of extreme wave dynamics, which is particularly relevant for validation against laboratory experiments.

\subsection{Akhmediev breather and the nonlinear interpretation of MI} 
The MI of weakly nonlinear waves was independently discovered by several pioneering researchers in the context of specific systems, such as water \cite{benjamin1967disintegration,zakharov1968stability} or electromagnetic waves \cite{bespalov1966filamentary}, as well as more general weakly nonlinear wave systems \cite{benney1967propagation}. The classical result demonstrates that a narrowband wave train becomes unstable to long-wave perturbations. This instability manifests as the excitation and exponential growth of a symmetric pair of sidebands around the carrier peak wavenumber; or equivalently, around the carrier frequency when considering the time-series evolution of the wave field.

To relate this to the NLS equation, we can begin by considering its dimensionless form, as previously outlined in Eq. (\ref{NLSEq}), with its plane wave solution, $\Psi(x,t) = \exp(it)$. It is straightforward to show that this solution is unstable to long-wave perturbations when the perturbation wavenumber $K$ satisfies the bound $0 < K < 2$. When we translate this to the hydrodynamic dimensional form, i.e., the solution $\psi(\xi,\tau) = a_0 \exp\left( i \frac{a_0^2 k_0^2 \omega_0}{2} \tau \right)$ of Eq. (\ref{spaceNLSEhydro}), the instability condition becomes $0 < K < 2\sqrt{2} k_0^2 a_0$. This result exactly aligns with the findings in \cite{benjamin1967disintegration}, which applied linear stability analysis to second-order Stokes waves, which are derived from the hydrodynamic Euler equations.

To summarize, the unifying results of the linear stability analysis provides only a criterion for triggering the instability. The specific waveforms, however, depend on numerical NLS  simulations, as demonstrated in the late 1970s \cite{yuen1978relationship}. With the discovery of the AB  \cite{akhmediev1985generation,akhmediev1987exact}, the understanding evolved while it took a few decades to fully realize the usefulness of these solutions, for instance for triggering MI in laboratory settings at any arbitrary stage. 
\begin{figure*}[t]
\centering
\includegraphics[width=0.45\textwidth]{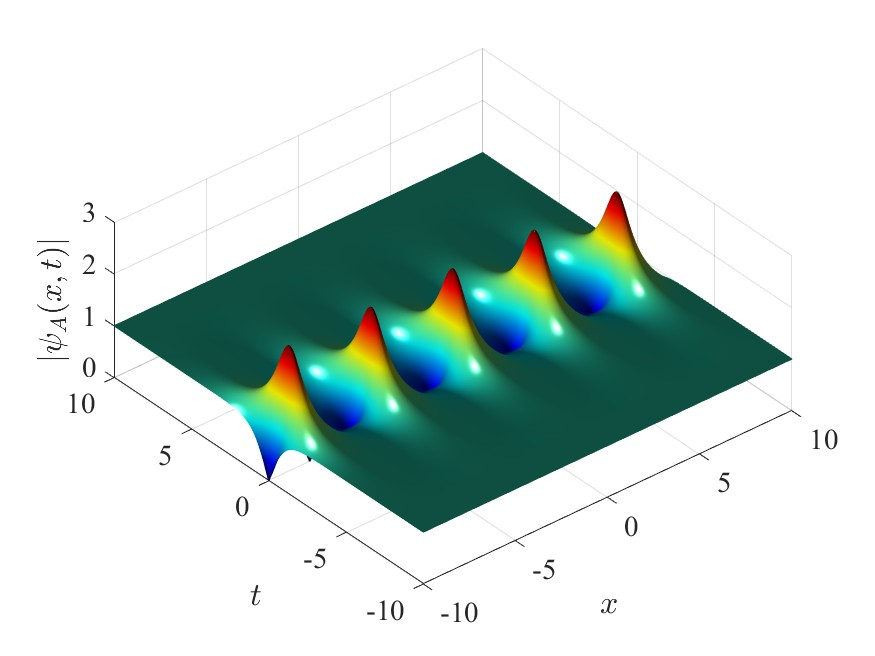}
\includegraphics[width=0.45\textwidth]{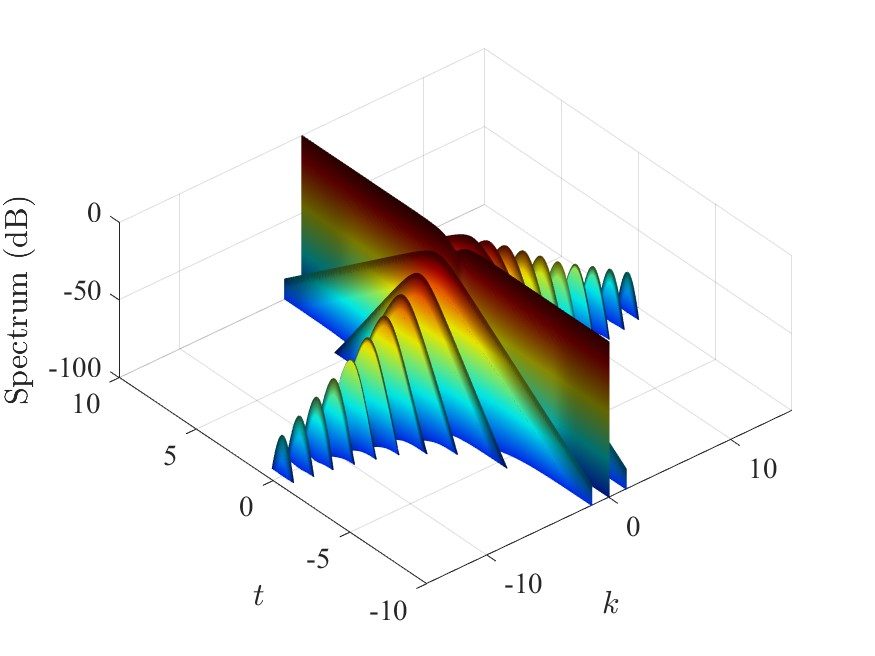}
\caption{Left: AB for 
$B=\sqrt{2}$ in (Eq. \ref{eq:AB}), corresponding to the case of maximal growth rate. Right: corresponding spectral evolution depicting the complex side-bands evolution. A similar figure and discussion have been reported in \cite{wetzel2011new}.}
\label{AB}
\end{figure*}
In fact, the use of ABs removes any constraints, which impose the initialization of an unstable wave using a small-amplitude sideband pair. Indeed, the AB evolution involves more than one sideband pair (sideband cascade) throughout the entire process and thus, providing a comprehensive description of the evolution of all infinite sidebands and unstable nonlinear waves overall at their nonlinear stage. This is illustrated in Fig. \ref{AB}. 

\begin{figure*}[t]
\centering
\includegraphics[width=0.9\textwidth]{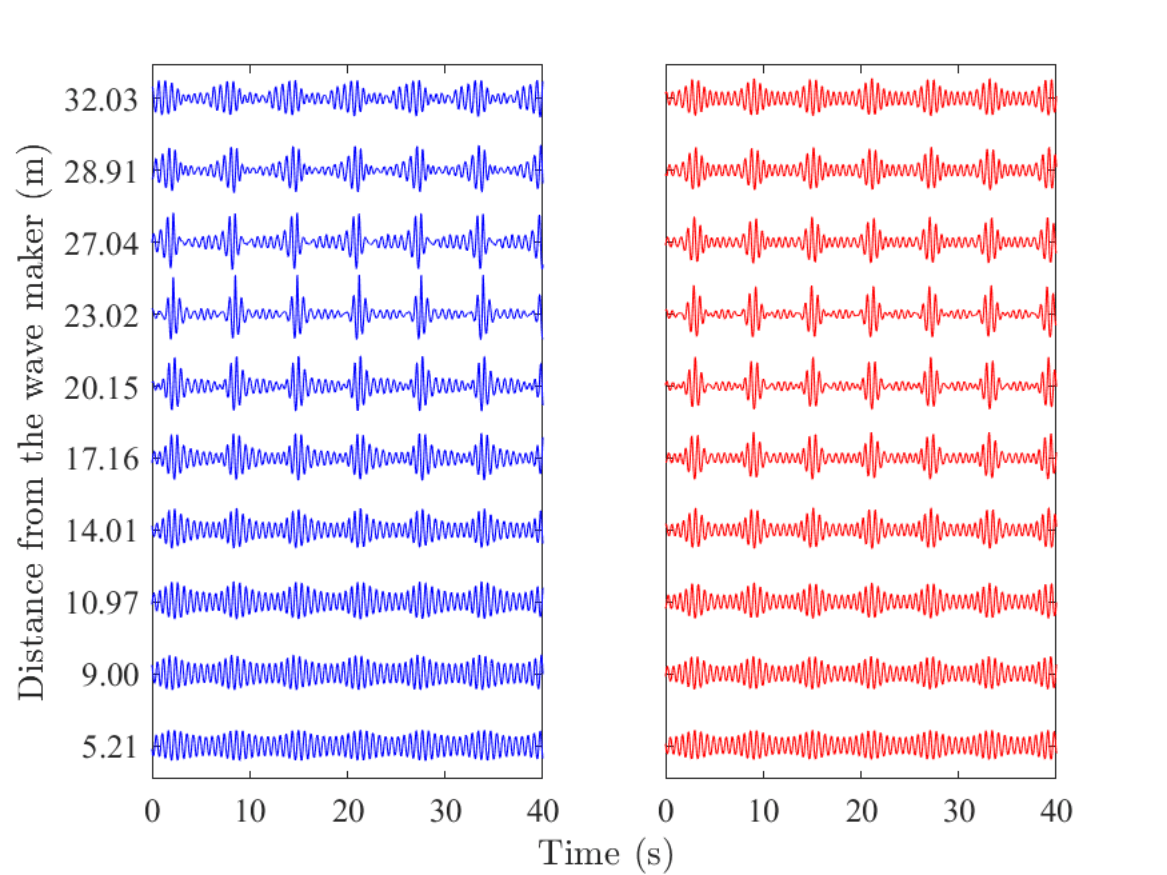}
\caption{AB evolution (case of maximal growth rate $B=\sqrt{2}$) for the carrier parameters $a_0=0.01$ m and $a_0k_0=0.1$, as measured from several discrete locations. Left: Experiments. Right: Hydrodynamic NLS Predictions.}
\label{AB_exp}
\end{figure*}

An example of an AB evolution for the case of maximal growth rate $B=\sqrt{2}$ in a water wave tank, and thus in a controlled setting, together with hydrodynamic NLS predictions are shown in Fig. \ref{AB_exp}.
To ensure generation within the frequency range of the wave maker and an evolution without wave breaking, the carrier wave parameters have been chosen to be $a_0=0.01$ m and $a_0k_0=0.1$. The wave frequency can then be determined from the linear dispersion relation $\omega_0=\sqrt{gk_0}=9.9$ s$^{-1}$. 

Although the hydrodynamic NLS framework describes wave evolution only up to the third-order of approximation (of the Euler equations of motion), while also accounting for its limitations assuming narrowband processes, the agreement between theory and experiments is quite remarkable. This is also true for the observation of the Peregrine breather \cite{Chabchoub_water}, which represents a special case with zero modulation wavenumber and maximum amplification within the context of classical MI theory. 

\subsection{Spontaneous MI} 

\begin{figure*}[t]
\centering
\includegraphics[width=0.49\textwidth]{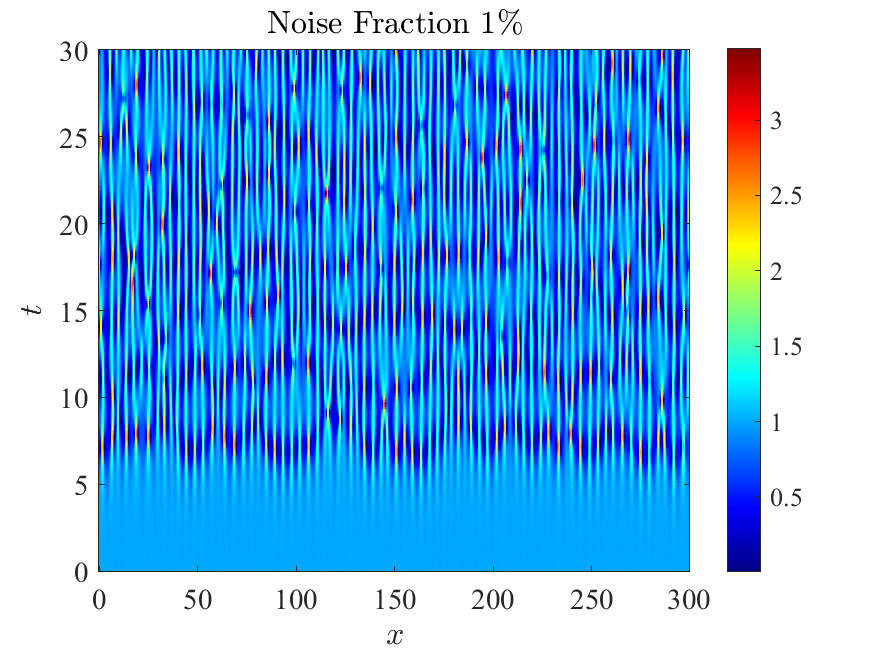}
\includegraphics[width=0.49\textwidth]{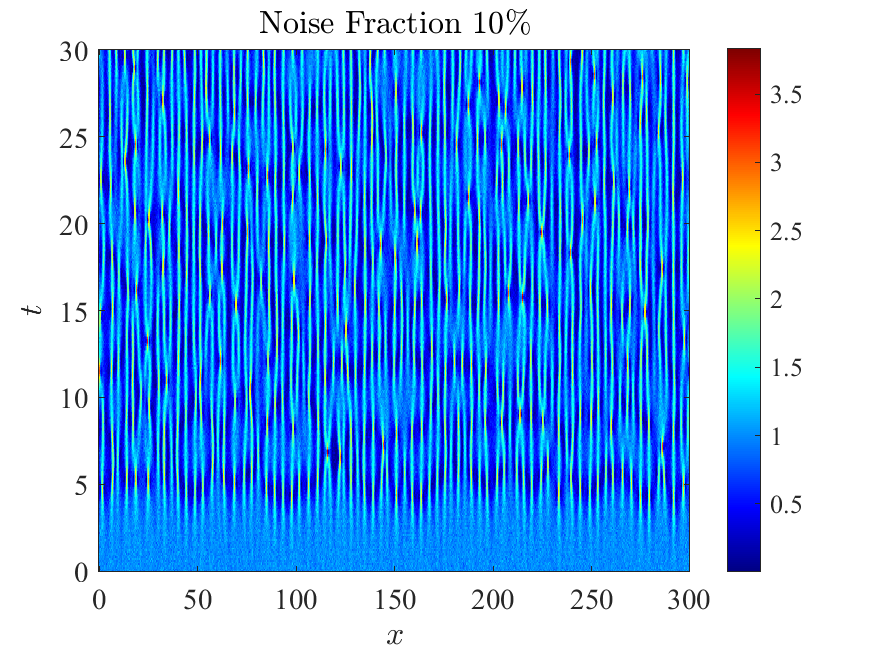}
\caption{Wave power evolution of an initial constant condensate with different noise fraction following Eq. (\ref{NLSEq}). Left: Noise fraction 1$\%$. Right: Noise fraction 10$\%$. Increasing the noise fraction accelerates the initial formation of extreme events.}
\label{SMI}
\end{figure*}

While breathers have now been observed in several nonlinear dispersive media, a natural first step toward understanding their spontaneous formation in more complex settings is to investigate their emergence from a noisy condensate \cite{toenger2015emergent}. Readers may have encountered the initiation of MI  in numerical simulations of a plane wave condensate/background governed by the NLS equation. In such simulations, numerical noise first excites the sideband with the maximal growth rate, followed by the growth of the opposite sideband, after which nonlinear interactions between the two sidebands develop. As demonstrated in that work, fundamental breather solutions, including ABs, KMs and PSs as well as higher-order breather envelopes can originate from such small white-noise perturbations, and the resulting complex envelope dynamics resemble chaotic wave propagation and interaction. This nonlinear and unstable wave evolution is commonly interpreted as a manifestation of the nonlinear stage of MI, which can be statistically studied. A rigorous theoretical framework for this phenomenon has also been developed within the context of integrable turbulence \cite{soto2016integrable}. Examples of such complex envelope dynamics are illustrated in Fig. \ref{SMI}.

This intrinsic nonlinear and recurrent extreme wave dynamics has been observed in controlled laboratory settings, most notably in optics \cite{narhi2016real,suret2016single} and hydrodynamics \cite{chabchoub2017experiments}, confirming statistical NLS predictions and demonstrating that collisions of fundamental breathers can generate higher-order breather events. It is important to note that water wave experiments are not as clean as their optical counterparts, since the evolution of hydrodynamic waves is influenced by wave breaking and subsequent energy loss, which becomes uncontrollable in a laboratory setting due to the large amplitude amplification that higher-order breathers can attain. 
Accordingly, the magnitude of nonlinearity and associated spectral bandwidth favor the observations of breathers in optics without major departure from NLS predictions compared to water waves, since the NLS  dispersion and nonlinearity parameters of typical optical fibers are two orders of magnitude smaller than their hydrodynamic counterparts \cite{waseda2019asymmetric}. 

\subsection{Unexplored Breather Dynamics in BECs}

Many types of breather solutions have been experimentally studied in controlled experiments, particularly in optics and hydrodynamics. These observations provide valuable insights into the validity and limitations of weakly nonlinear theory, as the solutions of such systems offer crucial information about key dynamical features of localized structures and nonlinear wave evolution in the respective medium. The first observation of a PS dynamics in BECs serves as a gateway to a multitude of future experiments, where proper manipulation of these localized structures in BECs presents not only opportunities but also challenges. As each new structure emerges, it requires theoretical triggering and subsequent translation into specific initial conditions for experimental realization.

A natural next step in these experiments involves exploring the fundamental periodic hierarchy of breather families, with the PS being the limiting case, i.e., AB and KM breathers, which provide a platform for the study of MI and more complex structures, such as higher-order ABs, KM breathers, and HORWs, resulting from intrinsic interaction of fundamental structures. Moreover, the evolution of breather structures on a periodic 
(cnoidal) CN or DN background could reveal new insights of more complicated MI features in BECs beyond the classical MI theory requiring the start of the instability from a long-wave perturbation of a regular wave. Last but certainly not least, observations of higher-order solitons on a zero background, including degenerate solitons, offer a promising avenue for advancing our understanding of highly nonlinear waves in BECs, which could potentially lead to supercontinuum generation and thus, strengthen the analogies to other nonlinear dispersive media.

\section{Outlook and Future Perspectives}\label{sec:conclusions}

RWs -- namely highly localized, large-amplitude excitations -- appear across many physical systems but ultracold atomic gases now offer unprecedented control and tunability for probing their quantum nature, dynamical birth, evolution and dissolvement. 
Their central theoretical framework is provided by the focusing nonlinear Schr\"odinger equation, that permits exact analytical rational  solutions within the integrable limit like the PS, the KM and AB breathers, and HORW structures with increasingly complex (multi-lobe) spatiotemporal profiles. 

A main theme delineated herein concerns the mechanisms related to the seeding of RWs. 
MI constitutes one of the primary pathways leading to RW generation where small amplitude perturbations atop a finite background grow and reorganize into localized extreme events. 
Complementary mechanisms revolve around gradient catastrophes caused, e.g., by Gaussian initial conditions in the semiclassical limit of dispersive hydrodynamics and dam-break flows initiated by step-like initial conditions. 
For instance, it is showcased that, depending on the width and amplitude of Gaussian initial states, a variety of RW structures such as the PS spike,  periodic breathers, or multi-peak “Christmas-tree’’ cascades can emerge. 
Moreover, attempts toward appreciating the stability of such entities are discussed based on Floquet analysis exploiting the periodicity of the KM breather, revealed that instabilities arise primarily from the ensuing unstable background, although they spatially emerge
in the spatial vicinity of the KM structure.

Another focal point of the Chapter concerned RW creation in  highly-population-imbalanced immiscible repulsive two-component BECs by leveraging their reduction to an effective attractively interacting single-component system. 
This enables the controlled nucleation of RWs despite all interactions being repulsive avoiding in this way the complexity of wave collapse inevitably arising in attractive gases. 
This has led to the first experimental observation of the PS in a two-component $^{87}$Rb BEC. 
In this context, an optically induced Gaussian potential well enforces the development of spatiotemporal localization accompanied by side density nodes characteristic of a PS signature before its eventual decay into solitonic fragments.

Additionally, HORWs -- i.e., multi-lobed, higher than the PS amplitude structures -- were also investigated. 
These can be dynamically generated using semicircular initial data naturally arising in non-integrable repulsive condensates  from Thomas-Fermi distributions leading to different HORWs in different dimensions upon considering interaction quenches towards the attractive regime with simultaneous trap release. 
Further extensions to quantum droplets described by extended Gross-Pitaevskii models encompassing quantum fluctuations reveal new families of RWs in competing repulsive-attractive interaction environments. 
Appearances of RWs in water tanks have also been outlined providing a broader perspective for their occurrence. In representative JONSWAP ocean wave fields, the influence of wave nonlinearity on rogue wave formation has been experimentally demonstrated to be critical, highlighting the importance of nonlinear mechanisms, particularly quasi-four-wave resonant interactions (or MI), as well as the suitability of the NLS framework for capturing rogue wave statistics \cite{onorato2006extreme}. Moreover, the multiple laboratory observations of fundamental breather solutions on the water surface \cite{chabchoub2016hydrodynamic} have further confirmed the relevance of these distinctive nonlinear dynamical features, including in realistic oceanic conditions \cite{chabchoub2016tracking,waseda2021directional}. Here, further work is required to clarify the role of MI under more complex, multi-component ocean wave conditions. 

This collection of topics highlights the flexibility of ultracold quantum simulators for exploring extreme events and nonlinear excitations in both integrable and non-integrable regimes, while linking fundamental nonlinear science with quantum many-body physics. 
At the same time, we have dedicated a significant portion of the Chapter illustrating
the pervasive nature of related ideas in a wide variety of different contexts featuring
dispersion and nonlinearity, including nonlinear optics and dispersive hydrodynamics.
Despite the progress achieved so far, the understanding of the quantum nature, controllability and observation of RWs is still at its initial stages with a plethora of outstanding open questions awaiting explanations. 
Indeed, the PS realization motivates further studies on engineering external potentials in order to nucleate more complex RW configurations such as the AB and the KM that are limiting cases of the PS but also HORWs and vectors variants thereof as in the recent experiment of~\cite{bougas2025observation}. 
In this vein, interactions between such wave structures (or between them and other
entities, such as defects or solitary waves) would be desirable too since currently it is unknown how these events couple. 
An additional intriguing pathway is to explore connections of RW structures and their catastrophe with the potential emergence of soliton gases~\cite{Suret_solitongas}. 
Deviating further from the integrable limit a promising direction would be to study RW formation in long-range interacting ultracold dipolar platforms hosting exotic phases-of-matter such as droplets and supersolids~\cite{chomaz2022dipolar}.  
Certainly, explorations of quantum variants of the aforementioned RWs is a highly intriguing avenue that is currently almost at its infancy and requires the utilization of sophisticated ab-initio numerical techniques~\cite{cao2017unified,mistakidis2023few} that are able to capture interparticle correlations and entanglement.  Such topics are
currently starting to be investigated~\cite{diplaris2025correlatedmanybodyquantumdynamics} and, undoubtedly, will lead to numerous
future publications.

\begin{acknowledgments}
This preprint will appear as a chapter in the Springer book entitled Short and Long
Range Quantum Atomic Platforms — Theoretical and Experimental Developments
(provisional title), edited by P. G. Kevrekidis, C. L. Hung, and S. I. Mistakidis. 
The authors are thankful to Professors Nail Akhmediev, Gino Biondini, George Bougas, Stathis Charalampidis, Jesus Cuevas, John M. Dudley, Dimitri Frantzeskakis,  Nikos Karachalios for extensive discussions and
collaboration on Rogue waves. A.C. acknowledges support from Okinawa Institute of Science and Technology (OIST) with subsidy funding from the Cabinet Office, Government of Japan. S.I.M acknowledges support by the Army Research Office under Award number: W911NF-26-1-A043.  
P.G.K. acknowledges support from the National Science Foundation (NSF) under the awards PHY-2408988 and DMS-2204702.
This research was partly conducted while P.G.K. was visiting the Okinawa Institute of Science and
Technology (OIST) through the Theoretical Sciences Visiting Program (TSVP). 
This work was also supported by a grant from the Simons Foundation
[SFI-MPS-SFM-00011048, P.G.K]. P.E. acknowledges support from the National Science Foundation (NSF) under the awards PHY-2207588 and PHY-2513366. M.E.M. and P.E. acknowledge NSF support under the award OSI-2427154.
\end{acknowledgments}

\bibliographystyle{apsrev4-1}
\bibliography{RougeWaves}

\begin{thebibliography}{154}%
\makeatletter
\providecommand \@ifxundefined [1]{%
 \@ifx{#1\undefined}
}%
\providecommand \@ifnum [1]{%
 \ifnum #1\expandafter \@firstoftwo
 \else \expandafter \@secondoftwo
 \fi
}%
\providecommand \@ifx [1]{%
 \ifx #1\expandafter \@firstoftwo
 \else \expandafter \@secondoftwo
 \fi
}%
\providecommand \natexlab [1]{#1}%
\providecommand \enquote  [1]{``#1''}%
\providecommand \bibnamefont  [1]{#1}%
\providecommand \bibfnamefont [1]{#1}%
\providecommand \citenamefont [1]{#1}%
\providecommand \href@noop [0]{\@secondoftwo}%
\providecommand \href [0]{\begingroup \@sanitize@url \@href}%
\providecommand \@href[1]{\@@startlink{#1}\@@href}%
\providecommand \@@href[1]{\endgroup#1\@@endlink}%
\providecommand \@sanitize@url [0]{\catcode `\\12\catcode `\$12\catcode `\&12\catcode `\#12\catcode `\^12\catcode `\_12\catcode `\%12\relax}%
\providecommand \@@startlink[1]{}%
\providecommand \@@endlink[0]{}%
\providecommand \url  [0]{\begingroup\@sanitize@url \@url }%
\providecommand \@url [1]{\endgroup\@href {#1}{\urlprefix }}%
\providecommand \urlprefix  [0]{URL }%
\providecommand \Eprint [0]{\href }%
\providecommand \doibase [0]{http://dx.doi.org/}%
\providecommand \selectlanguage [0]{\@gobble}%
\providecommand \bibinfo  [0]{\@secondoftwo}%
\providecommand \bibfield  [0]{\@secondoftwo}%
\providecommand \translation [1]{[#1]}%
\providecommand \BibitemOpen [0]{}%
\providecommand \bibitemStop [0]{}%
\providecommand \bibitemNoStop [0]{.\EOS\space}%
\providecommand \EOS [0]{\spacefactor3000\relax}%
\providecommand \BibitemShut  [1]{\csname bibitem#1\endcsname}%
\let\auto@bib@innerbib\@empty
\bibitem [{\citenamefont {Akhmediev}\ \emph {et~al.}(2009)\citenamefont {Akhmediev}, \citenamefont {Ankiewicz},\ and\ \citenamefont {Taki}}]{akhmediev2009waves}%
  \BibitemOpen
  \bibfield  {author} {\bibinfo {author} {\bibfnamefont {N.}~\bibnamefont {Akhmediev}}, \bibinfo {author} {\bibfnamefont {A.}~\bibnamefont {Ankiewicz}}, \ and\ \bibinfo {author} {\bibfnamefont {M.}~\bibnamefont {Taki}},\ }\href@noop {} {\bibfield  {journal} {\bibinfo  {journal} {Phys. Lett. A}\ }\textbf {\bibinfo {volume} {373}},\ \bibinfo {pages} {675} (\bibinfo {year} {2009})}\BibitemShut {NoStop}%
\bibitem [{\citenamefont {Kharif}\ \emph {et~al.}(2008)\citenamefont {Kharif}, \citenamefont {Pelinovsky},\ and\ \citenamefont {Slunyaev}}]{kharif2008rogue}%
  \BibitemOpen
  \bibfield  {author} {\bibinfo {author} {\bibfnamefont {C.}~\bibnamefont {Kharif}}, \bibinfo {author} {\bibfnamefont {E.}~\bibnamefont {Pelinovsky}}, \ and\ \bibinfo {author} {\bibfnamefont {A.}~\bibnamefont {Slunyaev}},\ }\href@noop {} {\emph {\bibinfo {title} {Rogue waves in the ocean}}}\ (\bibinfo  {publisher} {Springer Science \& Business Media},\ \bibinfo {year} {2008})\BibitemShut {NoStop}%
\bibitem [{\citenamefont {Chabchoub}\ \emph {et~al.}(2011)\citenamefont {Chabchoub}, \citenamefont {Hoffmann},\ and\ \citenamefont {Akhmediev}}]{Chabchoub_water}%
  \BibitemOpen
  \bibfield  {author} {\bibinfo {author} {\bibfnamefont {A.}~\bibnamefont {Chabchoub}}, \bibinfo {author} {\bibfnamefont {N.~P.}\ \bibnamefont {Hoffmann}}, \ and\ \bibinfo {author} {\bibfnamefont {N.}~\bibnamefont {Akhmediev}},\ }\href {\doibase 10.1103/PhysRevLett.106.204502} {\bibfield  {journal} {\bibinfo  {journal} {Phys. Rev. Lett.}\ }\textbf {\bibinfo {volume} {106}},\ \bibinfo {pages} {204502} (\bibinfo {year} {2011})}\BibitemShut {NoStop}%
\bibitem [{\citenamefont {Chabchoub}\ \emph {et~al.}(2012)\citenamefont {Chabchoub}, \citenamefont {Hoffmann}, \citenamefont {Onorato},\ and\ \citenamefont {Akhmediev}}]{Chabchoub_water1}%
  \BibitemOpen
  \bibfield  {author} {\bibinfo {author} {\bibfnamefont {A.}~\bibnamefont {Chabchoub}}, \bibinfo {author} {\bibfnamefont {N.}~\bibnamefont {Hoffmann}}, \bibinfo {author} {\bibfnamefont {M.}~\bibnamefont {Onorato}}, \ and\ \bibinfo {author} {\bibfnamefont {N.}~\bibnamefont {Akhmediev}},\ }\href {\doibase 10.1103/PhysRevX.2.011015} {\bibfield  {journal} {\bibinfo  {journal} {Phys. Rev. X}\ }\textbf {\bibinfo {volume} {2}},\ \bibinfo {pages} {011015} (\bibinfo {year} {2012})}\BibitemShut {NoStop}%
\bibitem [{\citenamefont {Kibler}\ \emph {et~al.}(2010)\citenamefont {Kibler}, \citenamefont {Fatome}, \citenamefont {Finot}, \citenamefont {Millot}, \citenamefont {Dias}, \citenamefont {Genty}, \citenamefont {Akhmediev},\ and\ \citenamefont {Dudley}}]{kibler2010peregrine}%
  \BibitemOpen
  \bibfield  {author} {\bibinfo {author} {\bibfnamefont {B.}~\bibnamefont {Kibler}}, \bibinfo {author} {\bibfnamefont {J.}~\bibnamefont {Fatome}}, \bibinfo {author} {\bibfnamefont {C.}~\bibnamefont {Finot}}, \bibinfo {author} {\bibfnamefont {G.}~\bibnamefont {Millot}}, \bibinfo {author} {\bibfnamefont {F.}~\bibnamefont {Dias}}, \bibinfo {author} {\bibfnamefont {G.}~\bibnamefont {Genty}}, \bibinfo {author} {\bibfnamefont {N.}~\bibnamefont {Akhmediev}}, \ and\ \bibinfo {author} {\bibfnamefont {J.~M.}\ \bibnamefont {Dudley}},\ }\href@noop {} {\bibfield  {journal} {\bibinfo  {journal} {Nature Phys.}\ }\textbf {\bibinfo {volume} {6}},\ \bibinfo {pages} {790} (\bibinfo {year} {2010})}\BibitemShut {NoStop}%
\bibitem [{\citenamefont {Solli}\ \emph {et~al.}(2007)\citenamefont {Solli}, \citenamefont {Ropers}, \citenamefont {Koonath},\ and\ \citenamefont {Jalali}}]{solli2007optical}%
  \BibitemOpen
  \bibfield  {author} {\bibinfo {author} {\bibfnamefont {D.~R.}\ \bibnamefont {Solli}}, \bibinfo {author} {\bibfnamefont {C.}~\bibnamefont {Ropers}}, \bibinfo {author} {\bibfnamefont {P.}~\bibnamefont {Koonath}}, \ and\ \bibinfo {author} {\bibfnamefont {B.}~\bibnamefont {Jalali}},\ }\href@noop {} {\bibfield  {journal} {\bibinfo  {journal} {Nature}\ }\textbf {\bibinfo {volume} {450}},\ \bibinfo {pages} {1054} (\bibinfo {year} {2007})}\BibitemShut {NoStop}%
\bibitem [{\citenamefont {Dudley}\ \emph {et~al.}(2014)\citenamefont {Dudley}, \citenamefont {Dias}, \citenamefont {Erkintalo},\ and\ \citenamefont {Genty}}]{dudley2014instabilities}%
  \BibitemOpen
  \bibfield  {author} {\bibinfo {author} {\bibfnamefont {J.~M.}\ \bibnamefont {Dudley}}, \bibinfo {author} {\bibfnamefont {F.}~\bibnamefont {Dias}}, \bibinfo {author} {\bibfnamefont {M.}~\bibnamefont {Erkintalo}}, \ and\ \bibinfo {author} {\bibfnamefont {G.}~\bibnamefont {Genty}},\ }\href@noop {} {\bibfield  {journal} {\bibinfo  {journal} {Nature Photon.}\ }\textbf {\bibinfo {volume} {8}},\ \bibinfo {pages} {755} (\bibinfo {year} {2014})}\BibitemShut {NoStop}%
\bibitem [{\citenamefont {Bailung}\ \emph {et~al.}(2011)\citenamefont {Bailung}, \citenamefont {Sharma},\ and\ \citenamefont {Nakamura}}]{Bailung_plasmas}%
  \BibitemOpen
  \bibfield  {author} {\bibinfo {author} {\bibfnamefont {H.}~\bibnamefont {Bailung}}, \bibinfo {author} {\bibfnamefont {S.~K.}\ \bibnamefont {Sharma}}, \ and\ \bibinfo {author} {\bibfnamefont {Y.}~\bibnamefont {Nakamura}},\ }\href {\doibase 10.1103/PhysRevLett.107.255005} {\bibfield  {journal} {\bibinfo  {journal} {Phys. Rev. Lett.}\ }\textbf {\bibinfo {volume} {107}},\ \bibinfo {pages} {255005} (\bibinfo {year} {2011})}\BibitemShut {NoStop}%
\bibitem [{\citenamefont {Ganshin}\ \emph {et~al.}(2008)\citenamefont {Ganshin}, \citenamefont {Efimov}, \citenamefont {Kolmakov}, \citenamefont {Mezhov-Deglin},\ and\ \citenamefont {McClintock}}]{Ganshin}%
  \BibitemOpen
  \bibfield  {author} {\bibinfo {author} {\bibfnamefont {A.~N.}\ \bibnamefont {Ganshin}}, \bibinfo {author} {\bibfnamefont {V.~B.}\ \bibnamefont {Efimov}}, \bibinfo {author} {\bibfnamefont {G.~V.}\ \bibnamefont {Kolmakov}}, \bibinfo {author} {\bibfnamefont {L.~P.}\ \bibnamefont {Mezhov-Deglin}}, \ and\ \bibinfo {author} {\bibfnamefont {P.~V.~E.}\ \bibnamefont {McClintock}},\ }\href {\doibase 10.1103/PhysRevLett.101.065303} {\bibfield  {journal} {\bibinfo  {journal} {Phys. Rev. Lett.}\ }\textbf {\bibinfo {volume} {101}},\ \bibinfo {pages} {065303} (\bibinfo {year} {2008})}\BibitemShut {NoStop}%
\bibitem [{\citenamefont {Shats}\ \emph {et~al.}(2010)\citenamefont {Shats}, \citenamefont {Punzmann},\ and\ \citenamefont {Xia}}]{Shats}%
  \BibitemOpen
  \bibfield  {author} {\bibinfo {author} {\bibfnamefont {M.}~\bibnamefont {Shats}}, \bibinfo {author} {\bibfnamefont {H.}~\bibnamefont {Punzmann}}, \ and\ \bibinfo {author} {\bibfnamefont {H.}~\bibnamefont {Xia}},\ }\href {\doibase 10.1103/PhysRevLett.104.104503} {\bibfield  {journal} {\bibinfo  {journal} {Phys. Rev. Lett.}\ }\textbf {\bibinfo {volume} {104}},\ \bibinfo {pages} {104503} (\bibinfo {year} {2010})}\BibitemShut {NoStop}%
\bibitem [{\citenamefont {Romero-Ros}\ \emph {et~al.}(2024)\citenamefont {Romero-Ros}, \citenamefont {Katsimiga}, \citenamefont {Mistakidis}, \citenamefont {Mossman}, \citenamefont {Biondini}, \citenamefont {Schmelcher}, \citenamefont {Engels},\ and\ \citenamefont {Kevrekidis}}]{Romero_Ros_2024}%
  \BibitemOpen
  \bibfield  {author} {\bibinfo {author} {\bibfnamefont {A.}~\bibnamefont {Romero-Ros}}, \bibinfo {author} {\bibfnamefont {G.~C.}\ \bibnamefont {Katsimiga}}, \bibinfo {author} {\bibfnamefont {S.~I.}\ \bibnamefont {Mistakidis}}, \bibinfo {author} {\bibfnamefont {S.}~\bibnamefont {Mossman}}, \bibinfo {author} {\bibfnamefont {G.}~\bibnamefont {Biondini}}, \bibinfo {author} {\bibfnamefont {P.}~\bibnamefont {Schmelcher}}, \bibinfo {author} {\bibfnamefont {P.}~\bibnamefont {Engels}}, \ and\ \bibinfo {author} {\bibfnamefont {P.~G.}\ \bibnamefont {Kevrekidis}},\ }\href {\doibase 10.1103/PhysRevLett.132.033402} {\bibfield  {journal} {\bibinfo  {journal} {Phys. Rev. Lett.}\ }\textbf {\bibinfo {volume} {132}},\ \bibinfo {pages} {033402} (\bibinfo {year} {2024})}\BibitemShut {NoStop}%
\bibitem [{\citenamefont {Bougas}\ \emph {et~al.}(2025)\citenamefont {Bougas}, \citenamefont {Katsimiga}, \citenamefont {Mossman}, \citenamefont {Engels}, \citenamefont {Kevrekidis},\ and\ \citenamefont {Mistakidis}}]{bougas2025observation}%
  \BibitemOpen
  \bibfield  {author} {\bibinfo {author} {\bibfnamefont {G.}~\bibnamefont {Bougas}}, \bibinfo {author} {\bibfnamefont {G.}~\bibnamefont {Katsimiga}}, \bibinfo {author} {\bibfnamefont {S.}~\bibnamefont {Mossman}}, \bibinfo {author} {\bibfnamefont {P.}~\bibnamefont {Engels}}, \bibinfo {author} {\bibfnamefont {P.}~\bibnamefont {Kevrekidis}}, \ and\ \bibinfo {author} {\bibfnamefont {S.}~\bibnamefont {Mistakidis}},\ }\href@noop {} {\bibfield  {journal} {\bibinfo  {journal} {arXiv:2510.24917}\ } (\bibinfo {year} {2025})}\BibitemShut {NoStop}%
\bibitem [{\citenamefont {Sulem}\ and\ \citenamefont {Sulem}(2007)}]{sulem2007nonlinear}%
  \BibitemOpen
  \bibfield  {author} {\bibinfo {author} {\bibfnamefont {C.}~\bibnamefont {Sulem}}\ and\ \bibinfo {author} {\bibfnamefont {P.-L.}\ \bibnamefont {Sulem}},\ }\href@noop {} {\emph {\bibinfo {title} {The nonlinear Schr{\"o}dinger equation: self-focusing and wave collapse}}},\ Vol.\ \bibinfo {volume} {139}\ (\bibinfo  {publisher} {Springer Science \& Business Media},\ \bibinfo {year} {2007})\BibitemShut {NoStop}%
\bibitem [{\citenamefont {Fibich}(2015)}]{fibich2015nonlinear}%
  \BibitemOpen
  \bibfield  {author} {\bibinfo {author} {\bibfnamefont {G.}~\bibnamefont {Fibich}},\ }\href@noop {} {\emph {\bibinfo {title} {The nonlinear Schr{\"o}dinger equation}}},\ Vol.\ \bibinfo {volume} {192}\ (\bibinfo  {publisher} {Springer},\ \bibinfo {year} {2015})\BibitemShut {NoStop}%
\bibitem [{\citenamefont {Peregrine}(1983)}]{Peregrine1983}%
  \BibitemOpen
  \bibfield  {author} {\bibinfo {author} {\bibfnamefont {D.~H.}\ \bibnamefont {Peregrine}},\ }\href {\doibase 10.1017/S0334270000003891} {\bibfield  {journal} {\bibinfo  {journal} {J. Aust. Math. Soc. Series B, Appl. Math.}\ }\textbf {\bibinfo {volume} {25}},\ \bibinfo {pages} {16} (\bibinfo {year} {1983})}\BibitemShut {NoStop}%
\bibitem [{\citenamefont {Kuznetsov}(1977)}]{kuznetsov1977solitons}%
  \BibitemOpen
  \bibfield  {author} {\bibinfo {author} {\bibfnamefont {E.~A.}\ \bibnamefont {Kuznetsov}},\ }in\ \href@noop {} {\emph {\bibinfo {booktitle} {Akademiia Nauk SSSR Doklady}}},\ Vol.\ \bibinfo {volume} {236}\ (\bibinfo {year} {1977})\ pp.\ \bibinfo {pages} {575--577}\BibitemShut {NoStop}%
\bibitem [{\citenamefont {Ma}(1979)}]{ma1979perturbed}%
  \BibitemOpen
  \bibfield  {author} {\bibinfo {author} {\bibfnamefont {Y.-C.}\ \bibnamefont {Ma}},\ }\href@noop {} {\bibfield  {journal} {\bibinfo  {journal} {Stud. Appl. Math.}\ }\textbf {\bibinfo {volume} {60}},\ \bibinfo {pages} {43} (\bibinfo {year} {1979})}\BibitemShut {NoStop}%
\bibitem [{\citenamefont {Dubard}\ \emph {et~al.}(2010)\citenamefont {Dubard}, \citenamefont {Gaillard}, \citenamefont {Klein},\ and\ \citenamefont {Matveev}}]{dubard2010multi}%
  \BibitemOpen
  \bibfield  {author} {\bibinfo {author} {\bibfnamefont {P.}~\bibnamefont {Dubard}}, \bibinfo {author} {\bibfnamefont {P.}~\bibnamefont {Gaillard}}, \bibinfo {author} {\bibfnamefont {C.}~\bibnamefont {Klein}}, \ and\ \bibinfo {author} {\bibfnamefont {V.}~\bibnamefont {Matveev}},\ }\href@noop {} {\bibfield  {journal} {\bibinfo  {journal} {Eur. Phys. J-Spec. Top.}\ }\textbf {\bibinfo {volume} {185}},\ \bibinfo {pages} {247} (\bibinfo {year} {2010})}\BibitemShut {NoStop}%
\bibitem [{\citenamefont {Dubard}\ and\ \citenamefont {Matveev}(2011)}]{dubard2011multi}%
  \BibitemOpen
  \bibfield  {author} {\bibinfo {author} {\bibfnamefont {P.}~\bibnamefont {Dubard}}\ and\ \bibinfo {author} {\bibfnamefont {V.}~\bibnamefont {Matveev}},\ }\href@noop {} {\bibfield  {journal} {\bibinfo  {journal} {Nat. Hazards Earth Syst. Sci.}\ }\textbf {\bibinfo {volume} {11}},\ \bibinfo {pages} {667} (\bibinfo {year} {2011})}\BibitemShut {NoStop}%
\bibitem [{\citenamefont {Kedziora}\ \emph {et~al.}(2012)\citenamefont {Kedziora}, \citenamefont {Ankiewicz},\ and\ \citenamefont {Akhmediev}}]{kedziora2012second}%
  \BibitemOpen
  \bibfield  {author} {\bibinfo {author} {\bibfnamefont {D.~J.}\ \bibnamefont {Kedziora}}, \bibinfo {author} {\bibfnamefont {A.}~\bibnamefont {Ankiewicz}}, \ and\ \bibinfo {author} {\bibfnamefont {N.}~\bibnamefont {Akhmediev}},\ }\href@noop {} {\bibfield  {journal} {\bibinfo  {journal} {Phys. Rev. E}\ }\textbf {\bibinfo {volume} {85}},\ \bibinfo {pages} {066601} (\bibinfo {year} {2012})}\BibitemShut {NoStop}%
\bibitem [{\citenamefont {Kedziora}\ \emph {et~al.}(2013)\citenamefont {Kedziora}, \citenamefont {Ankiewicz},\ and\ \citenamefont {Akhmediev}}]{kedziora2013classifying}%
  \BibitemOpen
  \bibfield  {author} {\bibinfo {author} {\bibfnamefont {D.~J.}\ \bibnamefont {Kedziora}}, \bibinfo {author} {\bibfnamefont {A.}~\bibnamefont {Ankiewicz}}, \ and\ \bibinfo {author} {\bibfnamefont {N.}~\bibnamefont {Akhmediev}},\ }\href@noop {} {\bibfield  {journal} {\bibinfo  {journal} {Phys. Rev. E}\ }\textbf {\bibinfo {volume} {88}},\ \bibinfo {pages} {013207} (\bibinfo {year} {2013})}\BibitemShut {NoStop}%
\bibitem [{\citenamefont {Bilman}\ \emph {et~al.}(2020)\citenamefont {Bilman}, \citenamefont {Ling},\ and\ \citenamefont {Miller}}]{bilman2020extreme}%
  \BibitemOpen
  \bibfield  {author} {\bibinfo {author} {\bibfnamefont {D.}~\bibnamefont {Bilman}}, \bibinfo {author} {\bibfnamefont {L.}~\bibnamefont {Ling}}, \ and\ \bibinfo {author} {\bibfnamefont {P.~D.}\ \bibnamefont {Miller}},\ }\href {\doibase 10.1215/00127094-2019-0066} {\bibfield  {journal} {\bibinfo  {journal} {Duke Mathematical Journal}\ }\textbf {\bibinfo {volume} {169}},\ \bibinfo {pages} {671 } (\bibinfo {year} {2020})}\BibitemShut {NoStop}%
\bibitem [{\citenamefont {Bilman}\ and\ \citenamefont {Miller}(2022)}]{bilman2022broader}%
  \BibitemOpen
  \bibfield  {author} {\bibinfo {author} {\bibfnamefont {D.}~\bibnamefont {Bilman}}\ and\ \bibinfo {author} {\bibfnamefont {P.~D.}\ \bibnamefont {Miller}},\ }\href@noop {} {\bibfield  {journal} {\bibinfo  {journal} {Physica D: Nonlinear Phenomena}\ }\textbf {\bibinfo {volume} {435}},\ \bibinfo {pages} {133289} (\bibinfo {year} {2022})}\BibitemShut {NoStop}%
\bibitem [{\citenamefont {Suleimanov}(2017)}]{suleimanov}%
  \BibitemOpen
  \bibfield  {author} {\bibinfo {author} {\bibfnamefont {B.}~\bibnamefont {Suleimanov}},\ }\href@noop {} {\bibfield  {journal} {\bibinfo  {journal} {JETP Letters}\ }\textbf {\bibinfo {volume} {106}},\ \bibinfo {pages} {400} (\bibinfo {year} {2017})}\BibitemShut {NoStop}%
\bibitem [{\citenamefont {Zakharov}\ and\ \citenamefont {Ostrovsky}(2009)}]{zakharov2009modulation}%
  \BibitemOpen
  \bibfield  {author} {\bibinfo {author} {\bibfnamefont {V.~E.}\ \bibnamefont {Zakharov}}\ and\ \bibinfo {author} {\bibfnamefont {L.~A.}\ \bibnamefont {Ostrovsky}},\ }\href@noop {} {\bibfield  {journal} {\bibinfo  {journal} {Physica D: Nonlinear Phenomena}\ }\textbf {\bibinfo {volume} {238}},\ \bibinfo {pages} {540} (\bibinfo {year} {2009})}\BibitemShut {NoStop}%
\bibitem [{\citenamefont {Zakharov}\ and\ \citenamefont {Gelash}(2013)}]{zakharov2013nonlinear}%
  \BibitemOpen
  \bibfield  {author} {\bibinfo {author} {\bibfnamefont {V.~E.}\ \bibnamefont {Zakharov}}\ and\ \bibinfo {author} {\bibfnamefont {A.}~\bibnamefont {Gelash}},\ }\href@noop {} {\bibfield  {journal} {\bibinfo  {journal} {Physical review letters}\ }\textbf {\bibinfo {volume} {111}},\ \bibinfo {pages} {054101} (\bibinfo {year} {2013})}\BibitemShut {NoStop}%
\bibitem [{\citenamefont {Biondini}\ and\ \citenamefont {Mantzavinos}(2016)}]{biondini2016universal}%
  \BibitemOpen
  \bibfield  {author} {\bibinfo {author} {\bibfnamefont {G.}~\bibnamefont {Biondini}}\ and\ \bibinfo {author} {\bibfnamefont {D.}~\bibnamefont {Mantzavinos}},\ }\href@noop {} {\bibfield  {journal} {\bibinfo  {journal} {Phys. Rev. Lett.}\ }\textbf {\bibinfo {volume} {116}},\ \bibinfo {pages} {043902} (\bibinfo {year} {2016})}\BibitemShut {NoStop}%
\bibitem [{\citenamefont {Benjamin}\ and\ \citenamefont {Feir}(1967)}]{benjamin1967disintegration}%
  \BibitemOpen
  \bibfield  {author} {\bibinfo {author} {\bibfnamefont {T.~B.}\ \bibnamefont {Benjamin}}\ and\ \bibinfo {author} {\bibfnamefont {J.~E.}\ \bibnamefont {Feir}},\ }\href@noop {} {\bibfield  {journal} {\bibinfo  {journal} {J. Fluid Mech.}\ }\textbf {\bibinfo {volume} {27}},\ \bibinfo {pages} {417} (\bibinfo {year} {1967})}\BibitemShut {NoStop}%
\bibitem [{\citenamefont {Dubrovin}(2008)}]{dubrovin2008universality}%
  \BibitemOpen
  \bibfield  {author} {\bibinfo {author} {\bibfnamefont {B.}~\bibnamefont {Dubrovin}},\ }\href@noop {} {\bibfield  {journal} {\bibinfo  {journal} {arXiv:0804.3790}\ } (\bibinfo {year} {2008})}\BibitemShut {NoStop}%
\bibitem [{\citenamefont {Bertola}\ and\ \citenamefont {Tovbis}(2013)}]{bertola2013universality}%
  \BibitemOpen
  \bibfield  {author} {\bibinfo {author} {\bibfnamefont {M.}~\bibnamefont {Bertola}}\ and\ \bibinfo {author} {\bibfnamefont {A.}~\bibnamefont {Tovbis}},\ }\href@noop {} {\bibfield  {journal} {\bibinfo  {journal} {Commun. Pure Appl. Math.}\ }\textbf {\bibinfo {volume} {66}},\ \bibinfo {pages} {678} (\bibinfo {year} {2013})}\BibitemShut {NoStop}%
\bibitem [{\citenamefont {El}\ \emph {et~al.}(2016)\citenamefont {El}, \citenamefont {Khamis},\ and\ \citenamefont {Tovbis}}]{el2016dam}%
  \BibitemOpen
  \bibfield  {author} {\bibinfo {author} {\bibfnamefont {G.~A.}\ \bibnamefont {El}}, \bibinfo {author} {\bibfnamefont {E.~G.}\ \bibnamefont {Khamis}}, \ and\ \bibinfo {author} {\bibfnamefont {A.}~\bibnamefont {Tovbis}},\ }\href@noop {} {\bibfield  {journal} {\bibinfo  {journal} {Nonlinearity}\ }\textbf {\bibinfo {volume} {29}},\ \bibinfo {pages} {2798} (\bibinfo {year} {2016})}\BibitemShut {NoStop}%
\bibitem [{\citenamefont {Mossman}\ \emph {et~al.}(2025)\citenamefont {Mossman}, \citenamefont {Mistakidis}, \citenamefont {Katsimiga}, \citenamefont {Romero-Ros}, \citenamefont {Biondini}, \citenamefont {Schmelcher}, \citenamefont {Engels},\ and\ \citenamefont {Kevrekidis}}]{Mossman_nonlinearMI}%
  \BibitemOpen
  \bibfield  {author} {\bibinfo {author} {\bibfnamefont {S.}~\bibnamefont {Mossman}}, \bibinfo {author} {\bibfnamefont {S.~I.}\ \bibnamefont {Mistakidis}}, \bibinfo {author} {\bibfnamefont {G.~C.}\ \bibnamefont {Katsimiga}}, \bibinfo {author} {\bibfnamefont {A.}~\bibnamefont {Romero-Ros}}, \bibinfo {author} {\bibfnamefont {G.}~\bibnamefont {Biondini}}, \bibinfo {author} {\bibfnamefont {P.}~\bibnamefont {Schmelcher}}, \bibinfo {author} {\bibfnamefont {P.}~\bibnamefont {Engels}}, \ and\ \bibinfo {author} {\bibfnamefont {P.~G.}\ \bibnamefont {Kevrekidis}},\ }\href {\doibase 10.1103/6jsr-f8q1} {\bibfield  {journal} {\bibinfo  {journal} {Phys. Rev. Lett.}\ }\textbf {\bibinfo {volume} {135}},\ \bibinfo {pages} {113401} (\bibinfo {year} {2025})}\BibitemShut {NoStop}%
\bibitem [{\citenamefont {Sharan}\ \emph {et~al.}(2025)\citenamefont {Sharan}, \citenamefont {Sorribes}, \citenamefont {Sprenger}, \citenamefont {Hoefer}, \citenamefont {Engels}, \citenamefont {Ilan},\ and\ \citenamefont {Mossman}}]{Shashwat}%
  \BibitemOpen
  \bibfield  {author} {\bibinfo {author} {\bibfnamefont {S.}~\bibnamefont {Sharan}}, \bibinfo {author} {\bibfnamefont {J.~G.}\ \bibnamefont {Sorribes}}, \bibinfo {author} {\bibfnamefont {P.}~\bibnamefont {Sprenger}}, \bibinfo {author} {\bibfnamefont {M.~A.}\ \bibnamefont {Hoefer}}, \bibinfo {author} {\bibfnamefont {P.}~\bibnamefont {Engels}}, \bibinfo {author} {\bibfnamefont {B.}~\bibnamefont {Ilan}}, \ and\ \bibinfo {author} {\bibfnamefont {M.~E.}\ \bibnamefont {Mossman}},\ }\href {\doibase 10.1103/38f7-ym2v} {\bibfield  {journal} {\bibinfo  {journal} {Phys. Rev. Lett.}\ }\textbf {\bibinfo {volume} {135}},\ \bibinfo {pages} {193402} (\bibinfo {year} {2025})}\BibitemShut {NoStop}%
\bibitem [{\citenamefont {Chin}\ \emph {et~al.}(2010)\citenamefont {Chin}, \citenamefont {Grimm}, \citenamefont {Julienne},\ and\ \citenamefont {Tiesinga}}]{chin2010feshbach}%
  \BibitemOpen
  \bibfield  {author} {\bibinfo {author} {\bibfnamefont {C.}~\bibnamefont {Chin}}, \bibinfo {author} {\bibfnamefont {R.}~\bibnamefont {Grimm}}, \bibinfo {author} {\bibfnamefont {P.}~\bibnamefont {Julienne}}, \ and\ \bibinfo {author} {\bibfnamefont {E.}~\bibnamefont {Tiesinga}},\ }\href@noop {} {\bibfield  {journal} {\bibinfo  {journal} {Rev. Mod. Phys.}\ }\textbf {\bibinfo {volume} {82}},\ \bibinfo {pages} {1225} (\bibinfo {year} {2010})}\BibitemShut {NoStop}%
\bibitem [{\citenamefont {Wolswijk}\ \emph {et~al.}(2025)\citenamefont {Wolswijk}, \citenamefont {Cavicchioli}, \citenamefont {Vinelli}, \citenamefont {Chiarotti}, \citenamefont {Donati}, \citenamefont {Fernandez}, \citenamefont {Rajkov}, \citenamefont {Mancini}, \citenamefont {Vezio}, \citenamefont {Zhou} \emph {et~al.}}]{wolswijk2025trapping}%
  \BibitemOpen
  \bibfield  {author} {\bibinfo {author} {\bibfnamefont {L.}~\bibnamefont {Wolswijk}}, \bibinfo {author} {\bibfnamefont {L.}~\bibnamefont {Cavicchioli}}, \bibinfo {author} {\bibfnamefont {G.}~\bibnamefont {Vinelli}}, \bibinfo {author} {\bibfnamefont {M.}~\bibnamefont {Chiarotti}}, \bibinfo {author} {\bibfnamefont {L.}~\bibnamefont {Donati}}, \bibinfo {author} {\bibfnamefont {M.~F.}\ \bibnamefont {Fernandez}}, \bibinfo {author} {\bibfnamefont {D.~H.}\ \bibnamefont {Rajkov}}, \bibinfo {author} {\bibfnamefont {C.}~\bibnamefont {Mancini}}, \bibinfo {author} {\bibfnamefont {P.}~\bibnamefont {Vezio}}, \bibinfo {author} {\bibfnamefont {T.}~\bibnamefont {Zhou}},  \emph {et~al.},\ }\href@noop {} {\bibfield  {journal} {\bibinfo  {journal} {arXiv:2510.20790}\ } (\bibinfo {year} {2025})}\BibitemShut {NoStop}%
\bibitem [{\citenamefont {Henderson}\ \emph {et~al.}(2009)\citenamefont {Henderson}, \citenamefont {Ryu}, \citenamefont {MacCormick},\ and\ \citenamefont {Boshier}}]{henderson2009experimental}%
  \BibitemOpen
  \bibfield  {author} {\bibinfo {author} {\bibfnamefont {K.}~\bibnamefont {Henderson}}, \bibinfo {author} {\bibfnamefont {C.}~\bibnamefont {Ryu}}, \bibinfo {author} {\bibfnamefont {C.}~\bibnamefont {MacCormick}}, \ and\ \bibinfo {author} {\bibfnamefont {M.}~\bibnamefont {Boshier}},\ }\href@noop {} {\bibfield  {journal} {\bibinfo  {journal} {New J. Phys.}\ }\textbf {\bibinfo {volume} {11}},\ \bibinfo {pages} {043030} (\bibinfo {year} {2009})}\BibitemShut {NoStop}%
\bibitem [{\citenamefont {Kawaguchi}\ and\ \citenamefont {Ueda}(2012)}]{Kawaguchi2012}%
  \BibitemOpen
  \bibfield  {author} {\bibinfo {author} {\bibfnamefont {Y.}~\bibnamefont {Kawaguchi}}\ and\ \bibinfo {author} {\bibfnamefont {M.}~\bibnamefont {Ueda}},\ }\href {\doibase 10.1016/j.physrep.2012.07.005} {\bibfield  {journal} {\bibinfo  {journal} {Phys. Rep.}\ }\textbf {\bibinfo {volume} {520}},\ \bibinfo {pages} {253} (\bibinfo {year} {2012})}\BibitemShut {NoStop}%
\bibitem [{\citenamefont {Strecker}\ \emph {et~al.}(2002)\citenamefont {Strecker}, \citenamefont {Partridge}, \citenamefont {Truscott},\ and\ \citenamefont {Hulet}}]{strecker2002formation}%
  \BibitemOpen
  \bibfield  {author} {\bibinfo {author} {\bibfnamefont {K.~E.}\ \bibnamefont {Strecker}}, \bibinfo {author} {\bibfnamefont {G.~B.}\ \bibnamefont {Partridge}}, \bibinfo {author} {\bibfnamefont {A.~G.}\ \bibnamefont {Truscott}}, \ and\ \bibinfo {author} {\bibfnamefont {R.~G.}\ \bibnamefont {Hulet}},\ }\href@noop {} {\bibfield  {journal} {\bibinfo  {journal} {Nature}\ }\textbf {\bibinfo {volume} {417}},\ \bibinfo {pages} {150} (\bibinfo {year} {2002})}\BibitemShut {NoStop}%
\bibitem [{\citenamefont {Khaykovich}\ \emph {et~al.}(2002)\citenamefont {Khaykovich}, \citenamefont {Schreck}, \citenamefont {Ferrari}, \citenamefont {Bourdel}, \citenamefont {Cubizolles}, \citenamefont {Carr}, \citenamefont {Castin},\ and\ \citenamefont {Salomon}}]{khaykovich2002formation}%
  \BibitemOpen
  \bibfield  {author} {\bibinfo {author} {\bibfnamefont {L.}~\bibnamefont {Khaykovich}}, \bibinfo {author} {\bibfnamefont {F.}~\bibnamefont {Schreck}}, \bibinfo {author} {\bibfnamefont {G.}~\bibnamefont {Ferrari}}, \bibinfo {author} {\bibfnamefont {T.}~\bibnamefont {Bourdel}}, \bibinfo {author} {\bibfnamefont {J.}~\bibnamefont {Cubizolles}}, \bibinfo {author} {\bibfnamefont {L.~D.}\ \bibnamefont {Carr}}, \bibinfo {author} {\bibfnamefont {Y.}~\bibnamefont {Castin}}, \ and\ \bibinfo {author} {\bibfnamefont {C.}~\bibnamefont {Salomon}},\ }\href@noop {} {\bibfield  {journal} {\bibinfo  {journal} {Science}\ }\textbf {\bibinfo {volume} {296}},\ \bibinfo {pages} {1290} (\bibinfo {year} {2002})}\BibitemShut {NoStop}%
\bibitem [{\citenamefont {Marchant}\ \emph {et~al.}(2013)\citenamefont {Marchant}, \citenamefont {Billam}, \citenamefont {Wiles}, \citenamefont {Yu}, \citenamefont {Gardiner},\ and\ \citenamefont {Cornish}}]{marchant2013controlled}%
  \BibitemOpen
  \bibfield  {author} {\bibinfo {author} {\bibfnamefont {A.}~\bibnamefont {Marchant}}, \bibinfo {author} {\bibfnamefont {T.}~\bibnamefont {Billam}}, \bibinfo {author} {\bibfnamefont {T.}~\bibnamefont {Wiles}}, \bibinfo {author} {\bibfnamefont {M.}~\bibnamefont {Yu}}, \bibinfo {author} {\bibfnamefont {S.}~\bibnamefont {Gardiner}}, \ and\ \bibinfo {author} {\bibfnamefont {S.}~\bibnamefont {Cornish}},\ }\href@noop {} {\bibfield  {journal} {\bibinfo  {journal} {Nature communications}\ }\textbf {\bibinfo {volume} {4}},\ \bibinfo {pages} {1865} (\bibinfo {year} {2013})}\BibitemShut {NoStop}%
\bibitem [{\citenamefont {Khaykovich}(2008)}]{khaykovich2008bright}%
  \BibitemOpen
  \bibfield  {author} {\bibinfo {author} {\bibfnamefont {L.}~\bibnamefont {Khaykovich}},\ }in\ \href@noop {} {\emph {\bibinfo {booktitle} {Emergent Nonlinear Phenomena in Bose-Einstein Condensates: Theory and Experiment}}}\ (\bibinfo  {publisher} {Springer},\ \bibinfo {year} {2008})\ pp.\ \bibinfo {pages} {45--61}\BibitemShut {NoStop}%
\bibitem [{\citenamefont {Strecker}\ \emph {et~al.}(2003)\citenamefont {Strecker}, \citenamefont {Partridge}, \citenamefont {Truscott},\ and\ \citenamefont {Hulet}}]{strecker2003bright}%
  \BibitemOpen
  \bibfield  {author} {\bibinfo {author} {\bibfnamefont {K.}~\bibnamefont {Strecker}}, \bibinfo {author} {\bibfnamefont {G.}~\bibnamefont {Partridge}}, \bibinfo {author} {\bibfnamefont {A.}~\bibnamefont {Truscott}}, \ and\ \bibinfo {author} {\bibfnamefont {R.~G.}\ \bibnamefont {Hulet}},\ }\href@noop {} {\bibfield  {journal} {\bibinfo  {journal} {New J. Phys.}\ }\textbf {\bibinfo {volume} {5}},\ \bibinfo {pages} {73} (\bibinfo {year} {2003})}\BibitemShut {NoStop}%
\bibitem [{\citenamefont {Nguyen}\ \emph {et~al.}(2014)\citenamefont {Nguyen}, \citenamefont {Dyke}, \citenamefont {Luo}, \citenamefont {Malomed},\ and\ \citenamefont {Hulet}}]{Nguyen2014Collisions}%
  \BibitemOpen
  \bibfield  {author} {\bibinfo {author} {\bibfnamefont {J.~H.~V.}\ \bibnamefont {Nguyen}}, \bibinfo {author} {\bibfnamefont {P.}~\bibnamefont {Dyke}}, \bibinfo {author} {\bibfnamefont {D.}~\bibnamefont {Luo}}, \bibinfo {author} {\bibfnamefont {B.~A.}\ \bibnamefont {Malomed}}, \ and\ \bibinfo {author} {\bibfnamefont {R.~G.}\ \bibnamefont {Hulet}},\ }\href {\doibase 10.1038/nphys3135} {\bibfield  {journal} {\bibinfo  {journal} {Nature Physics}\ }\textbf {\bibinfo {volume} {10}},\ \bibinfo {pages} {918} (\bibinfo {year} {2014})}\BibitemShut {NoStop}%
\bibitem [{\citenamefont {Everitt}\ \emph {et~al.}(2017{\natexlab{a}})\citenamefont {Everitt}, \citenamefont {Sooriyabandara}, \citenamefont {Guasoni}, \citenamefont {Wigley}, \citenamefont {Wei}, \citenamefont {McDonald}, \citenamefont {Hardman}, \citenamefont {Manju}, \citenamefont {Close}, \citenamefont {Kuhn}, \citenamefont {Szigeti}, \citenamefont {Kivshar},\ and\ \citenamefont {Robins}}]{Robbins}%
  \BibitemOpen
  \bibfield  {author} {\bibinfo {author} {\bibfnamefont {P.~J.}\ \bibnamefont {Everitt}}, \bibinfo {author} {\bibfnamefont {M.~A.}\ \bibnamefont {Sooriyabandara}}, \bibinfo {author} {\bibfnamefont {M.}~\bibnamefont {Guasoni}}, \bibinfo {author} {\bibfnamefont {P.~B.}\ \bibnamefont {Wigley}}, \bibinfo {author} {\bibfnamefont {C.~H.}\ \bibnamefont {Wei}}, \bibinfo {author} {\bibfnamefont {G.~D.}\ \bibnamefont {McDonald}}, \bibinfo {author} {\bibfnamefont {K.~S.}\ \bibnamefont {Hardman}}, \bibinfo {author} {\bibfnamefont {P.}~\bibnamefont {Manju}}, \bibinfo {author} {\bibfnamefont {J.~D.}\ \bibnamefont {Close}}, \bibinfo {author} {\bibfnamefont {C.~C.~N.}\ \bibnamefont {Kuhn}}, \bibinfo {author} {\bibfnamefont {S.~S.}\ \bibnamefont {Szigeti}}, \bibinfo {author} {\bibfnamefont {Y.~S.}\ \bibnamefont {Kivshar}}, \ and\ \bibinfo {author} {\bibfnamefont {N.~P.}\ \bibnamefont {Robins}},\ }\href {\doibase 10.1103/PhysRevA.96.041601} {\bibfield  {journal} {\bibinfo  {journal} {Phys. Rev. A}\ }\textbf {\bibinfo {volume}
  {96}},\ \bibinfo {pages} {041601} (\bibinfo {year} {2017}{\natexlab{a}})}\BibitemShut {NoStop}%
\bibitem [{\citenamefont {Nguyen}\ \emph {et~al.}(2017)\citenamefont {Nguyen}, \citenamefont {Luo},\ and\ \citenamefont {Hulet}}]{nguyen2017formation}%
  \BibitemOpen
  \bibfield  {author} {\bibinfo {author} {\bibfnamefont {J.~H.}\ \bibnamefont {Nguyen}}, \bibinfo {author} {\bibfnamefont {D.}~\bibnamefont {Luo}}, \ and\ \bibinfo {author} {\bibfnamefont {R.~G.}\ \bibnamefont {Hulet}},\ }\href@noop {} {\bibfield  {journal} {\bibinfo  {journal} {Science}\ }\textbf {\bibinfo {volume} {356}},\ \bibinfo {pages} {422} (\bibinfo {year} {2017})}\BibitemShut {NoStop}%
\bibitem [{\citenamefont {Di~Carli}\ \emph {et~al.}(2019)\citenamefont {Di~Carli}, \citenamefont {Colquhoun}, \citenamefont {Henderson}, \citenamefont {Flannigan}, \citenamefont {Oppo}, \citenamefont {Daley}, \citenamefont {Kuhr},\ and\ \citenamefont {Haller}}]{di_carli_excitation_2019}%
  \BibitemOpen
  \bibfield  {author} {\bibinfo {author} {\bibfnamefont {A.}~\bibnamefont {Di~Carli}}, \bibinfo {author} {\bibfnamefont {C.~D.}\ \bibnamefont {Colquhoun}}, \bibinfo {author} {\bibfnamefont {G.}~\bibnamefont {Henderson}}, \bibinfo {author} {\bibfnamefont {S.}~\bibnamefont {Flannigan}}, \bibinfo {author} {\bibfnamefont {G.-L.}\ \bibnamefont {Oppo}}, \bibinfo {author} {\bibfnamefont {A.~J.}\ \bibnamefont {Daley}}, \bibinfo {author} {\bibfnamefont {S.}~\bibnamefont {Kuhr}}, \ and\ \bibinfo {author} {\bibfnamefont {E.}~\bibnamefont {Haller}},\ }\href@noop {} {\bibfield  {journal} {\bibinfo  {journal} {Phys. Rev. Lett.}\ }\textbf {\bibinfo {volume} {123}},\ \bibinfo {pages} {123602} (\bibinfo {year} {2019})}\BibitemShut {NoStop}%
\bibitem [{\citenamefont {Luo}\ \emph {et~al.}(2020)\citenamefont {Luo}, \citenamefont {Jin}, \citenamefont {Nguyen}, \citenamefont {Malomed}, \citenamefont {Marchukov}, \citenamefont {Yurovsky}, \citenamefont {Dunjko}, \citenamefont {Olshanii},\ and\ \citenamefont {Hulet}}]{luo_creation_2020}%
  \BibitemOpen
  \bibfield  {author} {\bibinfo {author} {\bibfnamefont {D.}~\bibnamefont {Luo}}, \bibinfo {author} {\bibfnamefont {Y.}~\bibnamefont {Jin}}, \bibinfo {author} {\bibfnamefont {J.~H.~V.}\ \bibnamefont {Nguyen}}, \bibinfo {author} {\bibfnamefont {B.~A.}\ \bibnamefont {Malomed}}, \bibinfo {author} {\bibfnamefont {O.~V.}\ \bibnamefont {Marchukov}}, \bibinfo {author} {\bibfnamefont {V.~A.}\ \bibnamefont {Yurovsky}}, \bibinfo {author} {\bibfnamefont {V.}~\bibnamefont {Dunjko}}, \bibinfo {author} {\bibfnamefont {M.}~\bibnamefont {Olshanii}}, \ and\ \bibinfo {author} {\bibfnamefont {R.~G.}\ \bibnamefont {Hulet}},\ }\href@noop {} {\bibfield  {journal} {\bibinfo  {journal} {Phys. Rev. Lett.}\ }\textbf {\bibinfo {volume} {125}},\ \bibinfo {pages} {183902} (\bibinfo {year} {2020})}\BibitemShut {NoStop}%
\bibitem [{\citenamefont {Chen}\ and\ \citenamefont {Hung}(2020)}]{Chen_MI}%
  \BibitemOpen
  \bibfield  {author} {\bibinfo {author} {\bibfnamefont {C.-A.}\ \bibnamefont {Chen}}\ and\ \bibinfo {author} {\bibfnamefont {C.-L.}\ \bibnamefont {Hung}},\ }\href {\doibase 10.1103/PhysRevLett.125.250401} {\bibfield  {journal} {\bibinfo  {journal} {Phys. Rev. Lett.}\ }\textbf {\bibinfo {volume} {125}},\ \bibinfo {pages} {250401} (\bibinfo {year} {2020})}\BibitemShut {NoStop}%
\bibitem [{\citenamefont {Chen}\ and\ \citenamefont {Hung}(2021)}]{ChenHung2021ScaleInvPRL}%
  \BibitemOpen
  \bibfield  {author} {\bibinfo {author} {\bibfnamefont {C.-A.}\ \bibnamefont {Chen}}\ and\ \bibinfo {author} {\bibfnamefont {C.-L.}\ \bibnamefont {Hung}},\ }\href {\doibase 10.1103/PhysRevLett.127.023604} {\bibfield  {journal} {\bibinfo  {journal} {Physical Review Letters}\ }\textbf {\bibinfo {volume} {127}},\ \bibinfo {pages} {023604} (\bibinfo {year} {2021})}\BibitemShut {NoStop}%
\bibitem [{\citenamefont {Bakkali-Hassani}\ \emph {et~al.}(2021)\citenamefont {Bakkali-Hassani}, \citenamefont {Maury}, \citenamefont {Zou}, \citenamefont {Le~Cerf}, \citenamefont {Saint-Jalm}, \citenamefont {Castilho}, \citenamefont {Nascimbene}, \citenamefont {Dalibard},\ and\ \citenamefont {Beugnon}}]{Bakkali_realization_2021}%
  \BibitemOpen
  \bibfield  {author} {\bibinfo {author} {\bibfnamefont {B.}~\bibnamefont {Bakkali-Hassani}}, \bibinfo {author} {\bibfnamefont {C.}~\bibnamefont {Maury}}, \bibinfo {author} {\bibfnamefont {Y.-Q.}\ \bibnamefont {Zou}}, \bibinfo {author} {\bibfnamefont {E.}~\bibnamefont {Le~Cerf}}, \bibinfo {author} {\bibfnamefont {R.}~\bibnamefont {Saint-Jalm}}, \bibinfo {author} {\bibfnamefont {P.~C.~M.}\ \bibnamefont {Castilho}}, \bibinfo {author} {\bibfnamefont {S.}~\bibnamefont {Nascimbene}}, \bibinfo {author} {\bibfnamefont {J.}~\bibnamefont {Dalibard}}, \ and\ \bibinfo {author} {\bibfnamefont {J.}~\bibnamefont {Beugnon}},\ }\href {\doibase 10.1103/PhysRevLett.127.023603} {\bibfield  {journal} {\bibinfo  {journal} {Phys. Rev. Lett.}\ }\textbf {\bibinfo {volume} {127}},\ \bibinfo {pages} {023603} (\bibinfo {year} {2021})}\BibitemShut {NoStop}%
\bibitem [{\citenamefont {Bakkali-Hassani}\ and\ \citenamefont {Dalibard}(2022)}]{bakkali_townes_2022}%
  \BibitemOpen
  \bibfield  {author} {\bibinfo {author} {\bibfnamefont {B.}~\bibnamefont {Bakkali-Hassani}}\ and\ \bibinfo {author} {\bibfnamefont {J.}~\bibnamefont {Dalibard}},\ }\href@noop {} {\bibfield  {journal} {\bibinfo  {journal} {arXiv:2210.14045}\ } (\bibinfo {year} {2022})}\BibitemShut {NoStop}%
\bibitem [{\citenamefont {Pitaevskii}\ and\ \citenamefont {Stringari}(2016)}]{PitaevskiiStringari2016}%
  \BibitemOpen
  \bibfield  {author} {\bibinfo {author} {\bibfnamefont {L.~P.}\ \bibnamefont {Pitaevskii}}\ and\ \bibinfo {author} {\bibfnamefont {S.}~\bibnamefont {Stringari}},\ }\href@noop {} {\emph {\bibinfo {title} {Bose--Einstein Condensation and Superfluidity}}}\ (\bibinfo  {publisher} {Oxford University Press},\ \bibinfo {address} {Oxford},\ \bibinfo {year} {2016})\BibitemShut {NoStop}%
\bibitem [{\citenamefont {Karjanto}(2021)}]{karjanto2021peregrine}%
  \BibitemOpen
  \bibfield  {author} {\bibinfo {author} {\bibfnamefont {N.}~\bibnamefont {Karjanto}},\ }\href@noop {} {\bibfield  {journal} {\bibinfo  {journal} {Frontiers in Physics}\ }\textbf {\bibinfo {volume} {9}},\ \bibinfo {pages} {599767} (\bibinfo {year} {2021})}\BibitemShut {NoStop}%
\bibitem [{\citenamefont {Charalampidis}\ \emph {et~al.}(2018)\citenamefont {Charalampidis}, \citenamefont {Cuevas-Maraver}, \citenamefont {Frantzeskakis},\ and\ \citenamefont {Kevrekidis}}]{charalampidis2016rogue}%
  \BibitemOpen
  \bibfield  {author} {\bibinfo {author} {\bibfnamefont {E.~G.}\ \bibnamefont {Charalampidis}}, \bibinfo {author} {\bibfnamefont {J.}~\bibnamefont {Cuevas-Maraver}}, \bibinfo {author} {\bibfnamefont {D.~J.}\ \bibnamefont {Frantzeskakis}}, \ and\ \bibinfo {author} {\bibfnamefont {P.~G.}\ \bibnamefont {Kevrekidis}},\ }\href@noop {} {\bibfield  {journal} {\bibinfo  {journal} {Romanian Reports in Physics}\ }\textbf {\bibinfo {volume} {70}},\ \bibinfo {pages} {504} (\bibinfo {year} {2018})}\BibitemShut {NoStop}%
\bibitem [{\citenamefont {Cornish}\ \emph {et~al.}(2006)\citenamefont {Cornish}, \citenamefont {Thompson},\ and\ \citenamefont {Wieman}}]{Cornish_2006}%
  \BibitemOpen
  \bibfield  {author} {\bibinfo {author} {\bibfnamefont {S.~L.}\ \bibnamefont {Cornish}}, \bibinfo {author} {\bibfnamefont {S.~T.}\ \bibnamefont {Thompson}}, \ and\ \bibinfo {author} {\bibfnamefont {C.~E.}\ \bibnamefont {Wieman}},\ }\href {\doibase 10.1103/PhysRevLett.96.170401} {\bibfield  {journal} {\bibinfo  {journal} {Phys. Rev. Lett.}\ }\textbf {\bibinfo {volume} {96}},\ \bibinfo {pages} {170401} (\bibinfo {year} {2006})}\BibitemShut {NoStop}%
\bibitem [{\citenamefont {Wen}\ \emph {et~al.}(2011)\citenamefont {Wen}, \citenamefont {Li}, \citenamefont {Li}, \citenamefont {Song}, \citenamefont {Zhang},\ and\ \citenamefont {Liu}}]{wen2011matter}%
  \BibitemOpen
  \bibfield  {author} {\bibinfo {author} {\bibfnamefont {L.}~\bibnamefont {Wen}}, \bibinfo {author} {\bibfnamefont {L.}~\bibnamefont {Li}}, \bibinfo {author} {\bibfnamefont {Z.-D.}\ \bibnamefont {Li}}, \bibinfo {author} {\bibfnamefont {S.-W.}\ \bibnamefont {Song}}, \bibinfo {author} {\bibfnamefont {X.-F.}\ \bibnamefont {Zhang}}, \ and\ \bibinfo {author} {\bibfnamefont {W.}~\bibnamefont {Liu}},\ }\href@noop {} {\bibfield  {journal} {\bibinfo  {journal} {Eur. Phys. J. D}\ }\textbf {\bibinfo {volume} {64}},\ \bibinfo {pages} {473} (\bibinfo {year} {2011})}\BibitemShut {NoStop}%
\bibitem [{\citenamefont {He}\ \emph {et~al.}(2014)\citenamefont {He}, \citenamefont {Charalampidis}, \citenamefont {Kevrekidis},\ and\ \citenamefont {Frantzeskakis}}]{he2014rogue}%
  \BibitemOpen
  \bibfield  {author} {\bibinfo {author} {\bibfnamefont {J.}~\bibnamefont {He}}, \bibinfo {author} {\bibfnamefont {E.}~\bibnamefont {Charalampidis}}, \bibinfo {author} {\bibfnamefont {P.}~\bibnamefont {Kevrekidis}}, \ and\ \bibinfo {author} {\bibfnamefont {D.}~\bibnamefont {Frantzeskakis}},\ }\href@noop {} {\bibfield  {journal} {\bibinfo  {journal} {Phys. Lett. A}\ }\textbf {\bibinfo {volume} {378}},\ \bibinfo {pages} {577} (\bibinfo {year} {2014})}\BibitemShut {NoStop}%
\bibitem [{\citenamefont {Loomba}\ \emph {et~al.}(2014)\citenamefont {Loomba}, \citenamefont {Kaur}, \citenamefont {Gupta}, \citenamefont {Kumar},\ and\ \citenamefont {Raju}}]{LoombaRW}%
  \BibitemOpen
  \bibfield  {author} {\bibinfo {author} {\bibfnamefont {S.}~\bibnamefont {Loomba}}, \bibinfo {author} {\bibfnamefont {H.}~\bibnamefont {Kaur}}, \bibinfo {author} {\bibfnamefont {R.}~\bibnamefont {Gupta}}, \bibinfo {author} {\bibfnamefont {C.~N.}\ \bibnamefont {Kumar}}, \ and\ \bibinfo {author} {\bibfnamefont {T.~S.}\ \bibnamefont {Raju}},\ }\href {\doibase 10.1103/PhysRevE.89.052915} {\bibfield  {journal} {\bibinfo  {journal} {Phys. Rev. E}\ }\textbf {\bibinfo {volume} {89}},\ \bibinfo {pages} {052915} (\bibinfo {year} {2014})}\BibitemShut {NoStop}%
\bibitem [{\citenamefont {Manikandan}\ \emph {et~al.}(2014)\citenamefont {Manikandan}, \citenamefont {Muruganandam}, \citenamefont {Senthilvelan},\ and\ \citenamefont {Lakshmanan}}]{ManikandanRW}%
  \BibitemOpen
  \bibfield  {author} {\bibinfo {author} {\bibfnamefont {K.}~\bibnamefont {Manikandan}}, \bibinfo {author} {\bibfnamefont {P.}~\bibnamefont {Muruganandam}}, \bibinfo {author} {\bibfnamefont {M.}~\bibnamefont {Senthilvelan}}, \ and\ \bibinfo {author} {\bibfnamefont {M.}~\bibnamefont {Lakshmanan}},\ }\href {\doibase 10.1103/PhysRevE.90.062905} {\bibfield  {journal} {\bibinfo  {journal} {Phys. Rev. E}\ }\textbf {\bibinfo {volume} {90}},\ \bibinfo {pages} {062905} (\bibinfo {year} {2014})}\BibitemShut {NoStop}%
\bibitem [{\citenamefont {Bludov}\ \emph {et~al.}(2010)\citenamefont {Bludov}, \citenamefont {Konotop},\ and\ \citenamefont {Akhmediev}}]{bludov2010vector}%
  \BibitemOpen
  \bibfield  {author} {\bibinfo {author} {\bibfnamefont {Y.~V.}\ \bibnamefont {Bludov}}, \bibinfo {author} {\bibfnamefont {V.}~\bibnamefont {Konotop}}, \ and\ \bibinfo {author} {\bibfnamefont {N.}~\bibnamefont {Akhmediev}},\ }\href@noop {} {\bibfield  {journal} {\bibinfo  {journal} {Eur. Phys. J-Spec. Top.}\ }\textbf {\bibinfo {volume} {185}},\ \bibinfo {pages} {169} (\bibinfo {year} {2010})}\BibitemShut {NoStop}%
\bibitem [{\citenamefont {Baronio}\ \emph {et~al.}(2012)\citenamefont {Baronio}, \citenamefont {Degasperis}, \citenamefont {Conforti},\ and\ \citenamefont {Wabnitz}}]{BaronioRWs}%
  \BibitemOpen
  \bibfield  {author} {\bibinfo {author} {\bibfnamefont {F.}~\bibnamefont {Baronio}}, \bibinfo {author} {\bibfnamefont {A.}~\bibnamefont {Degasperis}}, \bibinfo {author} {\bibfnamefont {M.}~\bibnamefont {Conforti}}, \ and\ \bibinfo {author} {\bibfnamefont {S.}~\bibnamefont {Wabnitz}},\ }\href {\doibase 10.1103/PhysRevLett.109.044102} {\bibfield  {journal} {\bibinfo  {journal} {Phys. Rev. Lett.}\ }\textbf {\bibinfo {volume} {109}},\ \bibinfo {pages} {044102} (\bibinfo {year} {2012})}\BibitemShut {NoStop}%
\bibitem [{\citenamefont {Mareeswaran}\ \emph {et~al.}(2014)\citenamefont {Mareeswaran}, \citenamefont {Charalampidis}, \citenamefont {Kanna}, \citenamefont {Kevrekidis},\ and\ \citenamefont {Frantzeskakis}}]{Mareeswaran_bin_RWs}%
  \BibitemOpen
  \bibfield  {author} {\bibinfo {author} {\bibfnamefont {R.~B.}\ \bibnamefont {Mareeswaran}}, \bibinfo {author} {\bibfnamefont {E.~G.}\ \bibnamefont {Charalampidis}}, \bibinfo {author} {\bibfnamefont {T.}~\bibnamefont {Kanna}}, \bibinfo {author} {\bibfnamefont {P.~G.}\ \bibnamefont {Kevrekidis}}, \ and\ \bibinfo {author} {\bibfnamefont {D.~J.}\ \bibnamefont {Frantzeskakis}},\ }\href {\doibase 10.1103/PhysRevE.90.042912} {\bibfield  {journal} {\bibinfo  {journal} {Phys. Rev. E}\ }\textbf {\bibinfo {volume} {90}},\ \bibinfo {pages} {042912} (\bibinfo {year} {2014})}\BibitemShut {NoStop}%
\bibitem [{\citenamefont {Manikandan}\ \emph {et~al.}(2016)\citenamefont {Manikandan}, \citenamefont {Muruganandam}, \citenamefont {Senthilvelan},\ and\ \citenamefont {Lakshmanan}}]{Manikandan_bin_RWs}%
  \BibitemOpen
  \bibfield  {author} {\bibinfo {author} {\bibfnamefont {K.}~\bibnamefont {Manikandan}}, \bibinfo {author} {\bibfnamefont {P.}~\bibnamefont {Muruganandam}}, \bibinfo {author} {\bibfnamefont {M.}~\bibnamefont {Senthilvelan}}, \ and\ \bibinfo {author} {\bibfnamefont {M.}~\bibnamefont {Lakshmanan}},\ }\href {\doibase 10.1103/PhysRevE.93.032212} {\bibfield  {journal} {\bibinfo  {journal} {Phys. Rev. E}\ }\textbf {\bibinfo {volume} {93}},\ \bibinfo {pages} {032212} (\bibinfo {year} {2016})}\BibitemShut {NoStop}%
\bibitem [{\citenamefont {Mareeswaran}\ and\ \citenamefont {Kanna}(2016)}]{mareeswaran2016superposed}%
  \BibitemOpen
  \bibfield  {author} {\bibinfo {author} {\bibfnamefont {R.~B.}\ \bibnamefont {Mareeswaran}}\ and\ \bibinfo {author} {\bibfnamefont {T.}~\bibnamefont {Kanna}},\ }\href@noop {} {\bibfield  {journal} {\bibinfo  {journal} {Phys. Lett. A}\ }\textbf {\bibinfo {volume} {380}},\ \bibinfo {pages} {3244} (\bibinfo {year} {2016})}\BibitemShut {NoStop}%
\bibitem [{\citenamefont {Qin}\ and\ \citenamefont {Mu}(2012)}]{Qin_spinRWs}%
  \BibitemOpen
  \bibfield  {author} {\bibinfo {author} {\bibfnamefont {Z.}~\bibnamefont {Qin}}\ and\ \bibinfo {author} {\bibfnamefont {G.}~\bibnamefont {Mu}},\ }\href {\doibase 10.1103/PhysRevE.86.036601} {\bibfield  {journal} {\bibinfo  {journal} {Phys. Rev. E}\ }\textbf {\bibinfo {volume} {86}},\ \bibinfo {pages} {036601} (\bibinfo {year} {2012})}\BibitemShut {NoStop}%
\bibitem [{\citenamefont {Zhao}\ and\ \citenamefont {Liu}(2013)}]{Zhao_spinRWs}%
  \BibitemOpen
  \bibfield  {author} {\bibinfo {author} {\bibfnamefont {L.-C.}\ \bibnamefont {Zhao}}\ and\ \bibinfo {author} {\bibfnamefont {J.}~\bibnamefont {Liu}},\ }\href {\doibase 10.1103/PhysRevE.87.013201} {\bibfield  {journal} {\bibinfo  {journal} {Phys. Rev. E}\ }\textbf {\bibinfo {volume} {87}},\ \bibinfo {pages} {013201} (\bibinfo {year} {2013})}\BibitemShut {NoStop}%
\bibitem [{\citenamefont {Shen}\ \emph {et~al.}(2014)\citenamefont {Shen}, \citenamefont {Gao}, \citenamefont {Zuo}, \citenamefont {Sun}, \citenamefont {Feng},\ and\ \citenamefont {Xue}}]{Shen_spinRWS}%
  \BibitemOpen
  \bibfield  {author} {\bibinfo {author} {\bibfnamefont {Y.-J.}\ \bibnamefont {Shen}}, \bibinfo {author} {\bibfnamefont {Y.-T.}\ \bibnamefont {Gao}}, \bibinfo {author} {\bibfnamefont {D.-W.}\ \bibnamefont {Zuo}}, \bibinfo {author} {\bibfnamefont {Y.-H.}\ \bibnamefont {Sun}}, \bibinfo {author} {\bibfnamefont {Y.-J.}\ \bibnamefont {Feng}}, \ and\ \bibinfo {author} {\bibfnamefont {L.}~\bibnamefont {Xue}},\ }\href {\doibase 10.1103/PhysRevE.89.062915} {\bibfield  {journal} {\bibinfo  {journal} {Phys. Rev. E}\ }\textbf {\bibinfo {volume} {89}},\ \bibinfo {pages} {062915} (\bibinfo {year} {2014})}\BibitemShut {NoStop}%
\bibitem [{\citenamefont {Romero-Ros}\ \emph {et~al.}(2022)\citenamefont {Romero-Ros}, \citenamefont {Katsimiga}, \citenamefont {Mistakidis}, \citenamefont {Prinari}, \citenamefont {Biondini}, \citenamefont {Schmelcher},\ and\ \citenamefont {Kevrekidis}}]{Romero_theor}%
  \BibitemOpen
  \bibfield  {author} {\bibinfo {author} {\bibfnamefont {A.}~\bibnamefont {Romero-Ros}}, \bibinfo {author} {\bibfnamefont {G.~C.}\ \bibnamefont {Katsimiga}}, \bibinfo {author} {\bibfnamefont {S.~I.}\ \bibnamefont {Mistakidis}}, \bibinfo {author} {\bibfnamefont {B.}~\bibnamefont {Prinari}}, \bibinfo {author} {\bibfnamefont {G.}~\bibnamefont {Biondini}}, \bibinfo {author} {\bibfnamefont {P.}~\bibnamefont {Schmelcher}}, \ and\ \bibinfo {author} {\bibfnamefont {P.~G.}\ \bibnamefont {Kevrekidis}},\ }\href {\doibase 10.1103/PhysRevA.105.053306} {\bibfield  {journal} {\bibinfo  {journal} {Phys. Rev. A}\ }\textbf {\bibinfo {volume} {105}},\ \bibinfo {pages} {053306} (\bibinfo {year} {2022})}\BibitemShut {NoStop}%
\bibitem [{\citenamefont {Van~Gorder}(2014)}]{VanGorder2014Peregrine}%
  \BibitemOpen
  \bibfield  {author} {\bibinfo {author} {\bibfnamefont {R.~A.}\ \bibnamefont {Van~Gorder}},\ }\href {\doibase 10.7566/JPSJ.83.054005} {\bibfield  {journal} {\bibinfo  {journal} {Journal of the Physical Society of Japan}\ }\textbf {\bibinfo {volume} {83}} (\bibinfo {year} {2014}),\ 10.7566/JPSJ.83.054005}\BibitemShut {NoStop}%
\bibitem [{\citenamefont {Al~Khawaja}\ \emph {et~al.}(2014)\citenamefont {Al~Khawaja}, \citenamefont {Bahlouli}, \citenamefont {Asad-uz zaman},\ and\ \citenamefont {Al-Marzoug}}]{AlKhawaja2014PeregrineMI}%
  \BibitemOpen
  \bibfield  {author} {\bibinfo {author} {\bibfnamefont {U.}~\bibnamefont {Al~Khawaja}}, \bibinfo {author} {\bibfnamefont {H.}~\bibnamefont {Bahlouli}}, \bibinfo {author} {\bibfnamefont {M.}~\bibnamefont {Asad-uz zaman}}, \ and\ \bibinfo {author} {\bibfnamefont {S.~M.}\ \bibnamefont {Al-Marzoug}},\ }\href {\doibase 10.1016/j.cnsns.2014.01.029} {\bibfield  {journal} {\bibinfo  {journal} {Communications in Nonlinear Science and Numerical Simulation}\ }\textbf {\bibinfo {volume} {19}},\ \bibinfo {pages} {2706} (\bibinfo {year} {2014})}\BibitemShut {NoStop}%
\bibitem [{\citenamefont {Cuevas-Maraver}\ \emph {et~al.}(2017)\citenamefont {Cuevas-Maraver}, \citenamefont {Kevrekidis}, \citenamefont {Frantzeskakis}, \citenamefont {Karachalios}, \citenamefont {Haragus},\ and\ \citenamefont {James}}]{Cuevas_stability_PS}%
  \BibitemOpen
  \bibfield  {author} {\bibinfo {author} {\bibfnamefont {J.}~\bibnamefont {Cuevas-Maraver}}, \bibinfo {author} {\bibfnamefont {P.~G.}\ \bibnamefont {Kevrekidis}}, \bibinfo {author} {\bibfnamefont {D.~J.}\ \bibnamefont {Frantzeskakis}}, \bibinfo {author} {\bibfnamefont {N.~I.}\ \bibnamefont {Karachalios}}, \bibinfo {author} {\bibfnamefont {M.}~\bibnamefont {Haragus}}, \ and\ \bibinfo {author} {\bibfnamefont {G.}~\bibnamefont {James}},\ }\href {\doibase 10.1103/PhysRevE.96.012202} {\bibfield  {journal} {\bibinfo  {journal} {Phys. Rev. E}\ }\textbf {\bibinfo {volume} {96}},\ \bibinfo {pages} {012202} (\bibinfo {year} {2017})}\BibitemShut {NoStop}%
\bibitem [{\citenamefont {Guckenheimer}\ and\ \citenamefont {Holmes}(1983)}]{GuckenheimerHolmes1983}%
  \BibitemOpen
  \bibfield  {author} {\bibinfo {author} {\bibfnamefont {J.}~\bibnamefont {Guckenheimer}}\ and\ \bibinfo {author} {\bibfnamefont {P.}~\bibnamefont {Holmes}},\ }\href@noop {} {\emph {\bibinfo {title} {Nonlinear Oscillations, Dynamical Systems, and Bifurcations of Vector Fields}}}\ (\bibinfo  {publisher} {Springer},\ \bibinfo {year} {1983})\BibitemShut {NoStop}%
\bibitem [{\citenamefont {Coddington}\ \emph {et~al.}(1956)\citenamefont {Coddington}, \citenamefont {Levinson},\ and\ \citenamefont {Teichmann}}]{coddington1956theory}%
  \BibitemOpen
  \bibfield  {author} {\bibinfo {author} {\bibfnamefont {E.~A.}\ \bibnamefont {Coddington}}, \bibinfo {author} {\bibfnamefont {N.}~\bibnamefont {Levinson}}, \ and\ \bibinfo {author} {\bibfnamefont {T.}~\bibnamefont {Teichmann}},\ }\href@noop {} {\enquote {\bibinfo {title} {Theory of ordinary differential equations},}\ } (\bibinfo {year} {1956})\BibitemShut {NoStop}%
\bibitem [{\citenamefont {Arnold}(1992)}]{arnold1992ordinary}%
  \BibitemOpen
  \bibfield  {author} {\bibinfo {author} {\bibfnamefont {V.~I.}\ \bibnamefont {Arnold}},\ }\href@noop {} {\emph {\bibinfo {title} {Ordinary differential equations}}}\ (\bibinfo  {publisher} {Springer Science \& Business Media},\ \bibinfo {year} {1992})\BibitemShut {NoStop}%
\bibitem [{\citenamefont {Kevrekidis}\ \emph {et~al.}(2015)\citenamefont {Kevrekidis}, \citenamefont {Frantzeskakis},\ and\ \citenamefont {Carretero-Gonz{\'a}lez}}]{kevrekidis2015defocusing}%
  \BibitemOpen
  \bibfield  {author} {\bibinfo {author} {\bibfnamefont {P.~G.}\ \bibnamefont {Kevrekidis}}, \bibinfo {author} {\bibfnamefont {D.~J.}\ \bibnamefont {Frantzeskakis}}, \ and\ \bibinfo {author} {\bibfnamefont {R.}~\bibnamefont {Carretero-Gonz{\'a}lez}},\ }\href@noop {} {\emph {\bibinfo {title} {The defocusing nonlinear Schr{\"o}dinger equation: from dark solitons to vortices and vortex rings}}}\ (\bibinfo  {publisher} {SIAM},\ \bibinfo {year} {2015})\BibitemShut {NoStop}%
\bibitem [{\citenamefont {Kevrekidis}\ and\ \citenamefont {Frantzeskakis}(2016)}]{KevrekidisFrantzeskakis2016Review}%
  \BibitemOpen
  \bibfield  {author} {\bibinfo {author} {\bibfnamefont {P.~G.}\ \bibnamefont {Kevrekidis}}\ and\ \bibinfo {author} {\bibfnamefont {D.~J.}\ \bibnamefont {Frantzeskakis}},\ }\href {\doibase 10.1016/j.revip.2016.07.002} {\bibfield  {journal} {\bibinfo  {journal} {Reviews in Physics}\ }\textbf {\bibinfo {volume} {1}},\ \bibinfo {pages} {140} (\bibinfo {year} {2016})}\BibitemShut {NoStop}%
\bibitem [{\citenamefont {Ao}\ and\ \citenamefont {Chui}(1998)}]{AoChui1998Miscibility}%
  \BibitemOpen
  \bibfield  {author} {\bibinfo {author} {\bibfnamefont {P.}~\bibnamefont {Ao}}\ and\ \bibinfo {author} {\bibfnamefont {S.~T.}\ \bibnamefont {Chui}},\ }\href {\doibase 10.1103/PhysRevA.58.4836} {\bibfield  {journal} {\bibinfo  {journal} {Physical Review A}\ }\textbf {\bibinfo {volume} {58}},\ \bibinfo {pages} {4836} (\bibinfo {year} {1998})}\BibitemShut {NoStop}%
\bibitem [{\citenamefont {Dutton}\ and\ \citenamefont {Clark}(2005)}]{DuttonClark2005}%
  \BibitemOpen
  \bibfield  {author} {\bibinfo {author} {\bibfnamefont {Z.}~\bibnamefont {Dutton}}\ and\ \bibinfo {author} {\bibfnamefont {C.~W.}\ \bibnamefont {Clark}},\ }\href {\doibase 10.1103/PhysRevA.71.063618} {\bibfield  {journal} {\bibinfo  {journal} {Physical Review A}\ }\textbf {\bibinfo {volume} {71}},\ \bibinfo {pages} {063618} (\bibinfo {year} {2005})}\BibitemShut {NoStop}%
\bibitem [{\citenamefont {Kivshar}\ and\ \citenamefont {Agrawal}(2003)}]{KivsharAgrawal2003}%
  \BibitemOpen
  \bibfield  {author} {\bibinfo {author} {\bibfnamefont {Y.~S.}\ \bibnamefont {Kivshar}}\ and\ \bibinfo {author} {\bibfnamefont {G.~P.}\ \bibnamefont {Agrawal}},\ }\href@noop {} {\emph {\bibinfo {title} {Optical Solitons: From Fibers to Photonic Crystals}}}\ (\bibinfo  {publisher} {Academic Press},\ \bibinfo {address} {San Diego},\ \bibinfo {year} {2003})\BibitemShut {NoStop}%
\bibitem [{\citenamefont {Everitt}\ \emph {et~al.}(2017{\natexlab{b}})\citenamefont {Everitt}, \citenamefont {Sooriyabandara}, \citenamefont {Guasoni}, \citenamefont {Wigley}, \citenamefont {Wei}, \citenamefont {McDonald}, \citenamefont {Hardman}, \citenamefont {Manju}, \citenamefont {Close}, \citenamefont {Kuhn}, \citenamefont {Szigeti}, \citenamefont {Kivshar},\ and\ \citenamefont {Robins}}]{Everitt_MI}%
  \BibitemOpen
  \bibfield  {author} {\bibinfo {author} {\bibfnamefont {P.~J.}\ \bibnamefont {Everitt}}, \bibinfo {author} {\bibfnamefont {M.~A.}\ \bibnamefont {Sooriyabandara}}, \bibinfo {author} {\bibfnamefont {M.}~\bibnamefont {Guasoni}}, \bibinfo {author} {\bibfnamefont {P.~B.}\ \bibnamefont {Wigley}}, \bibinfo {author} {\bibfnamefont {C.~H.}\ \bibnamefont {Wei}}, \bibinfo {author} {\bibfnamefont {G.~D.}\ \bibnamefont {McDonald}}, \bibinfo {author} {\bibfnamefont {K.~S.}\ \bibnamefont {Hardman}}, \bibinfo {author} {\bibfnamefont {P.}~\bibnamefont {Manju}}, \bibinfo {author} {\bibfnamefont {J.~D.}\ \bibnamefont {Close}}, \bibinfo {author} {\bibfnamefont {C.~C.~N.}\ \bibnamefont {Kuhn}}, \bibinfo {author} {\bibfnamefont {S.~S.}\ \bibnamefont {Szigeti}}, \bibinfo {author} {\bibfnamefont {Y.~S.}\ \bibnamefont {Kivshar}}, \ and\ \bibinfo {author} {\bibfnamefont {N.~P.}\ \bibnamefont {Robins}},\ }\href {\doibase 10.1103/PhysRevA.96.041601} {\bibfield  {journal} {\bibinfo  {journal} {Phys. Rev. A}\ }\textbf {\bibinfo {volume}
  {96}},\ \bibinfo {pages} {041601} (\bibinfo {year} {2017}{\natexlab{b}})}\BibitemShut {NoStop}%
\bibitem [{\citenamefont {Whitham}(1965)}]{whitham1965non}%
  \BibitemOpen
  \bibfield  {author} {\bibinfo {author} {\bibfnamefont {G.~B.}\ \bibnamefont {Whitham}},\ }\href@noop {} {\bibfield  {journal} {\bibinfo  {journal} {Proc. R. Soc. Lond. A Math. Phys. Sci.}\ }\textbf {\bibinfo {volume} {283}},\ \bibinfo {pages} {238} (\bibinfo {year} {1965})}\BibitemShut {NoStop}%
\bibitem [{\citenamefont {Biondini}\ and\ \citenamefont {Kova{\v{c}}i{\v{c}}}(2014)}]{biondini2014inverse}%
  \BibitemOpen
  \bibfield  {author} {\bibinfo {author} {\bibfnamefont {G.}~\bibnamefont {Biondini}}\ and\ \bibinfo {author} {\bibfnamefont {G.}~\bibnamefont {Kova{\v{c}}i{\v{c}}}},\ }\href@noop {} {\bibfield  {journal} {\bibinfo  {journal} {J. Math. Phys.}\ }\textbf {\bibinfo {volume} {55}} (\bibinfo {year} {2014})}\BibitemShut {NoStop}%
\bibitem [{\citenamefont {Kraych}\ \emph {et~al.}(2019)\citenamefont {Kraych}, \citenamefont {Suret}, \citenamefont {El},\ and\ \citenamefont {Randoux}}]{Kraych_MI}%
  \BibitemOpen
  \bibfield  {author} {\bibinfo {author} {\bibfnamefont {A.~E.}\ \bibnamefont {Kraych}}, \bibinfo {author} {\bibfnamefont {P.}~\bibnamefont {Suret}}, \bibinfo {author} {\bibfnamefont {G.}~\bibnamefont {El}}, \ and\ \bibinfo {author} {\bibfnamefont {S.}~\bibnamefont {Randoux}},\ }\href {\doibase 10.1103/PhysRevLett.122.054101} {\bibfield  {journal} {\bibinfo  {journal} {Phys. Rev. Lett.}\ }\textbf {\bibinfo {volume} {122}},\ \bibinfo {pages} {054101} (\bibinfo {year} {2019})}\BibitemShut {NoStop}%
\bibitem [{\citenamefont {Copie}\ \emph {et~al.}(2020)\citenamefont {Copie}, \citenamefont {Randoux},\ and\ \citenamefont {Suret}}]{copie2020physics}%
  \BibitemOpen
  \bibfield  {author} {\bibinfo {author} {\bibfnamefont {F.}~\bibnamefont {Copie}}, \bibinfo {author} {\bibfnamefont {S.}~\bibnamefont {Randoux}}, \ and\ \bibinfo {author} {\bibfnamefont {P.}~\bibnamefont {Suret}},\ }\href@noop {} {\bibfield  {journal} {\bibinfo  {journal} {Rev. Phys.}\ }\textbf {\bibinfo {volume} {5}},\ \bibinfo {pages} {100037} (\bibinfo {year} {2020})}\BibitemShut {NoStop}%
\bibitem [{\citenamefont {Bonnefoy}\ \emph {et~al.}(2020)\citenamefont {Bonnefoy}, \citenamefont {Tikan}, \citenamefont {Copie}, \citenamefont {Suret}, \citenamefont {Ducrozet}, \citenamefont {Prabhudesai}, \citenamefont {Michel}, \citenamefont {Cazaubiel}, \citenamefont {Falcon}, \citenamefont {El},\ and\ \citenamefont {Randoux}}]{Bonnefoy}%
  \BibitemOpen
  \bibfield  {author} {\bibinfo {author} {\bibfnamefont {F.}~\bibnamefont {Bonnefoy}}, \bibinfo {author} {\bibfnamefont {A.}~\bibnamefont {Tikan}}, \bibinfo {author} {\bibfnamefont {F.~m.~c.}\ \bibnamefont {Copie}}, \bibinfo {author} {\bibfnamefont {P.}~\bibnamefont {Suret}}, \bibinfo {author} {\bibfnamefont {G.}~\bibnamefont {Ducrozet}}, \bibinfo {author} {\bibfnamefont {G.}~\bibnamefont {Prabhudesai}}, \bibinfo {author} {\bibfnamefont {G.}~\bibnamefont {Michel}}, \bibinfo {author} {\bibfnamefont {A.}~\bibnamefont {Cazaubiel}}, \bibinfo {author} {\bibfnamefont {E.}~\bibnamefont {Falcon}}, \bibinfo {author} {\bibfnamefont {G.}~\bibnamefont {El}}, \ and\ \bibinfo {author} {\bibfnamefont {S.}~\bibnamefont {Randoux}},\ }\href {\doibase 10.1103/PhysRevFluids.5.034802} {\bibfield  {journal} {\bibinfo  {journal} {Phys. Rev. Fluids}\ }\textbf {\bibinfo {volume} {5}},\ \bibinfo {pages} {034802} (\bibinfo {year} {2020})}\BibitemShut {NoStop}%
\bibitem [{\citenamefont {Hoefer}\ \emph {et~al.}(2006)\citenamefont {Hoefer}, \citenamefont {Ablowitz}, \citenamefont {Coddington}, \citenamefont {Cornell}, \citenamefont {Engels},\ and\ \citenamefont {Schweikhard}}]{Hoefer_DSWs}%
  \BibitemOpen
  \bibfield  {author} {\bibinfo {author} {\bibfnamefont {M.~A.}\ \bibnamefont {Hoefer}}, \bibinfo {author} {\bibfnamefont {M.~J.}\ \bibnamefont {Ablowitz}}, \bibinfo {author} {\bibfnamefont {I.}~\bibnamefont {Coddington}}, \bibinfo {author} {\bibfnamefont {E.~A.}\ \bibnamefont {Cornell}}, \bibinfo {author} {\bibfnamefont {P.}~\bibnamefont {Engels}}, \ and\ \bibinfo {author} {\bibfnamefont {V.}~\bibnamefont {Schweikhard}},\ }\href {\doibase 10.1103/PhysRevA.74.023623} {\bibfield  {journal} {\bibinfo  {journal} {Phys. Rev. A}\ }\textbf {\bibinfo {volume} {74}},\ \bibinfo {pages} {023623} (\bibinfo {year} {2006})}\BibitemShut {NoStop}%
\bibitem [{\citenamefont {Chang}\ \emph {et~al.}(2008)\citenamefont {Chang}, \citenamefont {Engels},\ and\ \citenamefont {Hoefer}}]{Chang_DSW}%
  \BibitemOpen
  \bibfield  {author} {\bibinfo {author} {\bibfnamefont {J.~J.}\ \bibnamefont {Chang}}, \bibinfo {author} {\bibfnamefont {P.}~\bibnamefont {Engels}}, \ and\ \bibinfo {author} {\bibfnamefont {M.~A.}\ \bibnamefont {Hoefer}},\ }\href {\doibase 10.1103/PhysRevLett.101.170404} {\bibfield  {journal} {\bibinfo  {journal} {Phys. Rev. Lett.}\ }\textbf {\bibinfo {volume} {101}},\ \bibinfo {pages} {170404} (\bibinfo {year} {2008})}\BibitemShut {NoStop}%
\bibitem [{\citenamefont {Gurevich}\ \emph {et~al.}(1993)\citenamefont {Gurevich}, \citenamefont {Khodorovskiǐ},\ and\ \citenamefont {Krylov}}]{gurevich1993modulational}%
  \BibitemOpen
  \bibfield  {author} {\bibinfo {author} {\bibfnamefont {A.}~\bibnamefont {Gurevich}}, \bibinfo {author} {\bibfnamefont {V.}~\bibnamefont {Khodorovskiǐ}}, \ and\ \bibinfo {author} {\bibfnamefont {A.}~\bibnamefont {Krylov}},\ }\href@noop {} {\bibfield  {journal} {\bibinfo  {journal} {Phys. Lett. A}\ }\textbf {\bibinfo {volume} {177}},\ \bibinfo {pages} {357} (\bibinfo {year} {1993})}\BibitemShut {NoStop}%
\bibitem [{\citenamefont {Whitham}()}]{whithamlinear}%
  \BibitemOpen
  \bibfield  {author} {\bibinfo {author} {\bibfnamefont {G.}~\bibnamefont {Whitham}},\ }\href@noop {} {\ }\BibitemShut {NoStop}%
\bibitem [{\citenamefont {He}\ \emph {et~al.}(2013)\citenamefont {He}, \citenamefont {Zhang}, \citenamefont {Wang}, \citenamefont {Porsezian},\ and\ \citenamefont {Fokas}}]{He_Darboux}%
  \BibitemOpen
  \bibfield  {author} {\bibinfo {author} {\bibfnamefont {J.~S.}\ \bibnamefont {He}}, \bibinfo {author} {\bibfnamefont {H.~R.}\ \bibnamefont {Zhang}}, \bibinfo {author} {\bibfnamefont {L.~H.}\ \bibnamefont {Wang}}, \bibinfo {author} {\bibfnamefont {K.}~\bibnamefont {Porsezian}}, \ and\ \bibinfo {author} {\bibfnamefont {A.~S.}\ \bibnamefont {Fokas}},\ }\href {\doibase 10.1103/PhysRevE.87.052914} {\bibfield  {journal} {\bibinfo  {journal} {Phys. Rev. E}\ }\textbf {\bibinfo {volume} {87}},\ \bibinfo {pages} {052914} (\bibinfo {year} {2013})}\BibitemShut {NoStop}%
\bibitem [{\citenamefont {Adriazola}\ and\ \citenamefont {Kevrekidis}(2025)}]{adriazola2024experimentally}%
  \BibitemOpen
  \bibfield  {author} {\bibinfo {author} {\bibfnamefont {J.}~\bibnamefont {Adriazola}}\ and\ \bibinfo {author} {\bibfnamefont {P.~G.}\ \bibnamefont {Kevrekidis}},\ }\href {\doibase 10.1103/1pn8-jhd2} {\bibfield  {journal} {\bibinfo  {journal} {Phys. Rev. E}\ }\textbf {\bibinfo {volume} {112}},\ \bibinfo {pages} {064207} (\bibinfo {year} {2025})}\BibitemShut {NoStop}%
\bibitem [{\citenamefont {Talanov}(1973)}]{talanov1973certain}%
  \BibitemOpen
  \bibfield  {author} {\bibinfo {author} {\bibfnamefont {V.~I.}\ \bibnamefont {Talanov}},\ }\href@noop {} {\bibfield  {journal} {\bibinfo  {journal} {Soviet Physics Uspekhi}\ }\textbf {\bibinfo {volume} {15}},\ \bibinfo {pages} {521} (\bibinfo {year} {1973})}\BibitemShut {NoStop}%
\bibitem [{\citenamefont {Demontis}\ \emph {et~al.}(2023)\citenamefont {Demontis}, \citenamefont {Ortenzi}, \citenamefont {Roberti},\ and\ \citenamefont {Sommacal}}]{Demontis}%
  \BibitemOpen
  \bibfield  {author} {\bibinfo {author} {\bibfnamefont {F.}~\bibnamefont {Demontis}}, \bibinfo {author} {\bibfnamefont {G.}~\bibnamefont {Ortenzi}}, \bibinfo {author} {\bibfnamefont {G.}~\bibnamefont {Roberti}}, \ and\ \bibinfo {author} {\bibfnamefont {M.}~\bibnamefont {Sommacal}},\ }\href {\doibase 10.1103/PhysRevE.108.024213} {\bibfield  {journal} {\bibinfo  {journal} {Phys. Rev. E}\ }\textbf {\bibinfo {volume} {108}},\ \bibinfo {pages} {024213} (\bibinfo {year} {2023})}\BibitemShut {NoStop}%
\bibitem [{\citenamefont {Gallo}\ and\ \citenamefont {Pelinovsky}(2009)}]{Gallo2009OnTT}%
  \BibitemOpen
  \bibfield  {author} {\bibinfo {author} {\bibfnamefont {C.}~\bibnamefont {Gallo}}\ and\ \bibinfo {author} {\bibfnamefont {D.~E.}\ \bibnamefont {Pelinovsky}},\ }\href {https://api.semanticscholar.org/CorpusID:5822634} {\bibfield  {journal} {\bibinfo  {journal} {Asymptot. Anal.}\ }\textbf {\bibinfo {volume} {73}},\ \bibinfo {pages} {53} (\bibinfo {year} {2009})}\BibitemShut {NoStop}%
\bibitem [{\citenamefont {Karali}\ and\ \citenamefont {Sourdis}(2015)}]{karali}%
  \BibitemOpen
  \bibfield  {author} {\bibinfo {author} {\bibfnamefont {G.}~\bibnamefont {Karali}}\ and\ \bibinfo {author} {\bibfnamefont {C.}~\bibnamefont {Sourdis}},\ }\href@noop {} {\bibfield  {journal} {\bibinfo  {journal} {Archive for Rational Mechanics and Analysis}\ }\textbf {\bibinfo {volume} {217}},\ \bibinfo {pages} {439} (\bibinfo {year} {2015})}\BibitemShut {NoStop}%
\bibitem [{\citenamefont {Ward}\ \emph {et~al.}(2019)\citenamefont {Ward}, \citenamefont {Kevrekidis},\ and\ \citenamefont {Whitaker}}]{WARD20192584}%
  \BibitemOpen
  \bibfield  {author} {\bibinfo {author} {\bibfnamefont {C.}~\bibnamefont {Ward}}, \bibinfo {author} {\bibfnamefont {P.}~\bibnamefont {Kevrekidis}}, \ and\ \bibinfo {author} {\bibfnamefont {N.}~\bibnamefont {Whitaker}},\ }\href {\doibase https://doi.org/10.1016/j.physleta.2019.05.030} {\bibfield  {journal} {\bibinfo  {journal} {Physics Letters A}\ }\textbf {\bibinfo {volume} {383}},\ \bibinfo {pages} {2584} (\bibinfo {year} {2019})}\BibitemShut {NoStop}%
\bibitem [{\citenamefont {Ward}\ and\ \citenamefont {Kevrekidis}(2019)}]{Ward2019}%
  \BibitemOpen
  \bibfield  {author} {\bibinfo {author} {\bibfnamefont {C.~B.}\ \bibnamefont {Ward}}\ and\ \bibinfo {author} {\bibfnamefont {P.~G.}\ \bibnamefont {Kevrekidis}},\ }\href@noop {} {\bibfield  {journal} {\bibinfo  {journal} {ArXiv190205494 Nlin}\ } (\bibinfo {year} {2019})},\ \Eprint {http://arxiv.org/abs/1902.05494} {arxiv:1902.05494 [nlin]} \BibitemShut {NoStop}%
\bibitem [{\citenamefont {Cheiney}\ \emph {et~al.}(2018)\citenamefont {Cheiney}, \citenamefont {Cabrera}, \citenamefont {Sanz}, \citenamefont {Naylor}, \citenamefont {Tanzi},\ and\ \citenamefont {Tarruell}}]{cheiney2018bright}%
  \BibitemOpen
  \bibfield  {author} {\bibinfo {author} {\bibfnamefont {P.}~\bibnamefont {Cheiney}}, \bibinfo {author} {\bibfnamefont {C.~R.}\ \bibnamefont {Cabrera}}, \bibinfo {author} {\bibfnamefont {J.}~\bibnamefont {Sanz}}, \bibinfo {author} {\bibfnamefont {B.}~\bibnamefont {Naylor}}, \bibinfo {author} {\bibfnamefont {L.}~\bibnamefont {Tanzi}}, \ and\ \bibinfo {author} {\bibfnamefont {L.}~\bibnamefont {Tarruell}},\ }\href {\doibase 10.1103/PhysRevLett.120.135301} {\bibfield  {journal} {\bibinfo  {journal} {Phys. Rev. Lett.}\ }\textbf {\bibinfo {volume} {120}},\ \bibinfo {pages} {135301} (\bibinfo {year} {2018})}\BibitemShut {NoStop}%
\bibitem [{\citenamefont {Semeghini}\ \emph {et~al.}(2018)\citenamefont {Semeghini}, \citenamefont {Ferioli}, \citenamefont {Masi}, \citenamefont {Mazzinghi}, \citenamefont {Wolswijk}, \citenamefont {Minardi}, \citenamefont {Modugno}, \citenamefont {Modugno}, \citenamefont {Inguscio},\ and\ \citenamefont {Fattori}}]{semeghini2018self}%
  \BibitemOpen
  \bibfield  {author} {\bibinfo {author} {\bibfnamefont {G.}~\bibnamefont {Semeghini}}, \bibinfo {author} {\bibfnamefont {G.}~\bibnamefont {Ferioli}}, \bibinfo {author} {\bibfnamefont {L.}~\bibnamefont {Masi}}, \bibinfo {author} {\bibfnamefont {C.}~\bibnamefont {Mazzinghi}}, \bibinfo {author} {\bibfnamefont {L.}~\bibnamefont {Wolswijk}}, \bibinfo {author} {\bibfnamefont {F.}~\bibnamefont {Minardi}}, \bibinfo {author} {\bibfnamefont {M.}~\bibnamefont {Modugno}}, \bibinfo {author} {\bibfnamefont {G.}~\bibnamefont {Modugno}}, \bibinfo {author} {\bibfnamefont {M.}~\bibnamefont {Inguscio}}, \ and\ \bibinfo {author} {\bibfnamefont {M.}~\bibnamefont {Fattori}},\ }\href {\doibase 10.1103/PhysRevLett.120.235301} {\bibfield  {journal} {\bibinfo  {journal} {Phys. Rev. Lett.}\ }\textbf {\bibinfo {volume} {120}},\ \bibinfo {pages} {235301} (\bibinfo {year} {2018})}\BibitemShut {NoStop}%
\bibitem [{\citenamefont {Cabrera}\ \emph {et~al.}(2018)\citenamefont {Cabrera}, \citenamefont {Tanzi}, \citenamefont {Sanz}, \citenamefont {Naylor}, \citenamefont {Thomas}, \citenamefont {Cheiney},\ and\ \citenamefont {Tarruell}}]{cabrera2018quantum}%
  \BibitemOpen
  \bibfield  {author} {\bibinfo {author} {\bibfnamefont {C.}~\bibnamefont {Cabrera}}, \bibinfo {author} {\bibfnamefont {L.}~\bibnamefont {Tanzi}}, \bibinfo {author} {\bibfnamefont {J.}~\bibnamefont {Sanz}}, \bibinfo {author} {\bibfnamefont {B.}~\bibnamefont {Naylor}}, \bibinfo {author} {\bibfnamefont {P.}~\bibnamefont {Thomas}}, \bibinfo {author} {\bibfnamefont {P.}~\bibnamefont {Cheiney}}, \ and\ \bibinfo {author} {\bibfnamefont {L.}~\bibnamefont {Tarruell}},\ }\href@noop {} {\bibfield  {journal} {\bibinfo  {journal} {Science}\ }\textbf {\bibinfo {volume} {359}},\ \bibinfo {pages} {301} (\bibinfo {year} {2018})}\BibitemShut {NoStop}%
\bibitem [{\citenamefont {D'Errico}\ \emph {et~al.}(2019)\citenamefont {D'Errico}, \citenamefont {Burchianti}, \citenamefont {Prevedelli}, \citenamefont {Salasnich}, \citenamefont {Ancilotto}, \citenamefont {Modugno}, \citenamefont {Minardi},\ and\ \citenamefont {Fort}}]{Errico_droplets}%
  \BibitemOpen
  \bibfield  {author} {\bibinfo {author} {\bibfnamefont {C.}~\bibnamefont {D'Errico}}, \bibinfo {author} {\bibfnamefont {A.}~\bibnamefont {Burchianti}}, \bibinfo {author} {\bibfnamefont {M.}~\bibnamefont {Prevedelli}}, \bibinfo {author} {\bibfnamefont {L.}~\bibnamefont {Salasnich}}, \bibinfo {author} {\bibfnamefont {F.}~\bibnamefont {Ancilotto}}, \bibinfo {author} {\bibfnamefont {M.}~\bibnamefont {Modugno}}, \bibinfo {author} {\bibfnamefont {F.}~\bibnamefont {Minardi}}, \ and\ \bibinfo {author} {\bibfnamefont {C.}~\bibnamefont {Fort}},\ }\href {\doibase 10.1103/PhysRevResearch.1.033155} {\bibfield  {journal} {\bibinfo  {journal} {Phys. Rev. Res.}\ }\textbf {\bibinfo {volume} {1}},\ \bibinfo {pages} {033155} (\bibinfo {year} {2019})}\BibitemShut {NoStop}%
\bibitem [{\citenamefont {Burchianti}\ \emph {et~al.}(2020)\citenamefont {Burchianti}, \citenamefont {D’Errico}, \citenamefont {Prevedelli}, \citenamefont {Salasnich}, \citenamefont {Ancilotto}, \citenamefont {Modugno}, \citenamefont {Minardi},\ and\ \citenamefont {Fort}}]{burchianti2020dual}%
  \BibitemOpen
  \bibfield  {author} {\bibinfo {author} {\bibfnamefont {A.}~\bibnamefont {Burchianti}}, \bibinfo {author} {\bibfnamefont {C.}~\bibnamefont {D’Errico}}, \bibinfo {author} {\bibfnamefont {M.}~\bibnamefont {Prevedelli}}, \bibinfo {author} {\bibfnamefont {L.}~\bibnamefont {Salasnich}}, \bibinfo {author} {\bibfnamefont {F.}~\bibnamefont {Ancilotto}}, \bibinfo {author} {\bibfnamefont {M.}~\bibnamefont {Modugno}}, \bibinfo {author} {\bibfnamefont {F.}~\bibnamefont {Minardi}}, \ and\ \bibinfo {author} {\bibfnamefont {C.}~\bibnamefont {Fort}},\ }\href@noop {} {\bibfield  {journal} {\bibinfo  {journal} {Cond. Matter}\ }\textbf {\bibinfo {volume} {5}},\ \bibinfo {pages} {21} (\bibinfo {year} {2020})}\BibitemShut {NoStop}%
\bibitem [{\citenamefont {Lee}\ \emph {et~al.}(1957)\citenamefont {Lee}, \citenamefont {Huang},\ and\ \citenamefont {Yang}}]{lee1957eigenvalues}%
  \BibitemOpen
  \bibfield  {author} {\bibinfo {author} {\bibfnamefont {T.~D.}\ \bibnamefont {Lee}}, \bibinfo {author} {\bibfnamefont {K.}~\bibnamefont {Huang}}, \ and\ \bibinfo {author} {\bibfnamefont {C.~N.}\ \bibnamefont {Yang}},\ }\href@noop {} {\bibfield  {journal} {\bibinfo  {journal} {Phys. Rev.}\ }\textbf {\bibinfo {volume} {106}},\ \bibinfo {pages} {1135} (\bibinfo {year} {1957})}\BibitemShut {NoStop}%
\bibitem [{\citenamefont {Petrov}\ and\ \citenamefont {Astrakharchik}(2016)}]{Petrov_2016}%
  \BibitemOpen
  \bibfield  {author} {\bibinfo {author} {\bibfnamefont {D.~S.}\ \bibnamefont {Petrov}}\ and\ \bibinfo {author} {\bibfnamefont {G.~E.}\ \bibnamefont {Astrakharchik}},\ }\href {\doibase 10.1103/PhysRevLett.117.100401} {\bibfield  {journal} {\bibinfo  {journal} {Phys. Rev. Lett.}\ }\textbf {\bibinfo {volume} {117}},\ \bibinfo {pages} {100401} (\bibinfo {year} {2016})}\BibitemShut {NoStop}%
\bibitem [{\citenamefont {Petrov}(2015)}]{Petrov_2015}%
  \BibitemOpen
  \bibfield  {author} {\bibinfo {author} {\bibfnamefont {D.~S.}\ \bibnamefont {Petrov}},\ }\href {\doibase 10.1103/PhysRevLett.115.155302} {\bibfield  {journal} {\bibinfo  {journal} {Phys. Rev. Lett.}\ }\textbf {\bibinfo {volume} {115}},\ \bibinfo {pages} {155302} (\bibinfo {year} {2015})}\BibitemShut {NoStop}%
\bibitem [{\citenamefont {Luo}\ \emph {et~al.}(2021)\citenamefont {Luo}, \citenamefont {Pang}, \citenamefont {Liu}, \citenamefont {Li},\ and\ \citenamefont {Malomed}}]{LuoPangLiuLiMalomed2021_Qdroplets}%
  \BibitemOpen
  \bibfield  {author} {\bibinfo {author} {\bibfnamefont {Z.-H.}\ \bibnamefont {Luo}}, \bibinfo {author} {\bibfnamefont {W.}~\bibnamefont {Pang}}, \bibinfo {author} {\bibfnamefont {B.}~\bibnamefont {Liu}}, \bibinfo {author} {\bibfnamefont {Y.-Y.}\ \bibnamefont {Li}}, \ and\ \bibinfo {author} {\bibfnamefont {B.~A.}\ \bibnamefont {Malomed}},\ }\href {\doibase 10.1007/s11467-020-1020-2} {\bibfield  {journal} {\bibinfo  {journal} {Frontiers of Physics}\ }\textbf {\bibinfo {volume} {16}},\ \bibinfo {pages} {32201} (\bibinfo {year} {2021})}\BibitemShut {NoStop}%
\bibitem [{\citenamefont {Chandramouli}\ \emph {et~al.}(2026)\citenamefont {Chandramouli}, \citenamefont {Mistakidis}, \citenamefont {Katsimiga}, \citenamefont {Ratliff}, \citenamefont {Frantzeskakis},\ and\ \citenamefont {Kevrekidis}}]{Chandramouli2025rogue}%
  \BibitemOpen
  \bibfield  {author} {\bibinfo {author} {\bibfnamefont {S.}~\bibnamefont {Chandramouli}}, \bibinfo {author} {\bibfnamefont {S.~I.}\ \bibnamefont {Mistakidis}}, \bibinfo {author} {\bibfnamefont {G.~C.}\ \bibnamefont {Katsimiga}}, \bibinfo {author} {\bibfnamefont {D.~J.}\ \bibnamefont {Ratliff}}, \bibinfo {author} {\bibfnamefont {D.~J.}\ \bibnamefont {Frantzeskakis}}, \ and\ \bibinfo {author} {\bibfnamefont {P.~G.}\ \bibnamefont {Kevrekidis}},\ }\href {\doibase 10.1103/7rlg-z74h} {\bibfield  {journal} {\bibinfo  {journal} {Physical Review A}\ }\textbf {\bibinfo {volume} {113}} (\bibinfo {year} {2026}),\ 10.1103/7rlg-z74h}\BibitemShut {NoStop}%
\bibitem [{\citenamefont {Tylutki}\ \emph {et~al.}(2020)\citenamefont {Tylutki}, \citenamefont {Astrakharchik}, \citenamefont {Malomed},\ and\ \citenamefont {Petrov}}]{tylutki}%
  \BibitemOpen
  \bibfield  {author} {\bibinfo {author} {\bibfnamefont {M.}~\bibnamefont {Tylutki}}, \bibinfo {author} {\bibfnamefont {G.~E.}\ \bibnamefont {Astrakharchik}}, \bibinfo {author} {\bibfnamefont {B.~A.}\ \bibnamefont {Malomed}}, \ and\ \bibinfo {author} {\bibfnamefont {D.~S.}\ \bibnamefont {Petrov}},\ }\href {\doibase 10.1103/PhysRevA.101.051601} {\bibfield  {journal} {\bibinfo  {journal} {Phys. Rev. A}\ }\textbf {\bibinfo {volume} {101}},\ \bibinfo {pages} {051601} (\bibinfo {year} {2020})}\BibitemShut {NoStop}%
\bibitem [{\citenamefont {Chandramouli}\ \emph {et~al.}(2024)\citenamefont {Chandramouli}, \citenamefont {Mistakidis}, \citenamefont {Katsimiga},\ and\ \citenamefont {Kevrekidis}}]{Chandramouli_DSWs}%
  \BibitemOpen
  \bibfield  {author} {\bibinfo {author} {\bibfnamefont {S.}~\bibnamefont {Chandramouli}}, \bibinfo {author} {\bibfnamefont {S.~I.}\ \bibnamefont {Mistakidis}}, \bibinfo {author} {\bibfnamefont {G.~C.}\ \bibnamefont {Katsimiga}}, \ and\ \bibinfo {author} {\bibfnamefont {P.~G.}\ \bibnamefont {Kevrekidis}},\ }\href {\doibase 10.1103/PhysRevA.110.023304} {\bibfield  {journal} {\bibinfo  {journal} {Phys. Rev. A}\ }\textbf {\bibinfo {volume} {110}},\ \bibinfo {pages} {023304} (\bibinfo {year} {2024})}\BibitemShut {NoStop}%
\bibitem [{\citenamefont {Panayotis G.~Kevrekidis}(2008)}]{kevrekidis2008emergent}%
  \BibitemOpen
  \bibfield  {author} {\bibinfo {author} {\bibfnamefont {R.~C.-G.}\ \bibnamefont {Panayotis G.~Kevrekidis}, \bibfnamefont {Dimitri J.~Frantzeskakis}},\ }\href {\doibase 10.1007/978-3-540-73591-5} {\emph {\bibinfo {title} {Emergent Nonlinear Phenomena in Bose-Einstein Condensates}}},\ \bibinfo {series} {Springer Series on Atomic, Optical, and Plasma Physics}, Vol.~\bibinfo {volume} {45}\ (\bibinfo  {publisher} {Springer},\ \bibinfo {address} {Berlin, Heidelberg},\ \bibinfo {year} {2008})\BibitemShut {NoStop}%
\bibitem [{\citenamefont {Straten}(2016)}]{Straten2016}%
  \BibitemOpen
  \bibfield  {author} {\bibinfo {author} {\bibfnamefont {P.}~\bibnamefont {Straten}},\ }\href@noop {} {\emph {\bibinfo {title} {Atoms and molecules interacting with light}}},\ edited by\ \bibinfo {editor} {\bibfnamefont {H.}~\bibnamefont {Metcalf}}\ (\bibinfo  {publisher} {Cambridge University Press},\ \bibinfo {address} {Cambridge},\ \bibinfo {year} {2016})\ \bibinfo {note} {includes bibliographical references and index}\BibitemShut {NoStop}%
\bibitem [{\citenamefont {Onorato}\ \emph {et~al.}(2013)\citenamefont {Onorato}, \citenamefont {Residori}, \citenamefont {Bortolozzo}, \citenamefont {Montina},\ and\ \citenamefont {Arecchi}}]{onorato2013rogue}%
  \BibitemOpen
  \bibfield  {author} {\bibinfo {author} {\bibfnamefont {M.}~\bibnamefont {Onorato}}, \bibinfo {author} {\bibfnamefont {S.}~\bibnamefont {Residori}}, \bibinfo {author} {\bibfnamefont {U.}~\bibnamefont {Bortolozzo}}, \bibinfo {author} {\bibfnamefont {A.}~\bibnamefont {Montina}}, \ and\ \bibinfo {author} {\bibfnamefont {F.}~\bibnamefont {Arecchi}},\ }\href@noop {} {\bibfield  {journal} {\bibinfo  {journal} {Physics Reports}\ }\textbf {\bibinfo {volume} {528}},\ \bibinfo {pages} {47} (\bibinfo {year} {2013})}\BibitemShut {NoStop}%
\bibitem [{\citenamefont {Dudley}\ \emph {et~al.}(2019)\citenamefont {Dudley}, \citenamefont {Genty}, \citenamefont {Mussot}, \citenamefont {Chabchoub},\ and\ \citenamefont {Dias}}]{dudley2019rogue}%
  \BibitemOpen
  \bibfield  {author} {\bibinfo {author} {\bibfnamefont {J.~M.}\ \bibnamefont {Dudley}}, \bibinfo {author} {\bibfnamefont {G.}~\bibnamefont {Genty}}, \bibinfo {author} {\bibfnamefont {A.}~\bibnamefont {Mussot}}, \bibinfo {author} {\bibfnamefont {A.}~\bibnamefont {Chabchoub}}, \ and\ \bibinfo {author} {\bibfnamefont {F.}~\bibnamefont {Dias}},\ }\href@noop {} {\bibfield  {journal} {\bibinfo  {journal} {Nature Reviews Physics}\ }\textbf {\bibinfo {volume} {1}},\ \bibinfo {pages} {675} (\bibinfo {year} {2019})}\BibitemShut {NoStop}%
\bibitem [{\citenamefont {Haver}(2004)}]{haver2004possible}%
  \BibitemOpen
  \bibfield  {author} {\bibinfo {author} {\bibfnamefont {S.}~\bibnamefont {Haver}},\ }in\ \href@noop {} {\emph {\bibinfo {booktitle} {Rogue waves}}},\ Vol.\ \bibinfo {volume} {2004}\ (\bibinfo {organization} {Ifremer Brest, France},\ \bibinfo {year} {2004})\ pp.\ \bibinfo {pages} {1--8}\BibitemShut {NoStop}%
\bibitem [{\citenamefont {Mori}\ \emph {et~al.}(2023)\citenamefont {Mori}, \citenamefont {Waseda},\ and\ \citenamefont {Chabchoub}}]{mori2023science}%
  \BibitemOpen
  \bibfield  {author} {\bibinfo {author} {\bibfnamefont {N.}~\bibnamefont {Mori}}, \bibinfo {author} {\bibfnamefont {T.}~\bibnamefont {Waseda}}, \ and\ \bibinfo {author} {\bibfnamefont {A.}~\bibnamefont {Chabchoub}},\ }\href@noop {} {\emph {\bibinfo {title} {Science and Engineering of Freak Waves}}}\ (\bibinfo  {publisher} {Elsevier},\ \bibinfo {year} {2023})\BibitemShut {NoStop}%
\bibitem [{\citenamefont {Tulin}(1996)}]{tulin1996breaking}%
  \BibitemOpen
  \bibfield  {author} {\bibinfo {author} {\bibfnamefont {M.~P.}\ \bibnamefont {Tulin}},\ }in\ \href@noop {} {\emph {\bibinfo {booktitle} {Waves and nonlinear processes in hydrodynamics}}}\ (\bibinfo  {publisher} {Springer},\ \bibinfo {year} {1996})\ pp.\ \bibinfo {pages} {177--190}\BibitemShut {NoStop}%
\bibitem [{\citenamefont {Fujimoto}\ \emph {et~al.}(2019)\citenamefont {Fujimoto}, \citenamefont {Waseda},\ and\ \citenamefont {Webb}}]{fujimoto2019impact}%
  \BibitemOpen
  \bibfield  {author} {\bibinfo {author} {\bibfnamefont {W.}~\bibnamefont {Fujimoto}}, \bibinfo {author} {\bibfnamefont {T.}~\bibnamefont {Waseda}}, \ and\ \bibinfo {author} {\bibfnamefont {A.}~\bibnamefont {Webb}},\ }\href@noop {} {\bibfield  {journal} {\bibinfo  {journal} {Ocean Dynamics}\ }\textbf {\bibinfo {volume} {69}},\ \bibinfo {pages} {101} (\bibinfo {year} {2019})}\BibitemShut {NoStop}%
\bibitem [{\citenamefont {Malila}\ \emph {et~al.}(2023)\citenamefont {Malila}, \citenamefont {Barbariol}, \citenamefont {Benetazzo}, \citenamefont {Breivik}, \citenamefont {Magnusson}, \citenamefont {Thomson},\ and\ \citenamefont {Ward}}]{malila2023statistical}%
  \BibitemOpen
  \bibfield  {author} {\bibinfo {author} {\bibfnamefont {M.~P.}\ \bibnamefont {Malila}}, \bibinfo {author} {\bibfnamefont {F.}~\bibnamefont {Barbariol}}, \bibinfo {author} {\bibfnamefont {A.}~\bibnamefont {Benetazzo}}, \bibinfo {author} {\bibfnamefont {{\O}.}~\bibnamefont {Breivik}}, \bibinfo {author} {\bibfnamefont {A.~K.}\ \bibnamefont {Magnusson}}, \bibinfo {author} {\bibfnamefont {J.}~\bibnamefont {Thomson}}, \ and\ \bibinfo {author} {\bibfnamefont {B.}~\bibnamefont {Ward}},\ }\href@noop {} {\bibfield  {journal} {\bibinfo  {journal} {Journal of Physical Oceanography}\ }\textbf {\bibinfo {volume} {53}},\ \bibinfo {pages} {509} (\bibinfo {year} {2023})}\BibitemShut {NoStop}%
\bibitem [{\citenamefont {Onorato}\ \emph {et~al.}(2006)\citenamefont {Onorato}, \citenamefont {Osborne}, \citenamefont {Serio}, \citenamefont {Cavaleri}, \citenamefont {Brandini},\ and\ \citenamefont {Stansberg}}]{onorato2006extreme}%
  \BibitemOpen
  \bibfield  {author} {\bibinfo {author} {\bibfnamefont {M.}~\bibnamefont {Onorato}}, \bibinfo {author} {\bibfnamefont {A.~R.}\ \bibnamefont {Osborne}}, \bibinfo {author} {\bibfnamefont {M.}~\bibnamefont {Serio}}, \bibinfo {author} {\bibfnamefont {L.}~\bibnamefont {Cavaleri}}, \bibinfo {author} {\bibfnamefont {C.}~\bibnamefont {Brandini}}, \ and\ \bibinfo {author} {\bibfnamefont {C.~T.}\ \bibnamefont {Stansberg}},\ }\href@noop {} {\bibfield  {journal} {\bibinfo  {journal} {European Journal of Mechanics-B/Fluids}\ }\textbf {\bibinfo {volume} {25}},\ \bibinfo {pages} {586} (\bibinfo {year} {2006})}\BibitemShut {NoStop}%
\bibitem [{\citenamefont {Janssen}(2003)}]{janssen2003nonlinear}%
  \BibitemOpen
  \bibfield  {author} {\bibinfo {author} {\bibfnamefont {P.~A.}\ \bibnamefont {Janssen}},\ }\href@noop {} {\bibfield  {journal} {\bibinfo  {journal} {Journal of Physical Oceanography}\ }\textbf {\bibinfo {volume} {33}},\ \bibinfo {pages} {863} (\bibinfo {year} {2003})}\BibitemShut {NoStop}%
\bibitem [{\citenamefont {Yuen}\ and\ \citenamefont {Lake}(1982)}]{yuen1982nonlinear}%
  \BibitemOpen
  \bibfield  {author} {\bibinfo {author} {\bibfnamefont {H.~C.}\ \bibnamefont {Yuen}}\ and\ \bibinfo {author} {\bibfnamefont {B.~M.}\ \bibnamefont {Lake}},\ }\href@noop {} {\bibfield  {journal} {\bibinfo  {journal} {Advances in applied mechanics}\ }\textbf {\bibinfo {volume} {22}},\ \bibinfo {pages} {67} (\bibinfo {year} {1982})}\BibitemShut {NoStop}%
\bibitem [{\citenamefont {Dysthe}(1979)}]{dysthe1979note}%
  \BibitemOpen
  \bibfield  {author} {\bibinfo {author} {\bibfnamefont {K.~B.}\ \bibnamefont {Dysthe}},\ }\href@noop {} {\bibfield  {journal} {\bibinfo  {journal} {Proceedings of the Royal Society of London. A. Mathematical and Physical Sciences}\ }\textbf {\bibinfo {volume} {369}},\ \bibinfo {pages} {105} (\bibinfo {year} {1979})}\BibitemShut {NoStop}%
\bibitem [{\citenamefont {Agrawal}(2013)}]{agrawal2013nonlinear}%
  \BibitemOpen
  \bibfield  {author} {\bibinfo {author} {\bibfnamefont {G.~P.}\ \bibnamefont {Agrawal}},\ }\href@noop {} {\emph {\bibinfo {title} {Nonlinear fiber optics}}}\ (\bibinfo  {publisher} {Elevier},\ \bibinfo {year} {2013})\BibitemShut {NoStop}%
\bibitem [{\citenamefont {Tlidi}\ and\ \citenamefont {Taki}(2022)}]{tlidi2022rogue}%
  \BibitemOpen
  \bibfield  {author} {\bibinfo {author} {\bibfnamefont {M.}~\bibnamefont {Tlidi}}\ and\ \bibinfo {author} {\bibfnamefont {M.}~\bibnamefont {Taki}},\ }\href@noop {} {\bibfield  {journal} {\bibinfo  {journal} {Advances in Optics and Photonics}\ }\textbf {\bibinfo {volume} {14}},\ \bibinfo {pages} {87} (\bibinfo {year} {2022})}\BibitemShut {NoStop}%
\bibitem [{\citenamefont {Chabchoub}\ \emph {et~al.}(2015)\citenamefont {Chabchoub}, \citenamefont {Kibler}, \citenamefont {Finot}, \citenamefont {Millot}, \citenamefont {Onorato}, \citenamefont {Dudley},\ and\ \citenamefont {Babanin}}]{chabchoub2015nonlinear}%
  \BibitemOpen
  \bibfield  {author} {\bibinfo {author} {\bibfnamefont {A.}~\bibnamefont {Chabchoub}}, \bibinfo {author} {\bibfnamefont {B.}~\bibnamefont {Kibler}}, \bibinfo {author} {\bibfnamefont {C.}~\bibnamefont {Finot}}, \bibinfo {author} {\bibfnamefont {G.}~\bibnamefont {Millot}}, \bibinfo {author} {\bibfnamefont {M.}~\bibnamefont {Onorato}}, \bibinfo {author} {\bibfnamefont {J.~M.}\ \bibnamefont {Dudley}}, \ and\ \bibinfo {author} {\bibfnamefont {A.}~\bibnamefont {Babanin}},\ }\href@noop {} {\bibfield  {journal} {\bibinfo  {journal} {Annals of Physics}\ }\textbf {\bibinfo {volume} {361}},\ \bibinfo {pages} {490} (\bibinfo {year} {2015})}\BibitemShut {NoStop}%
\bibitem [{\citenamefont {Osborne}(2010)}]{osborne2010nonlinear}%
  \BibitemOpen
  \bibfield  {author} {\bibinfo {author} {\bibfnamefont {A.~R.}\ \bibnamefont {Osborne}},\ }\href@noop {} {\emph {\bibinfo {title} {Nonlinear Ocean Waves and the Inverse Scattering Transform}}}\ (\bibinfo  {publisher} {Academic Press},\ \bibinfo {address} {Oxford},\ \bibinfo {year} {2010})\BibitemShut {NoStop}%
\bibitem [{\citenamefont {Mei}\ \emph {et~al.}(2005)\citenamefont {Mei}, \citenamefont {Stiassnie},\ and\ \citenamefont {Yue}}]{mei2005theory}%
  \BibitemOpen
  \bibfield  {author} {\bibinfo {author} {\bibfnamefont {C.~C.}\ \bibnamefont {Mei}}, \bibinfo {author} {\bibfnamefont {M.~A.}\ \bibnamefont {Stiassnie}}, \ and\ \bibinfo {author} {\bibfnamefont {D.~K.-P.}\ \bibnamefont {Yue}},\ }\href@noop {} {\emph {\bibinfo {title} {Theory and applications of ocean surface waves}}}\ (\bibinfo  {publisher} {World Scientific},\ \bibinfo {year} {2005})\BibitemShut {NoStop}%
\bibitem [{\citenamefont {Zakharov}(1968)}]{zakharov1968stability}%
  \BibitemOpen
  \bibfield  {author} {\bibinfo {author} {\bibfnamefont {V.~E.}\ \bibnamefont {Zakharov}},\ }\href@noop {} {\bibfield  {journal} {\bibinfo  {journal} {Journal of Applied Mechanics and Technical Physics}\ }\textbf {\bibinfo {volume} {9}},\ \bibinfo {pages} {190} (\bibinfo {year} {1968})}\BibitemShut {NoStop}%
\bibitem [{\citenamefont {Remoissenet}(2013)}]{remoissenet2013waves}%
  \BibitemOpen
  \bibfield  {author} {\bibinfo {author} {\bibfnamefont {M.}~\bibnamefont {Remoissenet}},\ }\href@noop {} {\emph {\bibinfo {title} {Waves called solitons: concepts and experiments}}}\ (\bibinfo  {publisher} {Springer Science \& Business Media},\ \bibinfo {year} {2013})\BibitemShut {NoStop}%
\bibitem [{\citenamefont {Trulsen}\ and\ \citenamefont {Stansberg}(2001)}]{trulsen2001spatial}%
  \BibitemOpen
  \bibfield  {author} {\bibinfo {author} {\bibfnamefont {K.}~\bibnamefont {Trulsen}}\ and\ \bibinfo {author} {\bibfnamefont {C.~T.}\ \bibnamefont {Stansberg}},\ }in\ \href@noop {} {\emph {\bibinfo {booktitle} {Isope international ocean and polar engineering conference}}}\ (\bibinfo {organization} {ISOPE},\ \bibinfo {year} {2001})\ pp.\ \bibinfo {pages} {ISOPE--I}\BibitemShut {NoStop}%
\bibitem [{\citenamefont {Kit}\ and\ \citenamefont {Shemer}(2002)}]{kit2002spatial}%
  \BibitemOpen
  \bibfield  {author} {\bibinfo {author} {\bibfnamefont {E.}~\bibnamefont {Kit}}\ and\ \bibinfo {author} {\bibfnamefont {L.}~\bibnamefont {Shemer}},\ }\href@noop {} {\bibfield  {journal} {\bibinfo  {journal} {Journal of Fluid Mechanics}\ }\textbf {\bibinfo {volume} {450}},\ \bibinfo {pages} {201} (\bibinfo {year} {2002})}\BibitemShut {NoStop}%
\bibitem [{\citenamefont {Goullet}\ and\ \citenamefont {Choi}(2011)}]{goullet2011numerical}%
  \BibitemOpen
  \bibfield  {author} {\bibinfo {author} {\bibfnamefont {A.}~\bibnamefont {Goullet}}\ and\ \bibinfo {author} {\bibfnamefont {W.}~\bibnamefont {Choi}},\ }\href@noop {} {\bibfield  {journal} {\bibinfo  {journal} {Physics of Fluids}\ }\textbf {\bibinfo {volume} {23}} (\bibinfo {year} {2011})}\BibitemShut {NoStop}%
\bibitem [{\citenamefont {Dudley}\ \emph {et~al.}(2006)\citenamefont {Dudley}, \citenamefont {Genty},\ and\ \citenamefont {Coen}}]{dudley2006supercontinuum}%
  \BibitemOpen
  \bibfield  {author} {\bibinfo {author} {\bibfnamefont {J.~M.}\ \bibnamefont {Dudley}}, \bibinfo {author} {\bibfnamefont {G.}~\bibnamefont {Genty}}, \ and\ \bibinfo {author} {\bibfnamefont {S.}~\bibnamefont {Coen}},\ }\href@noop {} {\bibfield  {journal} {\bibinfo  {journal} {Reviews of modern physics}\ }\textbf {\bibinfo {volume} {78}},\ \bibinfo {pages} {1135} (\bibinfo {year} {2006})}\BibitemShut {NoStop}%
\bibitem [{\citenamefont {Chabchoub}\ \emph {et~al.}(2013)\citenamefont {Chabchoub}, \citenamefont {Hoffmann}, \citenamefont {Onorato}, \citenamefont {Genty}, \citenamefont {Dudley},\ and\ \citenamefont {Akhmediev}}]{chabchoub2013hydrodynamic}%
  \BibitemOpen
  \bibfield  {author} {\bibinfo {author} {\bibfnamefont {A.}~\bibnamefont {Chabchoub}}, \bibinfo {author} {\bibfnamefont {N.}~\bibnamefont {Hoffmann}}, \bibinfo {author} {\bibfnamefont {M.}~\bibnamefont {Onorato}}, \bibinfo {author} {\bibfnamefont {G.}~\bibnamefont {Genty}}, \bibinfo {author} {\bibfnamefont {J.~M.}\ \bibnamefont {Dudley}}, \ and\ \bibinfo {author} {\bibfnamefont {N.}~\bibnamefont {Akhmediev}},\ }\href@noop {} {\bibfield  {journal} {\bibinfo  {journal} {Physical review letters}\ }\textbf {\bibinfo {volume} {111}},\ \bibinfo {pages} {054104} (\bibinfo {year} {2013})}\BibitemShut {NoStop}%
\bibitem [{\citenamefont {Bespalov}\ and\ \citenamefont {Talanov}(1966)}]{bespalov1966filamentary}%
  \BibitemOpen
  \bibfield  {author} {\bibinfo {author} {\bibfnamefont {V.~I.}\ \bibnamefont {Bespalov}}\ and\ \bibinfo {author} {\bibfnamefont {V.~I.}\ \bibnamefont {Talanov}},\ }\href@noop {} {\bibfield  {journal} {\bibinfo  {journal} {Soviet Journal of Experimental and Theoretical Physics Letters}\ }\textbf {\bibinfo {volume} {3}},\ \bibinfo {pages} {307} (\bibinfo {year} {1966})}\BibitemShut {NoStop}%
\bibitem [{\citenamefont {Benney}\ and\ \citenamefont {Newell}(1967)}]{benney1967propagation}%
  \BibitemOpen
  \bibfield  {author} {\bibinfo {author} {\bibfnamefont {D.}~\bibnamefont {Benney}}\ and\ \bibinfo {author} {\bibfnamefont {A.~C.}\ \bibnamefont {Newell}},\ }\href@noop {} {\bibfield  {journal} {\bibinfo  {journal} {Journal of mathematics and Physics}\ }\textbf {\bibinfo {volume} {46}},\ \bibinfo {pages} {133} (\bibinfo {year} {1967})}\BibitemShut {NoStop}%
\bibitem [{\citenamefont {Yuen}\ and\ \citenamefont {Ferguson~Jr}(1978)}]{yuen1978relationship}%
  \BibitemOpen
  \bibfield  {author} {\bibinfo {author} {\bibfnamefont {H.~C.}\ \bibnamefont {Yuen}}\ and\ \bibinfo {author} {\bibfnamefont {W.~E.}\ \bibnamefont {Ferguson~Jr}},\ }\href@noop {} {\bibfield  {journal} {\bibinfo  {journal} {The Physics of Fluids}\ }\textbf {\bibinfo {volume} {21}},\ \bibinfo {pages} {1275} (\bibinfo {year} {1978})}\BibitemShut {NoStop}%
\bibitem [{\citenamefont {Akhmediev}\ \emph {et~al.}(1985)\citenamefont {Akhmediev}, \citenamefont {Eleonskii},\ and\ \citenamefont {Kulagin}}]{akhmediev1985generation}%
  \BibitemOpen
  \bibfield  {author} {\bibinfo {author} {\bibfnamefont {N.}~\bibnamefont {Akhmediev}}, \bibinfo {author} {\bibfnamefont {V.}~\bibnamefont {Eleonskii}}, \ and\ \bibinfo {author} {\bibfnamefont {N.}~\bibnamefont {Kulagin}},\ }\href@noop {} {\bibfield  {journal} {\bibinfo  {journal} {Sov. Phys. JETP}\ }\textbf {\bibinfo {volume} {62}},\ \bibinfo {pages} {894} (\bibinfo {year} {1985})}\BibitemShut {NoStop}%
\bibitem [{\citenamefont {Akhmediev}\ \emph {et~al.}(1987)\citenamefont {Akhmediev}, \citenamefont {Eleonskii},\ and\ \citenamefont {Kulagin}}]{akhmediev1987exact}%
  \BibitemOpen
  \bibfield  {author} {\bibinfo {author} {\bibfnamefont {N.~N.}\ \bibnamefont {Akhmediev}}, \bibinfo {author} {\bibfnamefont {V.~M.}\ \bibnamefont {Eleonskii}}, \ and\ \bibinfo {author} {\bibfnamefont {N.}~\bibnamefont {Kulagin}},\ }\href@noop {} {\bibfield  {journal} {\bibinfo  {journal} {Theoretical and mathematical physics}\ }\textbf {\bibinfo {volume} {72}},\ \bibinfo {pages} {809} (\bibinfo {year} {1987})}\BibitemShut {NoStop}%
\bibitem [{\citenamefont {Wetzel}\ \emph {et~al.}(2011)\citenamefont {Wetzel}, \citenamefont {Erkintalo}, \citenamefont {Genty}, \citenamefont {Hammani}, \citenamefont {Kibler}, \citenamefont {Fatome}, \citenamefont {Finot}, \citenamefont {Dias}, \citenamefont {Akhmediev}, \citenamefont {Millot} \emph {et~al.}}]{wetzel2011new}%
  \BibitemOpen
  \bibfield  {author} {\bibinfo {author} {\bibfnamefont {B.}~\bibnamefont {Wetzel}}, \bibinfo {author} {\bibfnamefont {M.}~\bibnamefont {Erkintalo}}, \bibinfo {author} {\bibfnamefont {G.}~\bibnamefont {Genty}}, \bibinfo {author} {\bibfnamefont {K.}~\bibnamefont {Hammani}}, \bibinfo {author} {\bibfnamefont {B.}~\bibnamefont {Kibler}}, \bibinfo {author} {\bibfnamefont {J.}~\bibnamefont {Fatome}}, \bibinfo {author} {\bibfnamefont {C.}~\bibnamefont {Finot}}, \bibinfo {author} {\bibfnamefont {F.}~\bibnamefont {Dias}}, \bibinfo {author} {\bibfnamefont {N.}~\bibnamefont {Akhmediev}}, \bibinfo {author} {\bibfnamefont {G.}~\bibnamefont {Millot}},  \emph {et~al.},\ }\href@noop {} {\bibfield  {journal} {\bibinfo  {journal} {SPIE Newsroom}\ }\textbf {\bibinfo {volume} {5}} (\bibinfo {year} {2011})}\BibitemShut {NoStop}%
\bibitem [{\citenamefont {Toenger}\ \emph {et~al.}(2015)\citenamefont {Toenger}, \citenamefont {Godin}, \citenamefont {Billet}, \citenamefont {Dias}, \citenamefont {Erkintalo}, \citenamefont {Genty},\ and\ \citenamefont {Dudley}}]{toenger2015emergent}%
  \BibitemOpen
  \bibfield  {author} {\bibinfo {author} {\bibfnamefont {S.}~\bibnamefont {Toenger}}, \bibinfo {author} {\bibfnamefont {T.}~\bibnamefont {Godin}}, \bibinfo {author} {\bibfnamefont {C.}~\bibnamefont {Billet}}, \bibinfo {author} {\bibfnamefont {F.}~\bibnamefont {Dias}}, \bibinfo {author} {\bibfnamefont {M.}~\bibnamefont {Erkintalo}}, \bibinfo {author} {\bibfnamefont {G.}~\bibnamefont {Genty}}, \ and\ \bibinfo {author} {\bibfnamefont {J.~M.}\ \bibnamefont {Dudley}},\ }\href@noop {} {\bibfield  {journal} {\bibinfo  {journal} {Scientific reports}\ }\textbf {\bibinfo {volume} {5}},\ \bibinfo {pages} {10380} (\bibinfo {year} {2015})}\BibitemShut {NoStop}%
\bibitem [{\citenamefont {Soto-Crespo}\ \emph {et~al.}(2016)\citenamefont {Soto-Crespo}, \citenamefont {Devine},\ and\ \citenamefont {Akhmediev}}]{soto2016integrable}%
  \BibitemOpen
  \bibfield  {author} {\bibinfo {author} {\bibfnamefont {J.~M.}\ \bibnamefont {Soto-Crespo}}, \bibinfo {author} {\bibfnamefont {N.}~\bibnamefont {Devine}}, \ and\ \bibinfo {author} {\bibfnamefont {N.}~\bibnamefont {Akhmediev}},\ }\href@noop {} {\bibfield  {journal} {\bibinfo  {journal} {Physical review letters}\ }\textbf {\bibinfo {volume} {116}},\ \bibinfo {pages} {103901} (\bibinfo {year} {2016})}\BibitemShut {NoStop}%
\bibitem [{\citenamefont {N{\"a}rhi}\ \emph {et~al.}(2016)\citenamefont {N{\"a}rhi}, \citenamefont {Wetzel}, \citenamefont {Billet}, \citenamefont {Toenger}, \citenamefont {Sylvestre}, \citenamefont {Merolla}, \citenamefont {Morandotti}, \citenamefont {Dias}, \citenamefont {Genty},\ and\ \citenamefont {Dudley}}]{narhi2016real}%
  \BibitemOpen
  \bibfield  {author} {\bibinfo {author} {\bibfnamefont {M.}~\bibnamefont {N{\"a}rhi}}, \bibinfo {author} {\bibfnamefont {B.}~\bibnamefont {Wetzel}}, \bibinfo {author} {\bibfnamefont {C.}~\bibnamefont {Billet}}, \bibinfo {author} {\bibfnamefont {S.}~\bibnamefont {Toenger}}, \bibinfo {author} {\bibfnamefont {T.}~\bibnamefont {Sylvestre}}, \bibinfo {author} {\bibfnamefont {J.-M.}\ \bibnamefont {Merolla}}, \bibinfo {author} {\bibfnamefont {R.}~\bibnamefont {Morandotti}}, \bibinfo {author} {\bibfnamefont {F.}~\bibnamefont {Dias}}, \bibinfo {author} {\bibfnamefont {G.}~\bibnamefont {Genty}}, \ and\ \bibinfo {author} {\bibfnamefont {J.~M.}\ \bibnamefont {Dudley}},\ }\href@noop {} {\bibfield  {journal} {\bibinfo  {journal} {Nature communications}\ }\textbf {\bibinfo {volume} {7}},\ \bibinfo {pages} {13675} (\bibinfo {year} {2016})}\BibitemShut {NoStop}%
\bibitem [{\citenamefont {Suret}\ \emph {et~al.}(2016)\citenamefont {Suret}, \citenamefont {Koussaifi}, \citenamefont {Tikan}, \citenamefont {Evain}, \citenamefont {Randoux}, \citenamefont {Szwaj},\ and\ \citenamefont {Bielawski}}]{suret2016single}%
  \BibitemOpen
  \bibfield  {author} {\bibinfo {author} {\bibfnamefont {P.}~\bibnamefont {Suret}}, \bibinfo {author} {\bibfnamefont {R.~E.}\ \bibnamefont {Koussaifi}}, \bibinfo {author} {\bibfnamefont {A.}~\bibnamefont {Tikan}}, \bibinfo {author} {\bibfnamefont {C.}~\bibnamefont {Evain}}, \bibinfo {author} {\bibfnamefont {S.}~\bibnamefont {Randoux}}, \bibinfo {author} {\bibfnamefont {C.}~\bibnamefont {Szwaj}}, \ and\ \bibinfo {author} {\bibfnamefont {S.}~\bibnamefont {Bielawski}},\ }\href@noop {} {\bibfield  {journal} {\bibinfo  {journal} {Nature communications}\ }\textbf {\bibinfo {volume} {7}},\ \bibinfo {pages} {13136} (\bibinfo {year} {2016})}\BibitemShut {NoStop}%
\bibitem [{\citenamefont {Chabchoub}\ \emph {et~al.}(2017)\citenamefont {Chabchoub}, \citenamefont {Genty}, \citenamefont {Dudley}, \citenamefont {Kibler},\ and\ \citenamefont {Waseda}}]{chabchoub2017experiments}%
  \BibitemOpen
  \bibfield  {author} {\bibinfo {author} {\bibfnamefont {A.}~\bibnamefont {Chabchoub}}, \bibinfo {author} {\bibfnamefont {G.}~\bibnamefont {Genty}}, \bibinfo {author} {\bibfnamefont {J.~M.}\ \bibnamefont {Dudley}}, \bibinfo {author} {\bibfnamefont {B.}~\bibnamefont {Kibler}}, \ and\ \bibinfo {author} {\bibfnamefont {T.}~\bibnamefont {Waseda}},\ }in\ \href@noop {} {\emph {\bibinfo {booktitle} {ISOPE International Ocean and Polar Engineering Conference}}}\ (\bibinfo {organization} {ISOPE},\ \bibinfo {year} {2017})\ pp.\ \bibinfo {pages} {ISOPE--I}\BibitemShut {NoStop}%
\bibitem [{\citenamefont {Waseda}\ \emph {et~al.}(2019)\citenamefont {Waseda}, \citenamefont {Fujimoto},\ and\ \citenamefont {Chabchoub}}]{waseda2019asymmetric}%
  \BibitemOpen
  \bibfield  {author} {\bibinfo {author} {\bibfnamefont {T.}~\bibnamefont {Waseda}}, \bibinfo {author} {\bibfnamefont {W.}~\bibnamefont {Fujimoto}}, \ and\ \bibinfo {author} {\bibfnamefont {A.}~\bibnamefont {Chabchoub}},\ }\href@noop {} {\bibfield  {journal} {\bibinfo  {journal} {Fluids}\ }\textbf {\bibinfo {volume} {4}},\ \bibinfo {pages} {84} (\bibinfo {year} {2019})}\BibitemShut {NoStop}%
\bibitem [{\citenamefont {Chabchoub}\ \emph {et~al.}(2016)\citenamefont {Chabchoub}, \citenamefont {Onorato},\ and\ \citenamefont {Akhmediev}}]{chabchoub2016hydrodynamic}%
  \BibitemOpen
  \bibfield  {author} {\bibinfo {author} {\bibfnamefont {A.}~\bibnamefont {Chabchoub}}, \bibinfo {author} {\bibfnamefont {M.}~\bibnamefont {Onorato}}, \ and\ \bibinfo {author} {\bibfnamefont {N.}~\bibnamefont {Akhmediev}},\ }in\ \href@noop {} {\emph {\bibinfo {booktitle} {Rogue and Shock Waves in Nonlinear Dispersive Media}}}\ (\bibinfo  {publisher} {Springer},\ \bibinfo {year} {2016})\ pp.\ \bibinfo {pages} {55--87}\BibitemShut {NoStop}%
\bibitem [{\citenamefont {Chabchoub}(2016)}]{chabchoub2016tracking}%
  \BibitemOpen
  \bibfield  {author} {\bibinfo {author} {\bibfnamefont {A.}~\bibnamefont {Chabchoub}},\ }\href@noop {} {\bibfield  {journal} {\bibinfo  {journal} {Physical review letters}\ }\textbf {\bibinfo {volume} {117}},\ \bibinfo {pages} {144103} (\bibinfo {year} {2016})}\BibitemShut {NoStop}%
\bibitem [{\citenamefont {Waseda}\ \emph {et~al.}(2021)\citenamefont {Waseda}, \citenamefont {Watanabe}, \citenamefont {Fujimoto}, \citenamefont {Nose}, \citenamefont {Kodaira},\ and\ \citenamefont {Chabchoub}}]{waseda2021directional}%
  \BibitemOpen
  \bibfield  {author} {\bibinfo {author} {\bibfnamefont {T.}~\bibnamefont {Waseda}}, \bibinfo {author} {\bibfnamefont {S.}~\bibnamefont {Watanabe}}, \bibinfo {author} {\bibfnamefont {W.}~\bibnamefont {Fujimoto}}, \bibinfo {author} {\bibfnamefont {T.}~\bibnamefont {Nose}}, \bibinfo {author} {\bibfnamefont {T.}~\bibnamefont {Kodaira}}, \ and\ \bibinfo {author} {\bibfnamefont {A.}~\bibnamefont {Chabchoub}},\ }\href@noop {} {\bibfield  {journal} {\bibinfo  {journal} {Frontiers in Physics}\ }\textbf {\bibinfo {volume} {9}},\ \bibinfo {pages} {622303} (\bibinfo {year} {2021})}\BibitemShut {NoStop}%
\bibitem [{\citenamefont {Suret}\ \emph {et~al.}(2024)\citenamefont {Suret}, \citenamefont {Randoux}, \citenamefont {Gelash}, \citenamefont {Agafontsev}, \citenamefont {Doyon},\ and\ \citenamefont {El}}]{Suret_solitongas}%
  \BibitemOpen
  \bibfield  {author} {\bibinfo {author} {\bibfnamefont {P.}~\bibnamefont {Suret}}, \bibinfo {author} {\bibfnamefont {S.}~\bibnamefont {Randoux}}, \bibinfo {author} {\bibfnamefont {A.}~\bibnamefont {Gelash}}, \bibinfo {author} {\bibfnamefont {D.}~\bibnamefont {Agafontsev}}, \bibinfo {author} {\bibfnamefont {B.}~\bibnamefont {Doyon}}, \ and\ \bibinfo {author} {\bibfnamefont {G.}~\bibnamefont {El}},\ }\href {\doibase 10.1103/PhysRevE.109.061001} {\bibfield  {journal} {\bibinfo  {journal} {Phys. Rev. E}\ }\textbf {\bibinfo {volume} {109}},\ \bibinfo {pages} {061001} (\bibinfo {year} {2024})}\BibitemShut {NoStop}%
\bibitem [{\citenamefont {Chomaz}\ \emph {et~al.}(2022)\citenamefont {Chomaz}, \citenamefont {Ferrier-Barbut}, \citenamefont {Ferlaino}, \citenamefont {Laburthe-Tolra}, \citenamefont {Lev},\ and\ \citenamefont {Pfau}}]{chomaz2022dipolar}%
  \BibitemOpen
  \bibfield  {author} {\bibinfo {author} {\bibfnamefont {L.}~\bibnamefont {Chomaz}}, \bibinfo {author} {\bibfnamefont {I.}~\bibnamefont {Ferrier-Barbut}}, \bibinfo {author} {\bibfnamefont {F.}~\bibnamefont {Ferlaino}}, \bibinfo {author} {\bibfnamefont {B.}~\bibnamefont {Laburthe-Tolra}}, \bibinfo {author} {\bibfnamefont {B.~L.}\ \bibnamefont {Lev}}, \ and\ \bibinfo {author} {\bibfnamefont {T.}~\bibnamefont {Pfau}},\ }\href@noop {} {\bibfield  {journal} {\bibinfo  {journal} {Rep. Progr. Phys.}\ }\textbf {\bibinfo {volume} {86}},\ \bibinfo {pages} {026401} (\bibinfo {year} {2022})}\BibitemShut {NoStop}%
\bibitem [{\citenamefont {Cao}\ \emph {et~al.}(2017)\citenamefont {Cao}, \citenamefont {Bolsinger}, \citenamefont {Mistakidis}, \citenamefont {Koutentakis}, \citenamefont {Kr{\"o}nke}, \citenamefont {Schurer},\ and\ \citenamefont {Schmelcher}}]{cao2017unified}%
  \BibitemOpen
  \bibfield  {author} {\bibinfo {author} {\bibfnamefont {L.}~\bibnamefont {Cao}}, \bibinfo {author} {\bibfnamefont {V.}~\bibnamefont {Bolsinger}}, \bibinfo {author} {\bibfnamefont {S.~I.}\ \bibnamefont {Mistakidis}}, \bibinfo {author} {\bibfnamefont {G.}~\bibnamefont {Koutentakis}}, \bibinfo {author} {\bibfnamefont {S.}~\bibnamefont {Kr{\"o}nke}}, \bibinfo {author} {\bibfnamefont {J.}~\bibnamefont {Schurer}}, \ and\ \bibinfo {author} {\bibfnamefont {P.}~\bibnamefont {Schmelcher}},\ }\href@noop {} {\bibfield  {journal} {\bibinfo  {journal} {J. Chem. Phys.}\ }\textbf {\bibinfo {volume} {147}} (\bibinfo {year} {2017})}\BibitemShut {NoStop}%
\bibitem [{\citenamefont {Mistakidis}\ \emph {et~al.}(2023)\citenamefont {Mistakidis}, \citenamefont {Volosniev}, \citenamefont {Barfknecht}, \citenamefont {Fogarty}, \citenamefont {Busch}, \citenamefont {Foerster}, \citenamefont {Schmelcher},\ and\ \citenamefont {Zinner}}]{mistakidis2023few}%
  \BibitemOpen
  \bibfield  {author} {\bibinfo {author} {\bibfnamefont {S.~I.}\ \bibnamefont {Mistakidis}}, \bibinfo {author} {\bibfnamefont {A.}~\bibnamefont {Volosniev}}, \bibinfo {author} {\bibfnamefont {R.}~\bibnamefont {Barfknecht}}, \bibinfo {author} {\bibfnamefont {T.}~\bibnamefont {Fogarty}}, \bibinfo {author} {\bibfnamefont {T.}~\bibnamefont {Busch}}, \bibinfo {author} {\bibfnamefont {A.}~\bibnamefont {Foerster}}, \bibinfo {author} {\bibfnamefont {P.}~\bibnamefont {Schmelcher}}, \ and\ \bibinfo {author} {\bibfnamefont {N.}~\bibnamefont {Zinner}},\ }\href@noop {} {\bibfield  {journal} {\bibinfo  {journal} {Phys. Rep.}\ }\textbf {\bibinfo {volume} {1042}},\ \bibinfo {pages} {1} (\bibinfo {year} {2023})}\BibitemShut {NoStop}%
\bibitem [{\citenamefont {Diplaris}\ \emph {et~al.}(2025)\citenamefont {Diplaris}, \citenamefont {Bougas}, \citenamefont {Kevrekidis}, \citenamefont {Hung}, \citenamefont {Schmelcher},\ and\ \citenamefont {Mistakidis}}]{diplaris2025correlatedmanybodyquantumdynamics}%
  \BibitemOpen
  \bibfield  {author} {\bibinfo {author} {\bibfnamefont {D.}~\bibnamefont {Diplaris}}, \bibinfo {author} {\bibfnamefont {G.~A.}\ \bibnamefont {Bougas}}, \bibinfo {author} {\bibfnamefont {P.~G.}\ \bibnamefont {Kevrekidis}}, \bibinfo {author} {\bibfnamefont {C.~L.}\ \bibnamefont {Hung}}, \bibinfo {author} {\bibfnamefont {P.}~\bibnamefont {Schmelcher}}, \ and\ \bibinfo {author} {\bibfnamefont {S.~I.}\ \bibnamefont {Mistakidis}},\ }\href {https://arxiv.org/abs/2512.16031} {\enquote {\bibinfo {title} {Correlated many-body quantum dynamics of the peregrine soliton},}\ } (\bibinfo {year} {2025}),\ \Eprint {http://arxiv.org/abs/2512.16031} {arXiv:2512.16031 [cond-mat.quant-gas]} \BibitemShut {NoStop}%
\end{thebibliography}%

\end{document}